\documentclass[12pt,preprint]{aastex}
\slugcomment{Accepted and scheduled for publication  
in {\it The Astrophysical Journal},  for  the March 20, 2013, v 766, 1st issue}
\def\lax {\ifmmode{_<\atop^{\sim}}\else{${_<\atop^{\sim}}$}\fi}  
\def\gax {\ifmmode{_>\atop^{\sim}}\else{${_>\atop^{\sim}}$}\fi}  
\def\gtorder{\mathrel{\raise.3ex\hbox{$>$}\mkern-14mu
             \lower0.6ex\hbox{$\sim$}}}

\def\qcor{Q_{\rm cor}}

\def\qd{Q_{\rm disk}}

\def\cm2{cm$^{-2}$}
\def\s1{s$^{-1}$}

\def\sax{{\it BeppoSAX}}

\def\kte{kT_{\rm e}}

\begin{document}

\title{Stability of the photon indices in Z-source GX~340+0 for spectral states 
%transitions
}
% Observational Evidence for Neutron Star in GX~340+0

%\title{The stability of spectral index of the ``hard component'' as a function of mass accretion rate in Z-source GX~340+0}
% Observational Evidence for Neutron Star in GX~340+0

%\title{On the Constancy of the Photon Index of  X-ray spectra of 4U~1728-34 through all spectral states} 
%during outburst transitions}

\author{Elena Seifina\altaffilmark{1}, Lev Titarchuk\altaffilmark{2}  \& Filippo Frontera\altaffilmark{3} }
\altaffiltext{1}{Moscow State University/Sternberg Astronomical Institute, Universitetsky 
Prospect 13, Moscow, 119992, Russia; seif@sai.msu.ru}
\altaffiltext{2}{Dipartimento di Fisica, Universit\`a di Ferrara, Via Saragat 1, I-44122 Ferrara, Italy, email:titarchuk@fe.infn.it; 
%ICRANET, Piazza della Repubblica 10-12 65122 Pescara,  Italy; 
George Mason University Fairfax, VA 22030;   
Goddard Space Flight Center, NASA,  code 663, Greenbelt  
MD 20770, USA; email:lev@milkyway.gsfc.nasa.gov, USA}
\altaffiltext{3}{Dipartimento di Fisica, Universit\`a di Ferrara, Via Saragat 1, I-44122  Ferrara, Italy, email:frontera@fe.infn.it
}

\begin{abstract}
We show an analysis of the spectral and timing properties of  X-ray radiation from 
{\it Z-}source %Neutron Star X-ray binary 
GX~340+0 %(4U~???) 
during  its evolutions 
%transitions
 %along Z-track % three-branche track (on CCD)
%between {\it Horizontal} (HB)  and {\it Flaring} (FB) branches 
when the electron temperature %$kT_e$ 
of the  {\it Transition Layer} (TL) $kT_e$ monotonically decreases from 21 to 3 keV. 
We analyze 
%transition 
episodes observed  %all observations % number of  transition episodes 
%from this source obtaine
with the {\it Beppo}SAX  and {\it Rossi} X-ray Timing Explorer ({\it RXTE}).
%({\it RXTE}) 
 We reveal  that the X-ray broad-band energy 
spectra during all %these 
spectral  states 
can be 
%adequately 
reproduced by  a {physical} model, composed % sum % composition 
of  a 
%low-temperature 
soft $Blackbody$ component, 
%a Comptonized component ({\it COMPTB}) and  {\it Gaussian} component. 
% by 
{two Comptonized %({\it COMPTB}) 
components (both  due to the presence of that TL that up-scatters both seed photons of $T_{s1}\lax$1 keV coming from the disk (first component {\it Comptb1}), 
% made of ``seed'' photon temperature $T_{s1}\lax$1 keV], 
%which is presumably related to the TL , %(CC),   
and seed photons of temperature $T_{s2}\lax$1.5 keV coming from the neutron star (second component {\it Comptb2})} 
% $T_{s2}\sim$1.5 keV], which is associated with  the TL 
% inner part  (boundary layer),   
%a $hard$ component ({\it Comptb1}, photon index $\Gamma_1\approx$2) 
%with ``seed'' photon temperature $T_{s1}$=1.2 keV %with turnover at high energies 
%and {\it soft thermal} component ({\it Comptb2},  $\Gamma_2=$1.7 -- 3.1) 
%with characteristic color temperature $\sim$2 keV, 
and the iron-line ({\it Gaussian}) component.
Spectral analysis using this model indicates  that the photon power-law indices $\Gamma_{com1}$ and $\Gamma_{com2}$ 
of the Comptonized components are almost constant, $\Gamma_{com1}$ and $\Gamma_{com2}\sim 2$
%1.99$\pm$0.02, $\Gamma_2$=2.00$\pm$0.02) 
when 
%the TL electron temperature 
$kT_e$ 
%of {\bb TL} %Compton cloud (CC) 
changes from 3 to 21 keV %during %these 
along {\it Z-}track.
We %also 
interpret the detected quasi-stability of the indices 
%$\Gamma$ %$\Gamma_1$ and $\Gamma_2$ of  both 
of Comptonized components 
near a value of 2. %{\it Comptb1/Comptb2})
%in the framework of 
%using a model in which the spectrum consists 
%is determined by 
%of the  thermal Comptonizion 
%components formed  in the transition layer and its inner part.
%very  boundary layer.
%the {\it neutron star} surface. %in the transition layer (TL),  when the  energy release in the TL is much 
%higher than the flux  coming to the TL from the accretion disk.  
%The Comptonized emission from NS surface 
%is due to Comptonized component of CC and has the same spectral characterictics, except ``seed'' photon temperature. 
% located between the  accretion disk and NS %neutron star surface. 
%The index quasi-stability takes place when the  energy release in the TL is much higher than the flux  coming to the TL 
%from the accretion disk. 
Furthermore, this index stability 
%effect 
now found for the Comptonized spectral components  
%({\it COMPTB1}) 
of $\it Z-$source GX~340+0 
%during state %spectral 
%evolution of the source 
%from the low to high luminosity states 
is similar to {that}  was previously established  in  the  $atoll$ sources 4U~1728-34, GX~3+1 and  %suggested 
early proposed for a 
number  of 
%other 
%low mass 
X-ray neutron stars (NSs). 
%LMXB NS 
%binaries  
%(see Farinelli \& Titarchuk).
 This   
 %intrinsic 
 behavior of  NSs 
 %neutron stars, 
 both for $atoll$ and {\it Z-}sources,  is essentially 
 %fundamentally  
different   from that observed 
% seen 
in black hole binaries  where $\Gamma_{com}$ 
%the photon (spectral) index
% monotonically 
increases during 
a spectral  evolution 
%transition 
from the low state
% state  
to the high state and ultimately 
%then   
saturates at high 
%values 
 mass accretion rate.
\end{abstract}

\keywords{accretion, accretion disks---stars: neutron---black hole physics---stars: individual (GX 340+0):radiation mechanisms: nonthermal---physical data and processes}

\section{Introduction}

The  study of binaries hosting %properties of 
%similarities and the differences between
neutron stars (NSs) %in binaries
%compact objects within Neuntron star (NS) class 
allows one to investigate %matter 
%both 
properties of compact objects themselves and %under binary conditions at the late evolution stages and 
to better understand the differences between black hole (BH) and NS systems.
% as well as 
%between (for) $atoll$ and $Z$ type sources among NSs are discussed on many aspects (see for example, TS05, FT11, 
%Di Salvo et al. (2006), ST11).
Low-mass X-ray binaries (LMXBs) including  %hosting involving
weakly magnetized neutron stars may be observationally 
divided into two main sub-classes, so called {\it Z} and $atoll$ sources, based upon correlations between their 
spectral and timing properties during state evolution 
%(transitions) 
in X-ray wavelengths (Hasinger \& van der Klis 1989). 
%As a first trial 
The similarity and differences between $atoll$ and {\it Z} sources are clear visible 
on the {\it color-color diagram} (CCD) or {\it hardness-intensity diagram} (HID) 
%as well as %but 
%it can be seen in 
and determined by their 
%X-ray 
timing properties.

The {\it Z-}shaped track
% traced out 
shown by {\it Z-}sources in the 
%X-ray 
CCD or HID 
%{\it color-color diagram} (CCD) or {\it hardness-intensity diagram} (HID) 
%is divided 
consist of 
%into 
three branches: the {\it horizontal}, 
%branch 
%(HB), 
the {\it normal} 
%branch 
%(NB), 
and the {\it flaring} branches (HB, NB and  FB respectively). 
%It is worth  noting that
Note  the aforementioned names are merely historical 
[see e.g. \cite{vdKl89}] % van der Klis 1989] 
and were based on the fact that, in the first few sources studied, the horizontal branch 
%showed up 
appeared as a horizontally oriented limb in the HIDs, the normal branch as a state in which 
the sources 
% spend 
pass  most of their time, while in the flaring branch the sources demonstrated intensive 
flares. In turn, $atoll$ sources showed two kinds of branches:  a {\it banana} shaped branch and  
 an {\it island} state. %, in which the colours low change on a time scale of hours to days. 
Note that %atoll sources usially trace ``banana'' branch 
$atoll$ sources usually  
show 
%trace 
their  
%own 
pattern over a larger extent of luminosity and on 
%much 
longer time scales 
(weeks - month) than that of  {\it Z-}sources (hours - weeks). 
% While However, 
$Atoll$ sources always 
%trace 
 follow {\it banana}  branch  
on the same short time scale as {\it Z-}sources, hours - weeks \citep{muno02}.

The occurrence of the different X-ray states of the {\it Z-}sources and the {\it atoll} sources is presumably 
governed by  mass accretion rate in their innermost parts of the source. In the {\it Z-}sources this mass accretion rate increases from  
HB, through  NB, to  FB (Hasinger et al. 1990, % \cite{hasinger90}
 Vrtilek et al. 1990). On the other hand 
 %while 
 in the {\it atoll} sources 
 mass accretion rate increases from the  {\it island} state, through the lower {\it banana} to the upper 
{\it banana} (see e.g. Corbet et al. 1989; Di Salvo et al., 2001). 
%It was shown by 
Hasinger \& van der Klis (1989) showed that  each 
state 
%or branch 
has a characteristic 
fast timing variability, and therefore an extra diagnostic is % need it
{ available} to identify a source state, %(see e.g. Fig.~ N2), 
and to distinguish the {\it Z-}sources from the {\it atoll} sources. In fact, only a combined diagnostic 
including fast timing and spectral analysis (such as CCDs) can provide reliable identification 
of the source ({\it Z} or $atoll$) type . 
%it was only by using 
%this extra diagnostic in conjunction with the CDs that Hasinger \& van der Klis (1989) were able 
%too sort out the differences between Z and atoll sources and their various states; from just the 
%spectral data, without the fast timing, this was not possible (see e.g. Schulz et al. 1989). 

The  division of the bright LMXBs into {\it Z} and {\it atoll} sources was proposed because of 
%to be due to 
differences  in both  magnetic fields and  mass accretion rate (Hasinger \& van der Klis 1989). In fact,  
for {\it Z-}sources %are thought to have somewhat 
higher magnetic fields were suggested  ($10^9-10^{10}$ G) than that of  {\it atoll} sources 
($10^8-10^9$ G) and they probably accrete  at near-Eddington rates (e.g. Lamb 1989), while  
mass accretion rate $\dot M$ is well below Eddington rates in $atoll$ sources. 
This conjecture on the magnetic field is quite old-dated and no evidence has ever been given. 
Furthermore  {\it the discovery of the first transient Z source, turning to an} {\it atoll} 
{\it makes the B-field discriminant very unlikely} [see \cite{hom07}; \cite{lin09}].
A study of broad band 
spectral and  timing behavior turns to be  
%out to be 
a powerful  approach 
%tool 
to probe the 
%nature 
Phyisics of 
the brightest LMXBs. This study  is also used to invesigate the behavior of  other X-ray binaries, 
e.g.   black hole binaries (see e.g. van der Klis 1994).  An overall comparison of these various 
types of sources shows that the properties of the rapid fluctuations appear to depend on 
$\dot M$ in the innermost part of the source
% mass accretion rate 
and a compact object  type 
%of the 
(BH or NS).

Usually BH and NS sources demonstrate state transitions when the soft X-ray flux monotonically increases (or decreases)
%along the track (in the direction LHS$\to$IS$\to$HSS/$Island \to Banana$ states/HB$\to$NB$\to$FB for BH/$atoll$/Z sources, respectively), 
which  implies monotonic changes of $\dot M$ [e.g., \cite{tsei09}, \cite{ST11} and below].
%  along corresponding track.  
However, 
%for {\it Z} sources 
the total  X-ray flux is not at all times
%always 
a reliable indicator for  $\dot M$ 
%for the mass accretion rate 
(e.g., Mendez et al. 1999; %Mendez 2000; 
Ford et al. 2000).
% which fundamentally 
%determines the magnetospheric radius of the neutron star and thus the location of the inner edge 
%of the accretion disk. 
%Moreover, it is now well established that %the frequency of kHz QPOs is not 
%well correlated with X-ray flux over a long period of time, although short-term linear correlations 
%seem to be present (Zhang et al. 1998a; Mendez 2000 and references therein). 

A {\it Z-}type 
%of NS XRBs 
represents a category of seven Galactic LMXBs: %low-mass XRBs: 
Sco~X-1, GX~17+2, GX~349+2, Cyg~X-2, 
GX~5-1, GX~340+0 and XTE J1701-462 
 (with an addition of Cir~X-1 which is 
 %considered to be 
 a {\it peculiar} {\it Z-}source, see Shirey et al., 1999)  and the extra-galactic {\it Z}-source LMC X-2.
It has been suggested that there are two types of {\it Z-}sources, characterized by differences in 
the shape of their {\it Z-}patterns and their power spectral characteristics (Hasinger \& van der Klis 
1989, see also Penninx et al. 1991). 
%%Cyg~X-2, GX~5-1 and GX~340+0 have well developed {\it horizontal}  branches (HBs) 
%%with a strong HB  QPO (or  horizontal branch oscillations, HBO) and red low frequency noise (LFN), whereas Sco~X-1, GX~17+2 and GX~349+2 have small, vertical or absent HBs with a weak 
%%or absent HBO and peaked LFN. 
%Another difference between these two groups is the FB orientation: 
%of FB
%sources in the latter group have a well developed FB on which the intensity really 
%flares, whereas in the two sources of the former group 
%where FB had been seen, Cyg~X-2 and 
%GX~340+0,  this FB is oriented in different ways, and sometimes shows dips in the intensity 
%rather than flares. 

% -=========
%In turn, GX~340+0 is extensively investigated for more than forty years during many multiwave  
%campaigns, but up to now stays not well known. 
%a bright low-mass X-ray binary (LMXB) and a Z source~[\cite{hasinger89}].
%For more than fourty years since its discovery     inspite of extensive multivawelengh (including radio, 
%optical and X-ray domains) GX~340+03 ($\equiv$ Ara~XR-1) stays not well known. %(Hasinger \& van der Klis 1989). 
%Observations of GX~340+03 ($\equiv$ Ara~XR-1) have been carry out in all domains (X-ray, optical, NIR, UV and radio) for 
%more than fourty years since its discovery. % in X-ray~\cite{margon71}. 
%For the first time 
GX~340+0 was  discovered 
as a Galactic X-ray source by Friedman, Byram \& Chubb (1967) and later its position was found %determined 
by  Rappaport et. al (1971).
% with a precision of 1 arcmin. 
This source  was preliminary determined 
%identified  
as a NS binary 
%system by
by  Margon et al. (1971) using 
%on the basis of 
an $Aerobee$ rocket data. % carried out from an $Aerobee$ rocket. 
 $Ariel$ (Ponman, 1982), $EXOSAT$ (van Paradijs et al., 1988), HEAO-2 (Schulz et al. 1989) orbital 
observatories  provided unique high quality X-ray data of GX~340+0 to indicate three main tracks  
with characteristic correlation of hardness ratio, which  formed {\it Z-}pattern in CCD.
% $Ariel$ V data of 
%GX~340+0 showed a positive correlation of spectral hardness with intensity (Ponman, 1982), while the CD and 
%the hardness-intensity diagram of the source, obtained from $EXOSAT$ data (van Paradijs et al., 1988), 
%indicated 
%that GX~340+0 has a positive correlation between the hardness ratio (HR), defined as [6-20 keV]/[3-6 keV], and 
%the intensity in the 1-20 keV energy range when HR$<$0.4, while at HR$\sim$0.4 the intensity reduced with no 
%changes in HR (see Figs. 3c and 4c in Schulz et al., 1989). The change of correlation between the intensity 
%and HR happened roughly at the highest hardness values indicating a hard apex. Schulz et al. (1989) noted 
%that, in analogy to Cyg X-2, the data formed a horizontal branch (HB)-normal branch (NB) pattern. Taking also 
%in account that Garcia
%(1987) found indications of a third spectral branch at the soft end of the NB in an analysis of $HEAO-2$
%MPC data, Schulz et al. (1989) suggested that GX~340+0 showed a Z-shaped spectral variation pattern.
Despite of 
%Although 
no burst has yet been found,  
%observed, 
the very high values of QPO ($\gax 800$ Hz) 
%a correlation
%fa timing variability 
%(at more than 800 Hz),
%of the 
%correlated 
%power spectrum 
and X-ray energy spectrum
% behavior of the source 
(van Paradijs et al., 1988) and 
as well as the similarities with the different {\it Z-}sources 
%indicated that the compact object  
%The correlated power spectral and X-ray energy spectral behavior of the source, described by van Paradijs et al. (1988), 
allowed to classify GX~340+0 as a {\it Z-}source (Hasinger \& Van der Klis, 1989).

The spectral characteristics of this  source have not been fully studied 
%investigated 
up to now.
%so far. 
Schulz \& Wijers (1993), hereafter SW93, carried out a %unique 
spectral analysis of the source 
%studied the 2-12 keV spectrum using EXOSAT data: the spectrum could be well described 
%by a single component due to thermal Comptonization of soft photons, emerging from the NS surface, in a hot corona of moderately optical thickness (?  5-6).
%A unique spectral analysis of the source was carried out by Schulz et al. (1993) 
based on 
%using 
$EXOSAT$ data in the 2 -- 12 keV energy range. They found that the 
spectrum can be  fit by model which a sum of a blackbody of a temperature of 0.8 keV and  %Comptonized 
a  thermal Comptonization  component of soft NS seed photons.
% which emerge from the NS surface,
 SW93 also obtained an optical depth  of $\tau\sim$~5 -- 6 and 
an electron temperature of the hot corona changed  from 4 to 6 keV.
% and characterized by an absorption equivalent hydrogen column density
%of (4 -- 5)$\times$10$^{22}$ cm$^{-2}$. 
%It is worth noting that  %the residuals (see Fig. 4 in 
%Schulz \& Wijers  (1993) pointed out that %seemed to indicated that
%the spectrum of GX~340+0 is more complex for  the $normal$ branch (NB). 

%Penninx et al. (1993) identified the radio counterpart of 
%GX~340+0 with an accuracy of $\sim$3{\tt"}. Using the coordinates of the radio counterpart Miller et al. (1993) detected
%an infrared source having a magnitude of 17.3$\pm$0.5  in the K band as a possible infrared counterpart of
%GX~340+0. 
Fender
\& Hendry (2000) estimated the distance to the source of 11$\pm$3 kpc  using  the radio measurements .
%allowed to estimate (). 
Radio emission of GX~340+0 is highly variable, and probably  correlates  with the
 X-ray emission  on its 
 %when the source is in its 
horizontal branch.
 %$horizontal$ branch, 
 But the radio emission anti-correlates when   the source  is on the other  spectral branches 
(Oosterbroek et al. 1994).

Lavagetto et al. (2004) were the first 
who showed a broadband (0.1 -- 200 keV) spectrum of this source based  on the data using  a 
$Beppo$SAX observation on  August 9 -- 10, 2001 (thereafter Observation $S6$, see Table 1). They decomposed the spectrum  
of GX~340+0 %could be  
%into 
%a sum of 
 as 
 %an additive model of 
 a soft blackbody  of temperature $kT_{BB}\sim 
%thermal component of temperature 
0.5$ keV,   Comptonization component 
%an optically thick Comptonized component, 
and in addition an excess at photon energies more than 20 keV
was fit
% that they fitted 
using  an extended  
%simple 
power law  terminated by a cutoff at high energies. Ueda et al. (2005) studied a
% high-resolution 
spectrum of the source 
%was studied by , 
using  $Chandra$ observations. These $Chandra$ data indicated to 
%clearly showed 
the presence of an iron line emission. The shape of the line can be fit 
%fitted 
by  a %simple 
$Gaussian$ profile, which  center is at 6.6 keV and the line equivalent width is 40 eV.
Then, \cite{d'Ai09} detected a broad asymmetric emission line  in the Fe K$_{\alpha}$ energy band and tried to investigate the 
line evolution along HB of {\it Z-}track based on XMM-$Newton$ data.
%observations.
% of GX~340+0. 
However,  the line profile does not indicate to 
any 
%strong 
correlated variation although 
%despite in 
the continuum emission and luminosity underwent  significant changes.

%Lavagetto et al. (2004) and 
\cite{iaria06}, Iaria06 hereafter, also %Iaria et al. (2006) 
%studied 
examined the 0.1 -- 200 keV energy 
spectrum 
%of GX~340+0 
using $Beppo$SAX observations (thereafter Observations $S3-S6$, see Table 1) {and the Lavagetto model described above. 
%($S6$ and $S3$ -- $S5$ sets, respectively, in terms of 
%presend Paper listed in Table 1).  %carried out between 2001 August 9 and 10 (thereafter Observation 3). 
%Based on the Lavagetto model (Lavagetto et al. 2004),
For all observations of  different spectral states}, they found 
%obtained 
%an equivalent hydrogen column associated with the interstellar matter 
%of 
$N_H\sim 6.2\times 10^{22}$ cm$^{-2}$ and 
%found  
that the spectrum at 
%below 
less than 30 keV could be fit by a sum  a blackbody of 
%component 
%with a temperature of 
$kT_{BB}\sim$ 0.5 keV
%plus {the sum
 and  a Comptonization component $Comptt$ (Titarchuk, 1994) which  seed-photon temperature of $kT_s\sim$ 0.9 keV and
 plasma  temperature of $kT_e\sim$ 3 keV. In addition, Iaria06  used a power--law  component of 
   the photon index $\Gamma\sim 1.5$ and  a $Gaussian$ 
   %emission 
  line component  at 6.75 keV, related to Fe XXV.
They also found that a power-law component of  a photon index of 2.5  
 was required at higher energies.
%They revealed good agreement this model with  observations at the spectral index $\Gamma\sim 1.5$ for all observations of  different spectral states.

%Recently 
\cite{church06} 
investigated 
%studied 
the  {3--40 keV} 
%energy 
spectrum 
%of GX~340+0 
%using 
relied on  PCA/{\it RXTE} data 
($R3$ set in terms of our classification, 
%of the  present  paper 
see Table 2) and {applied a modified  Lavagetto model, with} a {\it cutoff power law}
% instead  of a simple power-law 
at  the energies above 30 keV. 
They showed %revealed 
a good agreement this model with {\it RXTE} observations at a stable value of the best-fit photon index $\Gamma\sim 1.7$ during all spectral states.
% transitions. %for all spectral states.

%Thus the broad-band spectra of GX~340+0 have been studed based on phenomenological model. 
%To investigate this source in a unified physical scenario for the less well studed energies above 20 keV...

Before the stability of the index at different luminosities was emphasized 
%noted 
 by \cite{ft11}
%Farinelli \& Titarchuk 
(hereafter FT11) for a
number of neutron star low mass X-ray binaries  (LMXBs). FT11 
%collected
gathered
X-ray {\it Beppo}SAX spectra 
%gained 
%obtained 
%by 
{and showed 
%demonstrated 
the
%relative 
stability of the  spectral index $\alpha$ at approximately 1 ($\Gamma=\alpha+1$) 
for many 
%quite a few 
NSs: 
%sources: 
X~1658-298, GX~354-0, GS~1826-238, 1E~1724-3045, Cyg~X-2, 
Sco~X-1, GX~17+2, and GX~349+2 at their different spectral states (or when the best-fit  electron temperatures for these sources 
are drastically different). 
%Recently Seifina \& Titarchuk (2011), (2012), hereafter ST11,  and ST12 respectively, presented
%results of their analysis of 
X-ray spectra 
for  {\it atoll} sources 4U 1728-34 and GX~3+1 detected with $Beppo$SAX and {\it RTXE}, respectively,  at different luminosities were studied in details by Seifina \& Titarchuk (2011), (2012), hereafter ST11,  and ST12 respectively.
%and values of the best-fit electron temperatures

 The ST11 and ST12 also pointed out 
%indicate 
that the value of $\Gamma$ 
%{photon index 
only varies slightly around  2},
% (or the photon index varies at about 2) 
independently of
%the electron temperature of the Compton cloud (CC) 
$kT_e$ and luminosity. This %unique 
stability of  $\Gamma$ 
%the photon index 
%may  
can be an universal 
%intrinsic 
signature 
%property 
for any kind NSs either the {\it atoll} or  {\it Z-}sources. 
It is presumably dictated
% determined 
by common Physics 
% physical conditions 
for this type of  
%class of 
sources. 
FT11, ST11 and ST12 demonstrated 
%interpreted 
this 
%quasi 
{\it stability} of 
%the index 
$\Gamma$ 
using 
%in the framework of 
a model in which
the emergent spectrum influenced  by a strong 
%{thermal 
Comptonizion component 
%formed 
is produced in the Transition Layer (TL) between NS surface and the {\it Keplerian} accretion disk.
%and .} 
In fact, {\it the  
%quasi-
stability of the index  occurs
when  $Q_{cor}$, which is the energy release in  TL, is much higher than $Q_{disk}$ which is the photon energy flux,  coming from the accretion disk and illuminating 
TL.}  

Namely, ST11 shows (see also FT11) that the TL optical depth 
%the transition layer (TL)  
 $\tau_0$, its electron temperature $kT_e$, the ratio 
 %of the soft photon flux 
 $Q_{disk}/Q_{cor}$ 
 %and the energy release  in the TL $Q_{cor}$ 
 are related to each other 
\begin{equation}
\frac{\kte \tau_0 (2+\tau_0)}{m_e c^2}=\frac{0.25}{1+\qd/\qcor},
\label{ktetau}
\end{equation}
 if one assumes  that  the 
 %energy 
 balance of energy  in the transition layer (TL) is established 
 %dictated 
 by 
 %Coulomb collisions with protons 
 gravitational energy release and cooling of electrons due to inverse Compton effect
and free-free emission 
%are the main cooling channels 
[see 
%a formulation of this problem in the pioneer work by 
\cite{zs69} 
%Zel'dovich \& Shakura 1969 
and later  a similar formulation of this problem 
%consideration 
in \cite{bisn80}].   
%Then one  should use a  formula  for  

The spectral index $\alpha$  of the Comptonization spectrum is determined by the formula 
\begin{equation}
\alpha=-\frac{3}{2}+\sqrt{\frac{9}{4}+\frac{\beta}{\Theta}}, 
\label{alpha_general}
\end{equation}
where the dimensionless plasma temperature $\Theta \equiv \kte/m_e c^2$ and  $\beta$-parameter
[see \cite{tl95}].
% (see also formula \ref{} here). 
%as it was assumed by the authors. 
If one replaces $\beta$ by  its diffusion approximation
%limit  
%$\beta_{\rm diff}$ 
\begin{equation}
\beta_{\rm diff}=\frac{1}{\tau_0 (2+\tau_0)}
\label{beta_diff}
\end{equation} 
and applies 
%using 
%(Eq. \ref{beta_diff}), and  
equation (\ref{ktetau}), we have  a formula for the diffusion spectral index $\alpha_{\rm diff}$ 
%as 
\begin{equation}
\alpha_{\rm diff}= -\frac{3}{2}+\sqrt{\frac{9}{4}+ \frac{1+\qd/\qcor}{0.25}},
\label{alpha_diff}
\end{equation}
%or $\alpha_{\rm diff}\approx1+0.8~ \qd/\qcor$  
and the photon index 
\begin{equation} 
\Gamma_{diff}\approx1+\alpha_{\rm diff}=2+0.8~\qd/\qcor
\label{gamma_diff}
\end{equation}
 for $\qd/\qcor<1$.
%  Thus, until 
Consequently  $\Gamma_{diff}\approx2$ as $\qd/\qcor\ll1$.

%The evolution of spectral parameters of compact objects in
%X-ray binaries is of great interest for understanding their nature.
Note, in contrast with the neutron star sources 
%It is well known that 
many black hole (BH) candidate binaries show a correlated behavior of the photon index $\Gamma$
%exhibit 
with 
%correlations between 
mass accretion rate $\dot M$ 
%and photon power-law index  
(see Shaposhnikov \& Titarchuk 2009 and
Titarchuk \& Seifina 2009, hereafter ST09 and TS09, respectively). 
%
%In the soft states of BHs these index-$\dot M$ correlations
%almost always show a saturation of $\Gamma$ at high values of %the $\dot M$.
%This saturation effect can be considered to be a BH signature %or
%equivalently as a signature of a converging flow into a BH 
%(see ST09
% and TS09).  The spectral index $\alpha$ 
%(equal to $\Gamma-1$) is inverse proportional to 
%Comptonizaion  parameter $Y$ which is a product of average 
%number of scattering  $N_{sc}$ and scatering efficiency  
%$\eta$  in the medium.
%  But in the case of the converging flow $N_{sc}$ is 
%proportional
  % to mass accretion rate  
%  and $\eta$ is  inverse proportional to 
% mass accretion rate $\dot M$ if that is much greater than 
%Eddington one $\dot M_{\rm Edd}$ and consequently $\alpha$ 
%should saturate when  $\dot M\gg \dot M_{\rm Edd}$.
%   Thus the question that
% naturally arises is how the spectral index behaves as a function
%of  mass accretion rate for the spectrum in {\it Z} (NS) sources.

%({\bf The following three paragraphs have been moved from the section on Timing properties. It seems to me that they should be in the introduction. I have also slightly revised the text.})

Previous investigations (see Penninx et al. 1991; Kuulkers \& van der Klis 1996; Jonker et al. 1998) show that %Generally, 
{\it Z-}sources, in particular GX~340+0, are characterized, in 
 HB or in 
an upper part of 
NB, by 
%quasi-periodic oscillations (
QPOs with frequencies  changing 
%varying 
from 20 to 50 Hz.
%: the horizontal branch quasi-periodic oscillations or HBOs
%Second harmonics of these HBOs were detected by Kuulkers \& van der Klis (1996) and
%Jonker et al. (1998) in the frequency range 73-76 Hz and 38-69 Hz, respectively. 
van Paradijs et al. (1988) revealed normal branch oscillations (NBOs) with a frequency of 5.6 Hz 
in the middle of NB.
 Jonker et al. (1998) found twin kHz QPOs 
%in GX~340+0, 
whose centroid frequencies are correlated 
with the HBO (HB oscillation) 
%peak 
frequency.
% in HB. 

%In turn, models describing the kHz 
%QPOs  predict QPOs 
%components 
%in the low-frequency part of the power spectrum  (PDS) and can
%be tested by investigating this low-frequency portion.  
Titarchuk \& Osherovich (2001)   
classified the power density spectrum features for each compact object class %(in particular, the Z source GX~340+0) 
using the {\it Transition Layer} model~[see Titarchuk et al. (1998)] and the basic properties of the 
accretion flow around compact object. They applied their analysis for the {\it RXTE} observations of 
%the {\it Z}-source 
GX~340+0.

It is interesting that other {\it Z}-sources  differ from GX~340+0 on the timing properties along {\it Z}-track. In particular, 
%As far another Z-sources, 
%Timing 1
Cyg~X-2 and GX~5-1 %and GX~340+0 
have well developed  HBs
%{\it horizontal}  branches (HBs) 
with a strong HBO  
%QPO (or  horizontal branch oscillations, HBO) 
and red low-frequency noise (LFN), 
whereas Sco~X-1, GX~17+2 and GX~349+2 have small, vertical or absent HBs with a weak 
or absent HBO and peaked LFN. However, in spite of various details, during {\it Z-}evolution %NB and HB states 
all these {\it Z-}sources indicate 6 -- 20 Hz N/FBO phenomenon and significant increasing VLFN component at FB.
Previous investigations detected no QPO in the FB of GX~340+0 (e.g., Penninx et al. 1991). 
This fact, along with the absence of  type-I X-ray bursts, make one  to think about  GX~340+0 as a mysterious 
object. However in this Paper we  reveal a  typical evolution of GX~340 with FBO events based on our investigation of {\it RXTE} data set,  which helps one to understand 
% to shed the light on 
the nature of FB in GX~340+0.

In this study 
%Paper 
we analyze 
% present 
%the analysis of 
the {\it Beppo}SAX  
%available 
observations 
%during  1998 -- 1999 years
 and  {\it RXTE} observations during  1997 -- 2009 years for 
 %{\it Z} source 
GX~340+0.  In \S 2 we show the list of observations for which we apply  our data analysis and   the details of X-ray spectral analysis are elaborated in \S 3.   We study  how 
X-ray spectral and timing  properties evolve during the different spectral states
% transition 
in \S 4.  We discuss our results  and make their comparisons with those known in the literature in \S 5. 
We  make our  final conclusions in  \S 6.

\section{Data Selection \label{data}}

We obtain broad band energy spectra of the source
%combining 
using  data of three {\it Beppo}SAX Narrow
Field Instruments (NFIs), namely the Low Energy Concentrator
Spectrometer (LECS) for 0.3 -- 4 keV, the Medium Energy Concentrator Spectrometer
(MECS)  for 1.8 -- 10 keV and the Phoswich Detection
System (PDS)
%[PDS; \citet{fron97}] 
for 15 -- 200 keV  [see \cite{parmar97}; \cite{boel97}; \cite{fron97} respectively]. 

 We use the SAXDAS data analysis package for  data processing. 
 We make 
% performed 
the data spectral analysis in the energy range 
for which  a response matrix of each of the instruments is well determined.  
% have been 
We renormalized the LECS data based on the MECS data. We treat  relative normalizations of the NFIs %were treated 
as free parameters when proceed with 
 model fitting, but  we fix 
 %except for 
 the MECS normalization
 % that was fixed 
 at a value
 of 1. 
 %We checked after that 
 %this  fitting procedure 
 We control if these normalizations
 are in a standard range for each
 instrument% (section 4.2 of Cookbook for the BeppoSAX NFI spectral analysis
\footnote{http://heasarc.nasa.gov/docs/sax/abc/saxabc/saxabc.html}.
% Specifically, LECS/MECS re-normalization ratio is 0.92 and PDS/MECS
% re-normalization ratio is 0.97. 
Furthermore,  we  rebinned   the spectra  
%accordingly to   energy resolution of the instruments 
in order to have 
%obtain
 % independent 
significant data points.  The LECS spectra are rebinned with a binning factor 
which is not constant over energy
(Sect.3.1.6 of Cookbook for the BeppoSAX NFI spectral analysis) implementing 
 rebinning template files
 in GRPPHA of
 XSPEC \footnote{http://heasarc.gsfc.nasa.gov/FTP/sax/cal/responses/grouping}. The PDS spectra are rebinned with a  linear binning
 factor 2,  namely we group two bins together leading to the resulting bin width of 1 keV.  We apply a systematic error of 1\%  to all of these spectra. 
We present a list of  the {\it Beppo}SAX observations implemented in our analysis in Table 1. 

 We   also use publicly available the {\it RXTE}  data sets 
 %observatory 
\citep{bradt93}  which were  obtained from April 1997 to March 2009. 
%in the  Paper. 
In total, they  include 92 observations taken at
different states of the source.
We apply standard tasks of the LHEASOFT/FTOOLS
5.3 software package 
%were utilized 
for data processing.
%For spectral analysis 
We use PCA {\it Standard 2} mode data, collected 
in the 3 -- 20~keV energy range and the most recent release of PCA response 
calibration (ftool pcarmf v11.1) for spectral analysis.
%processing 
We also apply standard dead time correction  to the data. 
%procedure 
%has been applied to the data. 
 %have also been used 
In order to construct broad-band spectra we include the data from HEXTE detectors.

A background corrected  in  off-source observations is subtracted from the data.
We use   only data  in the 20 -- 150~keV energy range
% for the spectral analysis 
in order to avoid 
%to account for 
the problems related to  the HEXTE response and 
background determination. %The HEXTE data have been re-normalized based on the PCA.
 We use the data which are available through the GSFC public archive 
(http://heasarc.gsfc.nasa.gov).  We present a full list  of  observations covering 
%a complete range of 
the source evolution during  different spectral  state events in  Table 2. 

%We have thus   made 
We accomplish an analysis of {\it RXTE} observations  
% of GX~340+0  
spanning twelve years 
for which  11 intervals indicated by  blue rectangles 
in Figure~1 in \cite{stf13} hereafter STF13.
%\ref{variability_97-09} ($top$).
%The {\it RXTE} %PCA 
We model the energy spectra 
%were modeled 
using XSPEC astrophysical fitting software and we apply 
%{\bf The HEXTE spectra were renormalized to the PCA data by adding to the other free parameters a normalization factor?????} 
%%%Spectral analysis was done using an approach similar to that adopted in ST11 for 4U~1728-34 data. 
systematic error of 0.5\% 
%have been applied 
to {all} of the analyzed spectra. 
It is  also worth noting that we use    
 public GX~340+0 data from the  All-Sky Monitor (ASM/{\it RXTE})  which 
%on-board \textit{RXTE} 
show long-term quasi-constancy behavior of mean soft flux 
during all observation scans. % $\sim$ six years cycle (Fig.~\ref{variability 96-10}). 
%In the following 
%We  use definitions of the  $fainter$ %low 
%and $brighter$ %high 
%on luminosity phases %states 
%to relate these phases %states 
%to the source luminosity and we demonstrate that during the bright/faint phase %low-high state 
%transition of GX~3+1  COMPTB Normalization
%%the electron temperature of Compton cloud 
%changes from 
%%2.3 keV to 15 keV 
%0.04 to 0.14 $L_{39}^{soft}/{D^2}_{10}$
%% erg/s/kpc$^2$ 
%where  $L_{39}^{soft}$ is the soft photon luminosity in units of $10^{39}$ erg/s  and  $D_{10}$ in units of  10 kpc  is distance to the source. 

%Before detailed spectral modeling it is reasonably to carry out a preliminary  analysis with 
%so called the color-color diagrams and the hard-intensity diagrams for qualitative evaluation 
% investigation 
%of state transition behavior of GX~340+0 within available observational set.

%\section{Phenomenological spectral diagrams \label{ccd}}

\section{Spectral Analysis \label{spectral analysis}}

\subsection{Choice of  the Spectral Model}
As a first step we proceed with 
%Initially, we tried  
a model which is a sum  of an absorbed thermal component ($Bbody$), a thermal Comptonization 
component ($Comptb$\footnote{{\it Comptb} is an XSPEC contributed model, \\ 
see http://heasarc.gsfc.nasa.gov/docs/software/lheasoft/xanadu/xspec/models/comptb.html}) 
and a $Gaussian$ line component.  But this model [{\it single COMPTB} model, $wabs*(Bbody+Comptb+Gauss$)] 
gave a poor description of about 
%($\sim$
40\% of the data. Significant 
negative residuals at low energies and greater than %$\sim$~
30 keV suggest
%and negative residuals near $\sim$~7 keV (and $\sim$~10 keV ????) 
the presence of additional  emission components. %and corresponding absorption edge.
For this reason we also attempt the so called {\it double Bbody} model  
[$wabs*(Bbody+Bbody+Comptb+Gauss)$], {however the best fit}
of both {\it RXTE} and {\it Beppo}SAX spectra 
has been  obtained %a so called 
by implementing  the {\it double COMPTB} model [$wabs*(Bbody+Comptb1+Comptb2+Gauss$)]
%As a result  we obtained a satisfactory %good 
%agreement  with  {\it double COMPTB} model for both {\it RXTE} and {\it Beppo}SAX observations 
%we then applied this model %%(see, for example, Table 3).
%For this reason we          applied "double COMPTB" model 
for all available set of the  data (see Table 3 for $Beppo$SAX data). 
% with good adjustment (agreement) with observations (goodness of fit).
%The addition of $Gaussian$ line component at 6.4 keV %and absorption edge component ($edge$) at 7.1 keV 
%considerably improves fit quality and provides a statistically acceptable $\chi^2_{red}$.

\begin{figure}[ptbptbptb]
\includegraphics[scale=0.98, angle=0]{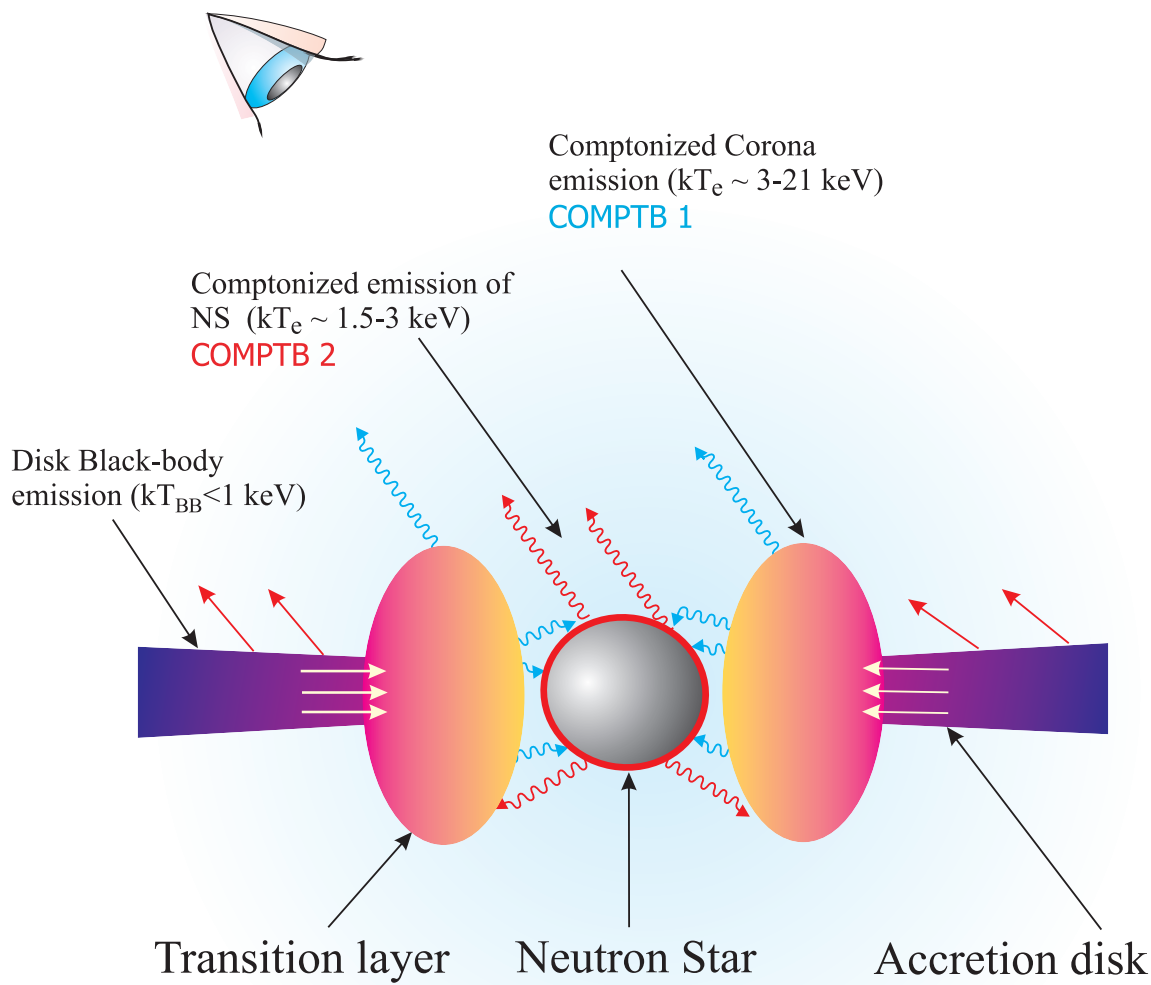}
\caption{A suggested  geometry of GX~340+0.   Disk and neutron star soft photons are 
upscattered off   hotter plasma of the Transition Layer (TL)  located 
between the accretion disk and NS surface.  Some fraction of these 
photons is seen directly by the Earth observer. Red and blue photon trajectories correspond to soft 
and hard (upscattered) photons respectively. 
In our model two  Comptonization components are considered: the first one ($Comptb1$), with 
``seed'' (disk) photon temperature $T_{s1}\lax 1$ keV and the CC electron temperature $kT^{(1)}_ e$ varies from
 3 keV to 21 keV.
%which is presumably related to {\it Compton cloud}, %(CC),   
The second Comptonized component ({\it Comptb2} with  $T_{s2}\sim$1.5 keV) and $kT^{(2)}_ e$ which is related to  NS surface and its inner part of the TL  (boundary)  layer respectively.
% component [{\it Comptb1}, ``seed'' photon temperature $T_{s1}$=1.2 keV], which presumably 
%related to {\it Compton cloud}, %(CC),   
%and second component [{\it Comptb2},  $T_{s2}$=1.5 keV], which associated with neutron star (NS) surface,   
%Upper panel presumably describes  the spectral formation 
%in the $faint$ phase %low state 
%when the TL temperature is higher than that in the bright phase. %high state. % (see bottom panel). 
}
\label{geometry}
\end{figure}

{This model describes} a scenario 
%related to our model 
in which a Keplerian disk  is connected to the {\it Neutron star} through  the  transition layer (TL) [see Titarchuk et al. (1998)].
 % at the inner edge thereof, which is innermost to NS surface.  %(see Fig.~2).
  We  display   our spectral model in Figure~\ref{geometry}.
% 
%as a basic model for  fitting  the {\it Beppo}SAX and {\it RXTE}  % spectral data for GX~340+0.
%
We suggest  that accretion into a neutron star occurs % when the material passes 
through the two main regions:  an 
%geometrically thin 
accretion disk [the standard Shakura-Sunyaev 
disk, see \citet{ss73}]
and TL. 
% transition layer (TL), 
Then NS   and  disk soft photons   are  upscattered off hot electrons of the transition layer (TL).
%In other words, 
Thus in our picture, the emergent  Comptonization spectrum is  formed in  TL, 
where thermal disk and NS seed photons of temperatures $kT_{s1}$ and $kT_{s2}$
 %and soft photons of temperature $kT_{s2}$ from the neutron star  
  upscatter off the  TL  hot electrons, giving rise to
two Comptonized components {\it Comptb1} and {\it Comptb2}, respectively.
%A fraction can be also seen directly
 The Earth observer can directly see some fraction of the seed soft photons that justifies an addition of  a soft blackbody component of temperature $T_{BB}$.
%  with  the normalization $N_{BB}$.} 
In Figure~\ref{geometry} we show red and blue photon trajectories which are related to   
seed and upscattered photons, respectively. 

Thus 
%It is worth noting that 
%the {\it Comptb} model describes 
the emergent X-ray spectrum can be presented as a convolution 
of  a soft 
%an input or {\it seed} 
black body spectrum of
normalization $N_{com}$ and color  temperature $kT_s$
 with a Comptonization  Green  function.
Normalization $N_{com}$ is a ratio of the source 
%(disk) 
luminosity
to square of the distance $D$ (similar to the ordinary {\it Bbody} XSPEC model)
\begin{equation}
N_{com}=\biggl(\frac{L}{10^{39}\mathrm{erg/s}}\biggr)\biggl(\frac{10\,\mathrm{kpc}}{D}\biggr)^2.
\label{comptb_norm}
\end{equation}  

%In result, 
%In the framework  of 
The applied model $wabs*(Bbody+Comptb1+Comptb2+Gauss)$ has the free parameters:
%To summarize the resulting spectral model parameters are 
the equivalent hydrogen
absorption column density $N_H$; the energy spectral indices $\alpha_1$, $\alpha_2$; 
%(photon index $\Gamma=\alpha+1$); 
{\it seed} color temperatures %color temperatures of the $Blackbody$-like photon spectra 
$T_{s1}$, $T_{s2}$;  parameters $log(A_1)$ and  $\log(A_2)$  related to the illumination fractions 
of the transition layer by disk and NS soft photons respectively, $f_1$, $f_2$,
 in other words $f=A/(1+A)$ is  
  the relative weight of the Comptonization component; 
electron temperatures $T^{(1)}_e$ and  $T^{(2)}_e$;  normalizations of the BB-like
components  $N_{Com1}$ and $N_{Com2}$ of the $Comptb1$ and $Comptb2$ components
respectively. 
%  The $COMPTB$ spectral component has the following parameters:
%temperature of the seed photons $T_s$,
%energy index of the Comptonization spectrum $\alpha$ ($=\Gamma-1$), 
%electron temperature $T_e$,   Comptonization  fraction $f$ [$f=A/(1+A)$, which is the relative weight of the Comptonization component] 
%and the normalization of the seed photon spectrum $N_{COMPTB}$. 
%We also include a soft blackbody component of the temperature 
 % $T_{BB}$ with  the normalization $N_{BB}$. 
 %was included. 
%
To fit the data in the 6 -- 8 keV  energy range  we add to the above model a Gaussian component, 
which has   as parameters: a centroid line energy $E_{line}$, the line
width  $\sigma_{line}$  
and its normalization $N_{line}$. 
%(see Fig. \ref{BeppoSAX_spectra}). 
%We also include  the interstellar absorption with a column density $N_H$ in the model.

% LLLLLLLLL

%It should be noted  that 
%we  fixed 
Note that some of the parameters of the $Comptb$ component: 
$\gamma=3$ is  related to the index of the seed  photon blackbody spectrum, $\alpha=\gamma-1=2$, 
%(low energy index of the seed photon spectrum) 
and $\delta$  characterizes 
%are fixed at values of 3 and 0 respectively. Namely, we disregard  
an efficiency  of the  bulk inflow effect vs the  thermal Comptonization. We neglect this  bulk motion effect  for NS and  we put   $\delta=0$.
%  should be equal to 0  in TL of   GX~340+0.  
The bulk inflow  Comptonization should take place very close to NS surface but if the radiation pressure from NS surface is high then the bulk motion is suppressed. On the other hand when 
%On the other hand 
mass accretion is  low then the bulk effect motion effect can be  ignored too. 
In general,  for NSs the bulk effect is negligible (compare with a BH case, see ST09). 

%We  use 
A value of hydrogen column $N_H=6.2\times 10^{22}$ cm$^{-2}$ found by~\cite{iaria06} is used  in our model.  We also fix the parameter $\log(A_2)$  at a value of 2, on the assumption %because 
that NS  surface is completely covered  by TL (see Fig.~\ref{geometry} for 
our model geometry). 
% which provide high Comptonization plasma of NS surface. 

\subsection{{\it Beppo}SAX data analysis}

%%We found   that the emission line feature  is  quite broad and it is much wider than 
%%the instrumental response whose width is smaller than 0.02 keV
%%\footnote{See ftp://heasarc.gsfc.nasa.gov/sax/cal/responses/98\_11}.
%%Thus  we   include a simple {\it Gaussian} component, 
%% whose  parameters are  a centroid line energy $E_{line}$, the width of the line $\sigma_{line}$  
%%and the normalization $N_{line}$,  in the model to fit the data in the 6 -- 8 keV  range.  
%%%(see Fig. \ref{BeppoSAX_spectra}). 
%%We also include  the interstellar absorption with a column density $N_H$ in the model.
%%It should be noted  that we  fixed certain parameters of the $COMPTB$ component: 
%%$\gamma=3$ (low energy index of the seed photon spectrum) and $\delta=0$ because we neglect an efficiency  of the  
%%bulk inflow effect vs the  thermal Comptonization   for  NS  GX~3+1.

%We show example of X-ray spectra in  Fig.~\ref{BeppoSAX_spectra} for $Beppo$SAX data. 
%%and in Fig.~\ref{rxte_hard_state_spectrum}
%%-$\ref{rxte_soft_state_spectrum}  
%%(for   {\it RXTE}  data). 
%Spectral analysis of {\it Beppo}SAX %and {\it RXTE}  
%observations indicates  that X-ray  spectra of GX~340+0 can be   
%described by a model with a  Comptonization component  represented by two $Comptb$ model. %(Farinelli et al, 2008)
%Moreover,  for  broad-band {\it Beppo}SAX  observations this spectral model  component is  modified  by  
%photoelectric absorption at low energies. 

In Figure~\ref{BeppoSAX_spectra} we present  two representative
% the best-fit {\it Beppo}SAX spectrum of GX~340+0 using  our   model for  {\it Beppo}SAX observation
% Two representative 
$EF_E$ spectral diagrams for  HB
%{\it Horizontal} 
and FB  
%{\it Flaring branch} 
events 
%($right$) 
along {\it Z} track of GX~340+0 using {\it Beppo}SAX data. %observation. 
We show  the best-fit spectral diagrams  of GX~340+0 %during {\it Horizontal branch} events 
% in $E*F(E)$ units 
using {\it Beppo}SAX observation  21375002 ($left$) %20261005 %21375002 carried out on 9 -- 10 August 2001.   
%carried out on 9 -- 10 August 2001. %carried out on 9 -- 10 September 1997.   
%$Top right:$ the best-fit spectrum of GX~340+0 during {\it Flaring branch} events 
% in $E*F(E)$ units using {\it Beppo}SAX observation  
and
20261006 ($right$) on the $top$ of this Figure. %21375002 carried out on 9 -- 10 August 2001.   
%carried out on 2 October 1997.   %carried out on 4 September 1997.   
We present the data  
%are shown 
by crosses and the best-fit spectral  model   {\it wabs*(Blackbody+Comptb1+Comptb2+Gaussian)} 
by light-blue line. The model components  are shown by dark-blue, red, green and crimson lines for {\it Blackbody}, 
{\it Comptb1}, {\it Comptb2}  and {\it Gaussian}, 
%components,
 respectively. 
%$Top:$  the best-fit spectrum of GX~3+1 %during {\it Intermediate state} events 
% in $E*F(E)$ units using {\it Beppo}SAX observation  20603001 carried out on 28 February -- 1 March 1999.  The data are presented by crosses and the best-fit spectral  model   {\it wabs*(blackbody+COMPTB+Gaussian)} by green line. The model components  are shown by  red, crimson and  blue lines for {\it blackbdody}, {\it COMPTB}  and {\it Gaussian} components respectively. 
%
{\it Bottom panels}: $\Delta \chi$ vs photon energy in keV. 
The best-fit model parameters for HB state ($left$ panel) are 
$\Gamma_1$=2.04$\pm$0.06, $kT^{(1)}_e$=19.50$\pm$0.07 keV, $\Gamma_2$=1.99$\pm$0.02, $kT^{(2)}_e$=2.76$\pm$0.09 keV 
and $E_{line}$=6.80$\pm$0.06 keV (reduced $\chi^2$=1.28 for 329 d.o.f),  
%(see more details in Table 3). 
%In turn, 
while the best-fit model parameters for FB state ($right$ panel) are 
$\Gamma_1$=2.00$\pm$0.08, $kT^{(1)}_e$=5.42$\pm$0.03 keV, $\Gamma_2$=2.04$\pm$0.05, $kT^{(2)}_e$=1.90$\pm$0.02 keV  
and $E_{line}$=6.68$\pm$0.08 keV (reduced $\chi^2$=0.98 for 319 d.o.f) 
(see more details in Table 3). 

{The {\it Blackbody} temperature $kT_{BB}$, independently of the source state, is consistent with 0.6 keV (2$\sigma$ upper limit).} 
%The addition of this component significantly ({\bf specify!!!!}) improves the fit quality of the \sax\ spectra.
The addition of this low temperature blackbody component makes   
%significantly 
%improves 
the fit quality noticeably higher. 
% of the {\it Beppo}SAX spectra. 
Specifically, the best-fit 
%spectral solution 
for HB events (id=21375002) is characterized by  reduced $\chi^2$ of 2.1 (331 d.o.f.) for the model {\it without the low temperature blackbody component}, while  $\chi^2$  is 0.99 (329 d.o.f.) for the model {\it with the low temperature blackbody component}. Moreover the best-fit spectral model for FB events (id=20261006) has  $\chi^2$ of 1.7 (321 d.o.f.) for the model without 
%low temperature
the  blackbody component, while   $\chi^2$ is 0.98 (319 d.o.f.) for the model with 
%a low temperature 
the blackbody component.

%On the $top$ of %this 
%Figure %~\ref{BeppoSAX_spectra} 
%we demonstrate  the best-fit {\it Beppo}SAX
%spectrum of GX~340+0 using  our   model for  {\it Beppo}SAX observation (id=20261005) carried out on 
%4 September 1997.   
%The data  are presented by crosses and the best-fit spectral  model   {\it wabs*(blackbody + Comptb1 + Comptb2 + Gaussian)} 
%by light-blue line. The model components  are shown by dark-blue, red, green and crimson lines for {\it Blackbdody}, 
%{\it Comptb1}, {\it Comptb2}  and {\it Gaussian} components respectively. 
%%$Top:$  the best-fit spectrum of GX~3+1 %during {\it Intermediate state} events 
%% in $E*F(E)$ units using {\it Beppo}SAX observation  20603001 carried out on 28 February -- 1 March 1999.   The data are presented by crosses and the best-fit spectral  model   {\it wabs*(blackbody+COMPTB+Gaussian)} by green line. The model components  are shown by  red, crimson and  blue lines for {\it blackbdody}, {\it COMPTB}  and {\it Gaussian} components respectively. 
%{\it Bottom panel}: $\Delta \chi$ vs photon energy in keV. 
%The best-fit model parameters are 
%$\Gamma_1$=2.04$\pm$0.06, $T^{(1)}_e$=19.50$\pm$0.07 keV, 
%$\Gamma_2$=1.99$\pm$0.02, $T^{(2)}_e$=2.76$\pm$0.09 keV and $E_{line}$=6.80$\pm$0.06 keV 
%(reduced $\chi^2$=1.28 for 332 d.o.f) 
%(see more details in Table 3). 

% In particular, we  find that  an addition of  the soft thermal   
%component with   temperature $kT_{BB}=$0.6 keV to the model  %significantly improves  the fit quality of the  
% {\it Beppo}SAX  spectra. 
For the {\it Beppo}SAX data 
%observations 
(see Tables 1, 3) we obtain  that {the best fit spectral indices $\alpha_1$ and $\alpha_2$ 
are %around 1 
given by  1.01$\pm$0.04 and 0.99$\pm$0.05, respectively, 
(or 
%the corresponding photon indices 
$\Gamma_1 = \alpha_1 + 1$ and $\Gamma_2 = \alpha_2+1$ are given by
%about 2
 2.01$\pm$0.04 and 1.99$\pm$0.05, respectively, for a $double~COMPTB$ model). 
We also find that, for the available {\it Beppo}SAX data, from HB to FB, the seed temperatures $kT_{s1}$ and $kT_{s2}$ change  in the ranges 0.7$-$1 keV and 1$-$1.7 keV, respectively.}
%    we fix these  values of $kT_{s1}$ and $kT_{s2}$ at  
%1.1 keV and 1.5 keV, respectively, for {\it  RXTE} data analysis (see next Section). 
%value of 1 keV in future calculations.
%An equivalent hydrogen absorption column density was fixed at the level of 
%$N_H=5\times 10^{22}$~cm$^{-2}$.
%The  best-fit parameters of  the source spectrum are presented in the Tables 1-6. 
We apply a systematic error of 1\% 
%has been applied 
to all  of the analyzed spectra.
%{\bf Lena, I want to make that this last statement is precisely true, is not it?  I hope  that you wrote in the previous version ``It is necessary to note that for  the spectral fitting we included 
%the systematic error 10\%'' is just typo.}.

%The general picture of $lower$NB-HB transition for $Beppo$SAX data is illustrated
%in Figure~\ref{sp_compar_SAX} where we bring together spectra of $lower$NB, $upper$NB and HB to demonstrate the source
% spectral evolution from hard to soft states. We should emphasize some different shapes of the
%spectra for the different spectral states.
%In this Figure %\ref{sp_compar_sax} 
%one can see differences (changes) of spectral 
%shape versus spectral branches. %relation to spectral aspect:
%During HB the X-ray spectra ({\it red line}) show clear (shaped, narrow, profiled) iron line signature %profile at soft energies 
%and significant hard emission %(``hard tail'') 
%up to 100 keV % 150 keV 
%(for instance, see $blue$ spectrum of Fig.~\ref{sp_compar_SAX}) % or $green$ spectrum of Fig.~\ref{sp_compar_xte}). 
%Note that this state can remind {\it low hard} state of BHs (TSA06, TS09). 
%%The hard emission at 50 -- 150 keV is relatively low % This ``hard tail'' is absent along the FB. 
%In turn, NB is characterized by some smeared iron line and significant decreasing of hard 
%emission at 50 -- 150 keV (for instance, see $pink$ spectrum of Fig.~\ref{sp_compar_SAX}), which 
%is remind so called  {\it high soft} state of BHs (TSA06, TS09).. %and weaker  $Blackbody$ component.

\subsection{{\it RXTE} data analysis}

\begin{figure}[ptbptbptb]
\includegraphics[scale=1.0,angle=0]{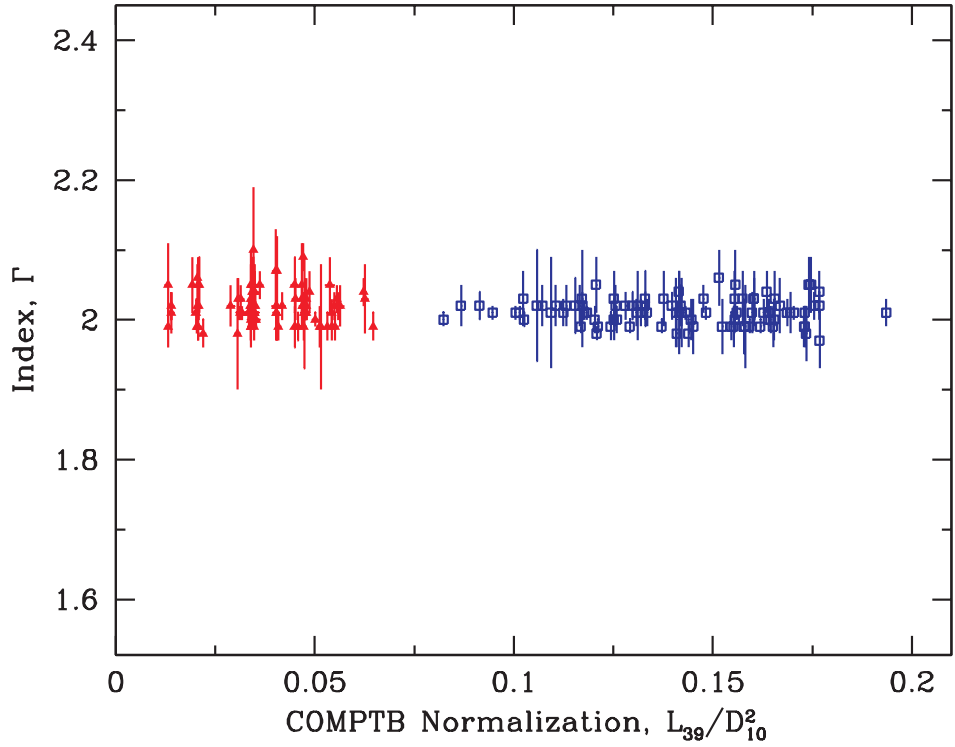}
\caption{
Photon indices $\Gamma_{com1}$ and $\Gamma_{com2}$ plotted vs.
the normalizations of the $Comptb1$ and $Comptb2$ components, respectively, measured in $L_{39}/D^2_{10}$ units 
% in the frame of  our spectral model $wabs*(blackbody+Comptb1+Comptb2+Gaussian)$ during {\it RXTE} observations 
%outburst transitions 
(see Table 4). 
$Red$ and $blue$ points correspond to $Comptb1$ and $Comptb2$ components, respectively, which are related to thermal upscattering of soft  photons by plasma electrons in CC and at  the NS boundary layer, respectively.
%{\it Beppo}SAX  and {\it RXTE} observations of GX~340+0 
%respectively. 
}
\label{index_norm}
\end{figure}

\begin{figure}[ptbptbptb]
\includegraphics[scale=1.0,angle=0]{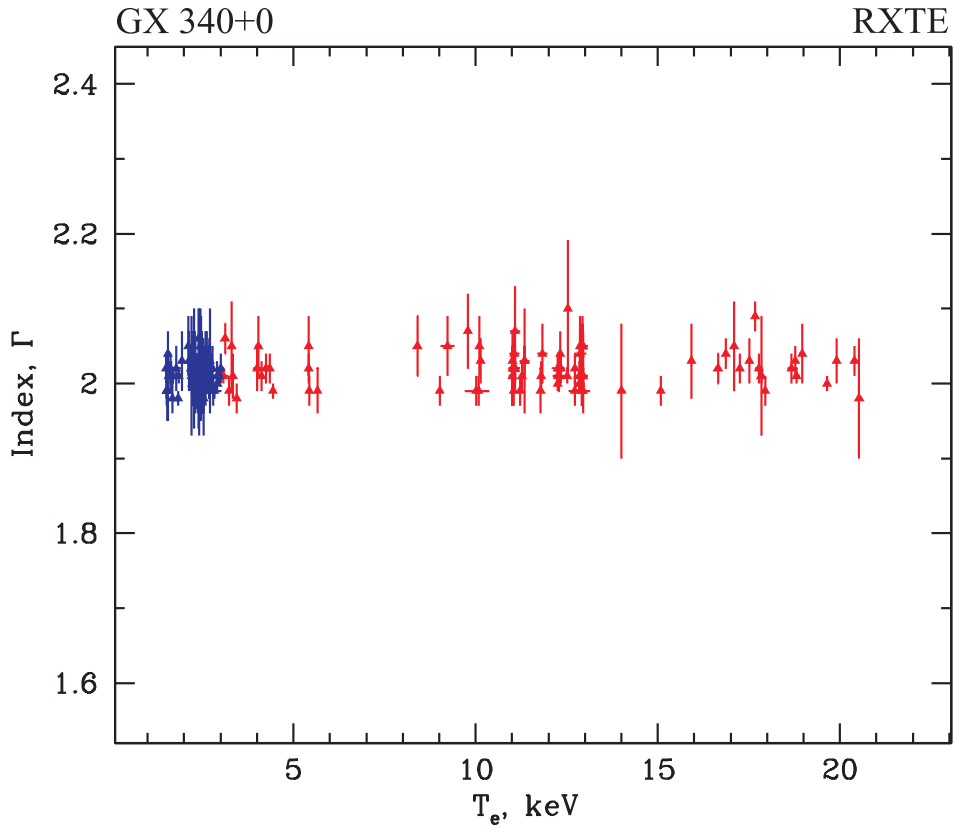}
\caption{
%Spectral indices $\alpha_1$/$\alpha_2$ 
Same as in Fig. \ref{index_norm} but plotted 
%Photon indices $\Gamma_1$ and $\Gamma_2$ 
%plotted vs 
vs the best-fit electron temperatures of the $Comptb1$ and $Comptb2$ components, respectively,  measured in keV  
% in the frame of  our spectral model $wabs*(blackbody+Comptb1+Comptb2+Gaussian)$ during RXTE %observations 
(see Table 4). }
\label{index_temperature12}
%\label{norm_T_e}
\end{figure}

%Insofar as %Unfortunately  
%Unfortunately {\it RXTE} detectors cannot provide well calibrated spectra  below 3 keV  but  the  
%broad energy band of {\it Beppo}SAX telescopes allows us to determine  the parameters 
%of {\it blackbody} components  at low energies.  
%Thus, in order to fit the {\it RXTE} data  we have to fix the  temperature of the low temperature  {\it blackbody} component at a value of 
%$kT_{BB}=$0.6 keV obtained 
%as an upper limit  in  
From our   analysis of   the \sax\ data we obtain the  temperature of the soft  {\it blackbody} component at a value of 
$kT_{BB}=0.6$ keV.  However  {\it RXTE} detectors cannot give us  well calibrated spectra  below 3 keV 
thus in order to fit the {\it RXTE} data  we should  fix this  {\it blackbody}  temperature at 0.6 keV.
% from  4U~1728-34. 
%As it was  argued above, 
We also  fix values of $kT_{s1}$ and $kT_{s2}$ at  
1.1 keV and 1.5 keV, respectively (see  \S 3.2).
in Table 4 we show the  best-fit spectral parameters   using {\it RXTE} data.
% observations.
% are presented  
Also  we find that the normalization %electron temperature $kT^({1})_e$ 
of the  $Comptb1$ component changes from 2 to 6 in units of $L_{37}/{D^2}_{10}$ (where $L_{37}$ is the luminosity of the seed blackbody component in units $10^{37}$ erg sec$^{-1}$ and $D_{10}$ is distance in units 10 kpc),  whereas 
%photon   index 
$\Gamma_1$  is 
%almost constant (
around  2
%$\Gamma_1=1.99\pm 0.02$, $\Gamma_2 = ???$,) 
for all observations (see Figs.~\ref{index_norm}-\ref{index_temperature12}).  Note  that the line width 
$\sigma_{line}$ of $Gaussian$
% component 
does not  change much  and numerous tests show that $\sigma_{line}=0.5-0.8$ keV.
% it is  in the range of 0.5 -- 0.8 keV.

Furthermore, a detailed analysis XMM-$Newton$ spectra of GX~340+0   recently carried out   
by \cite{d'Ai09} has revealed that 
 the line profile does not present  any strong  change 
 %correlated variation 
 for HB.  
Therefore we fix $\sigma_{line}$ of Gaussian component at a value 0.6 keV 
%for all spectra 
during our fitting procedure
while the electron temperature $kT^{(1)}_{e}$  of $Comptb1$ 
%component 
varies from  3 keV to 21 keV 
(Fig.~\ref{temperature_vs_hc}),
%{index_temp_cmpt}
%),
 which is similar to  that  obtained 
 using  the {\it Beppo}SAX data  
% analysis 
(see Table 3). % and previous studies by  \citet{lavagetto04}, \citet{iaria06} and \citet{church06}. 
%It is noted that the  temperature of the seed photons $kT_s$  of the $COMPTB$ component usually  increases up to 1.7 keV in the
%fainter phases %low luminosity state 
%and generally decreases to 1.2 keV in the $bright$ phases. %high luminosity state. 

%We use a value of hydrogen column $N_H=1.6\times 10^{22}$ cm$^{-2}$, which was found by~\citet{ooster01}.  
%%%Systematic error of 0.5\% has been applied to all analyzed {\it RXTE}  spectra.

In Figure~\ref{Zsp_compar_RXTE}  we  show three representative $EF_E$ spectral diagrams for different states along the {\it Z}-track.
% of GX~340+0. 
Data are taken from {\it RXTE} observations 
50016-01-01-11 ($Z1$ panel, {\it Horizontal branch}), 91125-03-01-000 ($Z2$ panel, {\it Normal branch}) and 
50016-01-02-18 ($Z3$ panel, {\it Flaring branch}). We show the data using black points while   
 the spectral model components are presented 
 %displayed  
 by red, green, blue and %dashed 
purple lines for $Comptb1$, $Comptb2$, 
 $Blackbody$ and $Gaussian$ respectively.  %\ref{sp_compar_xte} 
One can 
%clearly 
notice   changes of spectral 
shape for different  spectral branches in this  Figure. %({\it left panel}). %relation to spectral aspect:
To highlight the effect of variability of different components 
during evolution %transitions 
between the HB ($Z1$) and FB ($Z3$) states we mark the $Comptb1$ component by yellow shaded areas and trace its evolution 
when
% the electron temperature of the Compton
%cloud  
$kT^{(1)}_e$ monotonically drops 
%decreases 
from 21 keV ($Z1$ panel) to 3 keV ($Z3$ panel) (see also Fig.~\ref{index_temperature12}
for the electron temperature range).

In Figure~\ref{sp_compar_xte} three representative $EF_E$ spectral diagrams for different states along {\it Z} track 
%of GX~340+0 
({\it right panel}) are 
presented in combination with {\it color-color} diagram
% localization (indication) 
({\it left panel}).
% of Figure~\ref{sp_compar_xte}]. 
Data are taken from {\it RXTE} observations 
20053-05-01-01 ($green$, NB) 
%{\it Normal branch}), 
20053-05-01-02 ($violet$, FB)
%{\it Flaring branch}) 
and 
20053-05-01-00 ($red$,  HB).
%{\it Horizontal branch}).
On  the {\it right panel} of this  Figure %\ref{sp_compar_xte} 
one can see changes of spectral 
shape for different spectral branches which are presented on the {\it left panel}. %relation to spectral aspect:
%Analogously to $Beppo$SAX analysis (see Fig.~\ref{sp_compar_SAX}), during HB the X-ray spectra ({\it red line}) show clear 
%(shaped, narrow, profiled) iron line signature %profile at soft energies 
%and significant hard emission %(``hard tail'') 
%up to 150 keV 
%(for instance, see $blue$ spectrum of Fig.~\ref{sp_compar_SAX} or $green$ spectrum of Fig.~\ref{sp_compar_xte}). Note that 
%this state can remind {\it low hard} state of BHs (TSA06, TS09). 
Furthermore, the hard emission at 50 -- 150 keV ({\it hard tail}), {even if not strongly constrained, appears} relatively low % This ``hard tail'' is absent 
along  FB, while during HB and NB the {\it hard tail} becomes  stronger.% In turn, NB is characterized by smeared iron line and weaker  
Our spectral model demonstrates
 %shows 
 a very good quality 
 %fidelity 
 for
 %throughout
all data sets analyzed  in our study. 
%analysis. 
In fact, 
%a value of reduced
$\chi^2_{red}=\chi^2/N_{dof}$, where $N_{dof}$ is a number of degree of freedom, 
is  less than or about 1  for most observations.    With high counting statistics  for a small 
fraction of the spectra  (less than 2\%) $\chi^2_{red}$ can reaches 1.4. On the other hand, 
$\chi^2_{red}$  is never above  a rejection threshold 
of 1.5 (for 90\% confidence level). 
%%%%BEGIN
%{\it
We should note 
%that the energy range for the cases, 
 the energy range, in which   we obtain  the  poor fits 
 %statistics 
(one among 92 %91 
spectra with $\chi^2$=1.46 for 67 d.o.f),    is associated with  the iron line region.  
%Possibly, it is caused with complexity of iron line shape and relatively poor 
%energy resolution of RXTE. 
Possibly  that the iron line feature in the spectrum  is more complicated  
than just  one  Gaussian  line. 
%(i.e. a  blend of different 
%energies, presence of the edge, or broadening by Comptonization). 
In fact, for  XMM-$Newton$ observation 
%of GX~340+0 
\cite{d'Ai09} detected an asymmetric profile of 
Fe K$_{\alpha}$ emission line with an extensive red wing. The authors  interpret  that this line asymmetry is  caused by relativistic smearing in an accretion disk extending close to the NS.  However Laurent \& Titarchuk (2007)  showed that the line asymmetry  can be also formed 
%by recoil of line photons 
in an optically thick medium expanding with subrelativistic velocities.
%Recently, Laurent & Titarchuk (2007) proposed an alternative scenario for the formation of 
%broad iron lines in LMXBs, where extensive red wings could be formed by recoil of line photons 
%in an optically thick medium expanding, or converging, at relativistic velocities. Since a spectral line model, 
%adapted to this scenario, is not yet available for fitting X-ray data, we are not able to test this scenario 
%using our data. However, as pointed out in Pandel et al. (2008) for the case of 4U 1636-536, we note that 
%the lack of a narrow core and the presence of a blue wing in the iron line profile of GX 340+0 is 
%a strong indication of the disk reflection origin. Moreover, the anticorrelation which is observed between the continuum flux and the line equivalent is difficult to reconcile with this scenario, given that the source of the line photons must also contribute to the continuum emission, either in the form of disk emission or in the blackbody harder component.
%The fits tend to favor a broad line  (see Table 4), which might be caused by Comptonization. 
We must admit that this complex line  structure cannot be resolved by the {\it RXTE} data.
% However, this possible complexity is not well  constrained by our data. 

We should also point out some differences  between our best-fit values of the photon index $\Gamma$ and those found by us in the literature.
%It is worth noting that we find some differences 
%between our values of the best-fit model parameters and those in the literature.
%  for the same set of the  $Beppo$SAX observations. 
  In particular, the photon index  $\Gamma$, estimated by 
  \citet{iaria06} 
  %evaluate $\Gamma$ ,
  %Di Salvo et al. (2000)
   for $Beppo$SAX observations  ($S3$ -- $S6$)  gives the values of 
   % they find that %id=21240001, 212400011, 212400012, 21375002
 $ \Gamma$ is about  2.5 whereas 
  $\Gamma$ is about 1.7  
  %Di Salvo et al. (2000)
   for {\it RXTE} observations  ($S5$)  [see for  details \citet{church06}]. 
%while our value of $\Gamma=1.9\pm0.2$. 
However, this  difference of $\Gamma$  values is a result of using different spectral models. 
%discrepancy of  index values  can be explained  by  using  different models. 
{For example,  Church et al.  used  an additive model of  a blackbody, a power law components and   a single Comptonized component ({\it Comptt}).} 

%We note that the BB temperature at HB stage
%is a factor of about two higher than the value reported by \citet{iaria06} %F08 
%for
%the same data set. This is because we included a Gaussian emission line at 1 keV, 
%which, as already stated in Sect. 2, may affect
%both the estimated interstellar absorption $N_H$ and the parameters
%of low-energy continuum features.

Thus   using broad band 
{\it Beppo}SAX observations 
we 
%can
% accurately 
obtain 
%determine 
%all parameters 
our spectral model  parameters
%the model components of GX~3+1 spectrum. 
and we can evaluate the total spectral evolution of  GX~340+0   from 3 to 150 keV
%we  are able to  
%study 
%the overall pattern of the source  during the different spectral states 
%in the 3 -- 150 keV energy range while 
 %due to
due to the elongated observations
%because   
%the extensive  
by the {\it RXTE}.

\section{Overall pattern of X-ray properties \label{overall evolution}}

\subsection{Color-color diagrams \label{ccd}}

%\subsection{Comparison of spectral hardness diagrams for atoll-, Z and BHC binaries \label{ccd}}

To study 
%investigate 
%transition 
the 
%spectral 
properties of GX~340+0 during different spectral states 
%when the luminosity changes
%in terms of flux (or luminosity), 
%we define set 
{we make use} of  hard colors (HCs) and soft colors (SCs) to demonstrate different configurations  
%shapes
%tracks
% (branch) 
%(shapes) 
in HC versus SC.
% in the spectral diagrams. 
Namely, in Figure~9 in STF13
%\ref{ccd_evolution_sp rxte vs HB-NB-FB} 
we collected  color-color diagrams (CCDs)
({\it left column}) and hardness-intensity diagrams (HID) ({\it right column}) of 
GX~340+0. The ordinate and  abscissa of color-color diagrams show the flux ratios: 
(a) [16-50 keV/3-16 keV] and  [3-5 keV/7-10 keV]; 
(c) [20-50 keV/3-10 keV] and  [10-20 keV/3-10 keV]; 
(e) [10-20 keV/3-10 keV] and  [9-16 keV/3-5 keV]
while hardness-intensity diagrams demonstrate  flux ratios:  
(b) [16-50 keV/3-16 keV];
(d) [20-50 keV/3-10 keV];
(f) [7-10 keV/3-5 keV] 
versus flux (3-50 keV) measured in units of $10^{-9}$ erg/s/cm$^{2}$.  
%All these diagrams %contain 
%use  the same data but presented  in % shown with 
%different %test 
%energy bands. 
In this way, one can observe different shapes %behaviour 
of the {\it Z} pattern of GX~340+0. 
%As it appears from this Figure, the detailed shape on the spectral branches 
%in these diagrams depends on the choise of photon energy bands which define the flux 
%and the flux ratios. 
For the presented observations (see Table 1 and 2) we found the source in three spectral states: 
the flaring branch (FB), the normal branch (NB) and the horizontal branch (HB). 

The identification of CCD states was made using  simultaneous timing and spectral analysis, see \S 4.3  and  we have revised 
%was in accordance with 
the previous similar  investigation obtained using  EXOSAT % analysis 
and {\it RXTE} data, see Penninx et al., 1991;  Jonker et al., 1998, 2000, respectively (see \S 3). %of X-ray observations 
%for GX~340+0.
%[{\bf I would shorten the discussion stating that:] 
Both the color--color and color--flux behaviors depend on the photon energy bands in which either color or flux are evaluated. In particular plots, see panels (a), (b) and (d), there is evidence of HB, while in other ones this branch is not apparent.  The evidence of  FB appears only from panels (a) and (b). Instead, independently of the energy band,  NB is always apparent. A comparison of the color--flux behavior of different compact X--ray binaries that host  NS or a BH is shown in Fig.~\ref{HID_4object}.
%10.
%}
%
%In particular, as the source moves on  the horizontal branch % towards 
% the normal branch, 
%while 
%the hard color in (a) panel [16-50 keV/3-16 keV] is 
%approximately  constant while the hard color in (d) panel 
%[20-50 keV/3-10 keV] seems to decrease and  the hard color of %(b)  [16-50 keV/3-16 keV] and (c) [20-50 keV/3-10 keV] panels % increases.  
%For (e) [10-20 keV/3-10 keV] and (f) [7-10 keV/3-5 keV] 
%diagrams the
%horizontal branches overlap with NB.
%For (f) [7-10 keV/3-5 keV] diagram the horizontal branch  
% overlaps   with NB. 
%
%We also found different types
%kinds 
%of behavior in FB depending on the choice of photon energy 
%bands 
%which define the flux and the flux ratios. In the (a) and (b) %panels we see a  {\it horizontal}  FB 
%on the $left$ side from  the NB in the CCD. 
%While in the (e) panel we see a {\it raising} FB.
% at the $right$ side of NB in the CCD.  
%In  (c), (d), (e) and (f) panels one does not see  clear FB, %which is  possibly blended with NB. 
%
%Therefore, we want
%tend 
%to use a diagram with a good separation of the  branches as 
%it is shown in  Figure~\ref{HID_4object}.
%{PDS_FBO}.

%{\bf The following paragraph is not clear to me, I would change it as follows:}
 
It should  be noted that FB flux behavior of GX~340+0 corresponds to dips in the light curve 
(see Figs.~11-12 in STF13).
%\ref{lc_2000}-\ref{lc_2001}
%).
 {Indeed, when the source intensity dips, both the soft  and  hard colors increase (see panel (a) in Fig. 9 in STF13
%\ref{ccd_evolution_sp rxte vs HB-NB-FB}).  
This behavior is 
completely different from FB one  observed in Sco~X-1, GX~17+2 and GX~349+2, while it is similar to the
FB behavior seen in Cyg~X-2. A discussion of this phenomenon is given by Hasinger et al. (1989) and 
\cite{KK96}.} 

\subsection{Evolution of X-ray  spectral properties 
%during the spectral transitions  in the 3$-$150 energy  
in different spectral states \label{evolution}}

We find properties 
%have  established  
%common 
%characteristics 
of the  HB-NB-FB state  of GX~340+0 
based on their  spectral evolution in the 3$-$150 keV energy range 
%of X-ray emission  
%in  the energy range from 3 to 150 keV  
using {\it RXTE} data. 

The general picture of HB-NB-FB states is shown
%illustrated
in Figure~\ref{sp_compar_xte} %and \ref{sp_compar_SAX} 
where we put  
%bring 
together 
%spectra of 
the HB, NB and FB spectra
 to  display
% demonstrate 
the GX~340+0  spectral evolution (see also   panels for  HB and NB spectra in Fig.~\ref{Zsp_compar_RXTE}).
 % from hard to soft states. 
 One should point out 
 %We should emphasize 
different spectral shapes related to 
%of the spectra
 %for 
 the different spectral states. 
 Figures 3 and  7 in STF13. 
 %\ref{BeppoSAX_spectra},  
%\ref{Zsp_compar_RXTE} 
illustrate the model components while  Figs~11$-$12 in STF13
%\ref{lc_2000} -- \ref{lc_2001}
 demonstrate 
the evolution of these components. As we  indicate  below  (see \S  4.2.1),  {using of the best-fit model}, in all  spectra independently of the state 
%for  all states 
%are dominated by 
a strong Comptonized component { ({\em Comptb2})} related to 
%a boundary layer 
{the upscattering of the NS seed photons  off the TL innermost part  is dominant 
%the  coming from the
%NS surface 
(see illustration of this effect in  Fig. \ref{geometry})}. 
%located near NS surface 

\begin{figure}[ptbptbptb]
\includegraphics[scale=1.0, angle=0]{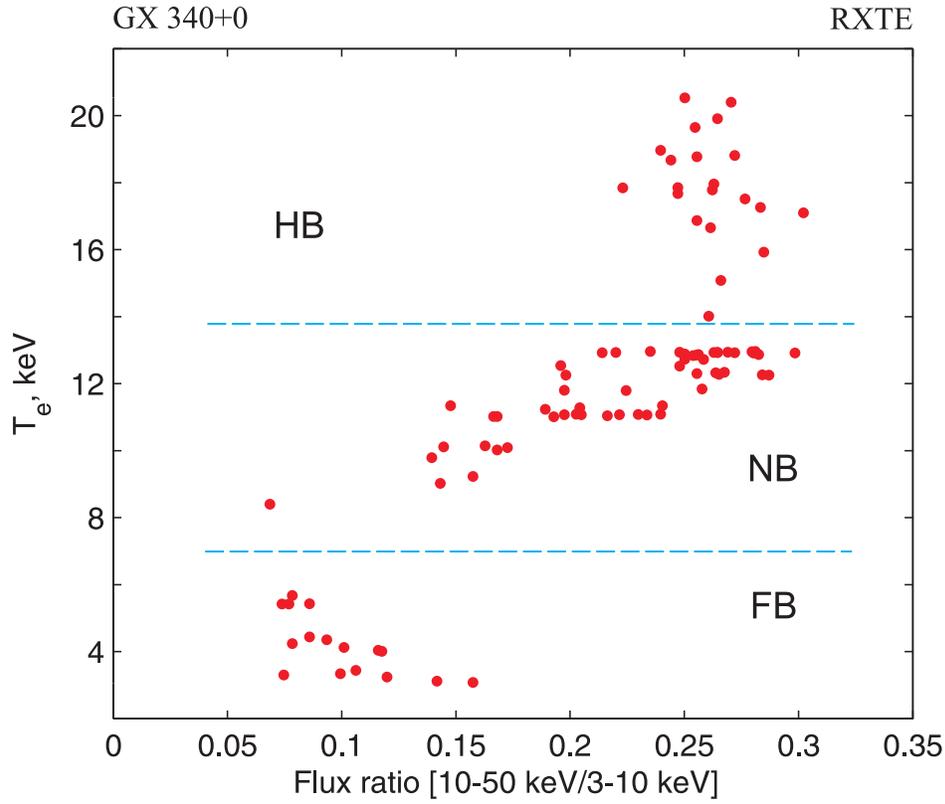}
\caption{
The best-fit  electron temperature $kT_e$  of $Comptb1$ component (in keV), using  our spectral model 
$wabs*(blackbody+Comptb1+Comptb2+Gaussian)$,   is plotted versus flux ratio [10 -- 50 keV/3 -- 10 keV]
during Z-state %transition 
events. The horizontal dashed lines 
mark the branch regions  of {\it Z} pattern.
% for which spectral analysis was carried out are also shown.
}
\label{temperature_vs_hc}
\end{figure}

\begin{figure}[ptbptbptb]
\includegraphics[scale=1.05, angle=0]{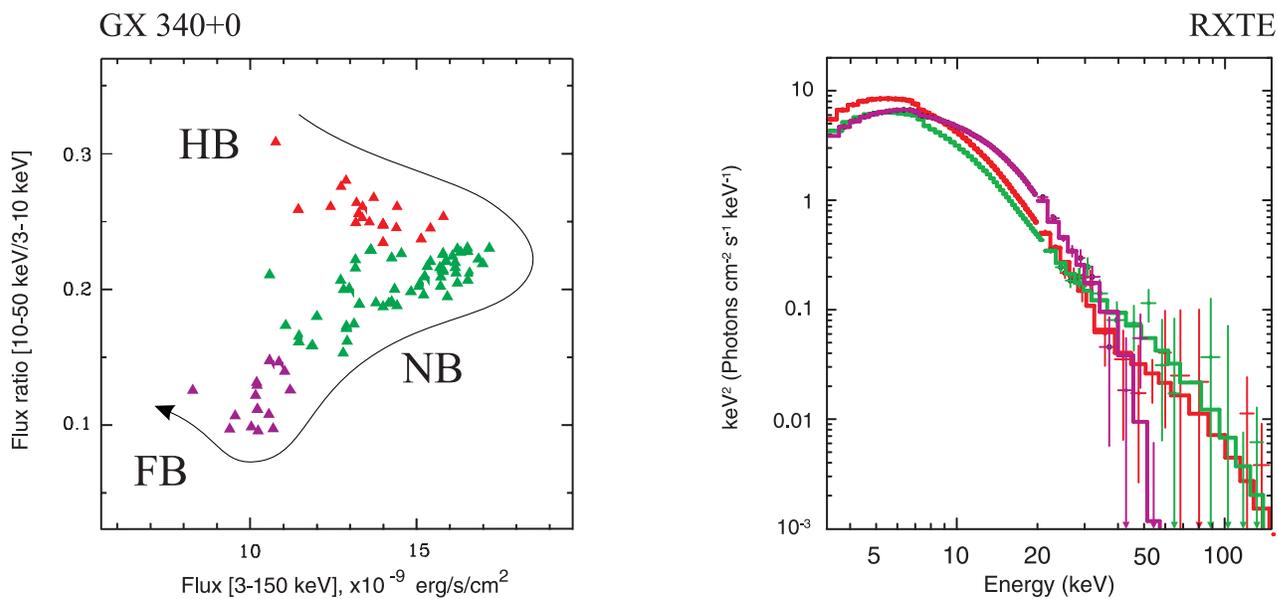}
\caption{{\it Left panel:}
%Flux ratio [9-16 keV/3-5 keV] versus flux (intensity) [6-9 keV] 
Flux ratio [10-50 keV/3-10 keV] versus flux 
%(intensity) 
in the range of  3-50 keV 
%diagram of GX~340+0 
for {\it RXTE} data. The spectral branches have been indicated by $pink$ points  for FB, $green$ points  for NB and $red$ points for HB. The direction of HB$\to$NB$\to$FB evolution %transition 
is indicated by arrow.
{\it Right panel:} Three representative $EF_E$  diagrams for different states along Z track of GX~340+0. 
Data are taken from {\it RXTE} observations 
20053-05-01-01 ($green$, NB)
%{\it Normal branch}), 
20053-05-01-02 ($violet$, FB)
%{\it Flaring branch}) 
and  20053-05-01-00 ($red$, HB).
%{\it Horizontal branch}).
}
\label{sp_compar_xte}
\end{figure}

In these two states, the
Comptonization component, {\em Comptb1}, related to {
upscattering of the disk seed photons 
%coming from the disk, 
 by TL electrons }
%a {\it Compton cloud}} 
%is associated with the
of the temperature in the range of 8--21 keV,  extends up to 150 keV.
% in X-ray spectrum.
Instead, in FB spectra  (Fig.~7 in STF13
%\ref{Zsp_compar_RXTE}, 
panel for the FB spectrum)  the Comptonized component {{\em Comptb1}} is 
characterized by a decrease of the electron temperature to
%$kT^{(1)}_e$=
3 -- 6 keV and thus the spectrum only extends to 50 keV.
% and below.
Generally,  the $Blackbody$ and $Comptb1$ components are relatively weak with respect to $Comptb2$.
We find that values for the electron temperature $kT^{(2)}_e$ % of NS boundary layer  %(layer) 
is close to that of % of CC, 
$kT^{(1)}_e$,  
%the electron temperature $kT^1_e$ of CC, %Compton cloud, 
only during {\it flaring branch}, 
when  $kT^{(1)}_e$ monotonically decreases to 3 keV, 
while $kT^{(2)}_e$  shows low variability about  3 keV during all spectral states. 
HC %HR  
ratio  [10-50 keV]/[3-10 keV] 
%decreases 
drops from 0.3 to 0.1 
during HB-FB  spectral evolution  
%and thus the HR  ratio definitely 
%traces these  HB-FB  spectral  branches 
% a transition  from HB to FB 
(see Fig.~\ref{temperature_vs_hc}). % and vice versa associated with dip in the X-ray light curve (see details in Sect.~5.1).  
%In fact, 

Thus using   Figure \ref{temperature_vs_hc}, % \ref{rxte_soft_state_spectrum}$-$\ref{rxte_hard_state_spectrum}  
%while 
%the electron temperature of the Comptonized  component $Comptb1$ is not 
%a strong monotonic  function,that  
one can  {\it find} one-to-one correspondence  between 
spectral states  of GX~340+0 and  value of $kT^{(1)}_e$ within 
%predetermined 
the limits, marked by horizontal 
blue dashed lines.
%in Figure~\ref{temperature_vs_hc}. 
The electron temperature $kT^{(1)}_e$ 
varies by a factor of 10 during these states   %higher in the HB %$bright$ phase %high state  than that in the FB %faint phase, %low state, 
%although
while 
%  the photon indices { (
$\Gamma_1$ and $\Gamma_2$,
%),}
%$\Gamma_{1/2}$ 
for each of these spectra, only vary 
%in the range  
 from 1.9 to 2.1
 % and  they are  slightly spread  around $\Gamma$=2 (see 
 %that distribution of  
%$\Gamma$ on the {\it  left-hand} panel of
 (see Fig.~\ref{index_temperature12}).  

{Using  $\chi^2$-statistic criterion, we test the hypothesis that the approximate value $\Gamma_{appr}$  of both indices $\Gamma_1$ and $\Gamma_2$ is  $\Gamma_{appr} \approx 2 $.}
% by minimization of corresponding 
%residuals for all measurements. 
%{\bf For each of the two data sets R5 and R6,} 
We calculate the distribution of $\chi^2_{red}(\Gamma_{appr})=\frac{1}{N}\sum_{i=1}^N\left(
\frac{\Gamma_i-\Gamma_{appr}}{\Delta\Gamma_i}
\right)^2$  versus of $\Gamma_{appr}$. 
% on the {\it right-hand} panel of  Figure~\ref{hist}.
We  find  a sharp minimum  of the function $\chi^2_{red}(\Gamma_{appr})$ of about 1 
for two  Comptonized components ($Comptb1$ and  $Comptb2$). Namely
% which takes place in the range of 
%$\Gamma_{appr}=1.99\pm0.02$ for d.o.f=126 with null hypothesis probability $10^{-8}$.
%It is worth noting that 
%{\bf for  both data sets: 
$(\Gamma_1/\Gamma_2)_{approx} =1.99\pm0.01/2.00\pm0.01$ with a confidence level of 67\% 
%for data set R5, 
and $(\Gamma_1/\Gamma_2)_{approx}=1.99\pm0.02/2.00\pm0.02$ with a confidence level 90\% % 99\% 
for 92 d.o.f.
%for data set R6} 
(see also   Figure of  $\chi^2_{red}(\Gamma_{appr})$ for 4U 1728-34 in ST11).
We should point out 
%emphasize 
that the indices  $\Gamma_1$, $\Gamma_2$  %index  $\Gamma$ is  
are also independent of the normalizations of 
%$Comptb1$,  $L_{39}/d^2_{10}$ and 
$Comptb1$, $Comptb2$;  $L_{39}/d^2_{10}$ and 
%the plasma temperature of Compton cloud 
%$kT^{(1)}_e$
$kT^{(1)}_e$, $kT^{(2)}_e$
% when both of these parameters change by a factor 5 at least 
(see Figs.~\ref{index_norm}$-$\ref{index_temperature12}).
% \ref{index_temp_cmpt} -- \ref{index_norm}).
 Relying on {\it Beppo}SAX  observations and their analysis   FT11  propose 
 %suggested  
 that $\Gamma$ varies  about 2 for many   
NSs
% sources 
which are seen from the hard to soft states.
%observed 
%in the various  spectral states.  
FT11 relate  the spectral state to a value of electron temperature $T_e$  and they demonstrate
 that  the photon index $\Gamma=2\pm 0.2$ (or the spectral index 
$\alpha=1\pm 0.2$)  while   $kT_e$ varies in the range from 2 to 25 keV. 

\subsubsection{Light curves and related spectral characteristics }

We find  a number of X-ray flaring events of GX~340+0  using  {\it RXTE}  data obtained 
during  2000$-$2001 ($R5$, $R6$ sets) with a full  rise-decay evolution.  We examine
%have searched 
%for  
generic  spectral and timing properties which can be found  during these spectral state events.  
In   Figures 11 in STF13
%\ref{lc_2000}$-$\ref{lc_2001} 
we show 
%present 
%the 
%combined 
%results of  
the details of our spectral analysis of the  {\it RXTE} observations  applying 
 our model  $wabs*(bbody+Comptb1+Comptb2+Gaussian)$.

%A number of X-ray flaring episodes of GX~340+0 has been  detected with {\it RXTE} during  2000$-$2001
%($R5$, $R6$ sets) with a good rise-decay coverage.  We have searched for  common 
%spectral and timing features which can be revealed during these spectral transition episodes.  
%We present the combined results of  the spectral analysis of these observations  using 
% our  spectral model  $wabs*(blackbody+Comptb1+Comptb2+Gaussian)$  in   Figures \ref{lc_2000}$-$\ref{lc_2001}.   

In  Figure~11 in STF13
%\ref{lc_2000}, 
%from top to bottom, 
we show   the changes of count rate (2-9 keV) in counts s$^{-1}$ with 16~s time resolution,
%evolutions of count rate  
 the {\it hardness ratio} coefficient HC [10-50 keV]/[3-10 keV],  the 
%model flux in 3 -- 10 keV  and 10 -- 50 keV energy ranges ({\it black and green} points respectively), 
electron temperatures $kT^{(1)}_e$ ($red$)/$kT^{(2)}_e$ 
($blue$) of $Comptb1$/$Comptb2$ components,  
%in keV, 
the normalizations of  $Comptb1$ and $Comptb2$ components  %$blackbody$  
($red$ and $blue$ respectively)  
%Comptonized fraction $f=A/(1+A)$ of $Comptb1$ component 
and  the  index $\alpha=\Gamma-1$ for $Comptb1$ and $Comptb2$ components ($red$ and $blue$ points respectively) 
during the 2000 
%transition 
events ($R5$ set) of  time period from MJD 51722 to MJD 51777. 
The dipping phases  of the light curve, which are  %presumably 
related to  the {\it flaring branch} (based on timing analysis, see also Fig.~\ref{PDS} and \S~\ref{transitions})  %screening from Earth observer % of 
%inner (central) area by pushed up {\it Transition Layer} pars of disk body, 
are marked with blue vertical strips.  
Note that during these dipping phases of the light curve %, which correspond to {\it flaring branch} in GX~340+0, 
the {\it hardness ratio} coefficient (HC) is  usually less then 0.15.

%Detailed evolution of $Comptb1$, %and 
%$Comptb2$ and $blackbody$ normalizations %($red$ and $blue$ respectively)  
%%Comptonized fraction $f=A/(1+A)$ of $Comptb1$ component 
%%and  spectral index $\alpha=\Gamma-1$ for $Comptb1$ component 
%during these (2000) transition events ($R5$ set) is presented with trhee bottom panels of Fig.~\ref{lc_norm_2000}. 
%To match  dipping phases of the light curve we again indicate corresponding dates with  blue vertical strips 
%in addition to corresponding model (unabsorbed) total X-ray flux in 3 -- 50 keV energy ranges 
%and above mentioned evolutions of the {\it harndess ratio} coefficient HC 
%[10-50 keV]/[3-10 keV]. 

Additional  flaring events are presented in  Fig.~12 (STF13)
%\ref{lc_2001}, 
where we display  details 
similar to  those shown in Fig.~11 in STF13
%\ref{lc_2000} 
but  for %all the {\it RXTE} 
2001 %transition 
events ($R6$ set) % for demonstration (illustration) of $slow~(long-term)$ variability.
%,   from top to bottom, 
%evolutions of the {\it harndess ratio} coefficient HC [10-50 keV]/[3-10 keV], 
%model flux in 3 -- 10 keV  and 10 -- 50 keV energy ranges 
%({\it black and green} points respectively), electron temperatures $kT^{1}_e$ ($red$)/$kT^{2}_e$ ($blue$) 
%of $Comptb1$/$Comptb2$ components in keV, 
% $Comptb1$, $Comptb2$ and $blackbody$ 
%normalizations ($red$, $blue$ and $black$ respectively)  and  spectral index $\alpha=\Gamma-1$ for $Comptb1$ component 
%during 2001 transition events ($R6$ set)
during the MJD 51920$-$MJD 51925 period.
%  interval  the time period from . 
Note the  temperature $kT^{(1)}_e$ also %only 
correlates with the {\it hardness ratio} coefficient HC [10-50 keV]/[3-10 keV] and it monotonically diminishes
%decreases 
from 20 keV to 3 keV during the HB-FB transition.
% from HB to FB.  
 
%The dipping phases  of the light curve, which presumably related to screening of inner (cenral) area 
%by pushed up {\it Transition Layer} pars of disk body, are marked with blue vertical strips. 
%
%Similar to  that presented in Fig.~\ref{lc_2000} but  for %all the {\it RXTE} 
%2001 transition events ($R6$ set). % for demonstration (illustration) of $slow~(long-term)$ variability.
%  {\it RXTE}/ASM  count rate is shown on the top panel of these Figures. 
%Further, from the  top to the bottom,  we show the  model flux in two energy bands  3$-$10 keV 
%({\it blue points})   
%and 10 -- 50  keV  ({\it crimson  points}).  In the next panel  we show 
%a change of the  transition layer  electron temperature $kT_e$. One can clearly see the 
%{\it low amplitude} spectral transition {\it on time scales of$\sim$ 1 -- 2 days} from the $brighter$ phase %high state 
%to the $faint$ phase %low state   
%during the time period from MJD 52000 to  
%MJD 52200 while electron temperature $kT_e$ only varies from 2.3 keV to 4.5 keV during this transition.  

%Normalization of  the $COMPTB$ ({\it crimson} points) and   $blackbody$ component 
%({\it blue} points) 
% are shown in the next panel  of 
%Figs.~\ref{lc_1998} and \ref{evolution_lc_all}. 

%In particular, one can see
It is clear   to see, using Figs. 11- in STF13
%\ref{lc_2000}$-$\ref{lc_2001},  
that  the electron temperature $kT^{(1)}_e$ ($red$ points) steadily decreases from 20 keV 
%to 8 keV 
during the {transition from HB to NB, reaching} its minimum 
%with minimum 
at 3 keV in flaring FB [MJD 51774 -- 51774.3, 51776.5 -- 51776.7 (Fig. 11 in STF13
%Fig.~\ref{lc_2000}
); 
MJD 51921.4 -- 51921.7, 51922.4 -- 51922.7 (Fig.~12 in STF13)]
%\ref{lc_2001})], 
while the electron temperature $kT^{(2)}_e$ ($blue$ points) 
is almost constant %within errorbars 
at the mean value of about 3 keV.
On the other hand, the $Comptb$ normalizations $N_{Com1}$ and $N_{Com2}$ are weakly  correlated
% (or not correlated at all) 
with  the variations of  {\it hardness ratio} coefficient HC and  
%the model fluxes in 3-10 keV ($black$) and 10-50 keV ($green$) energy bands. 
 count rate in 2-9 keV energy band, {with the
X-ray emission related to {\it neutron star} that
 dominates during all spectral states (see Fig.~\ref{index_temp_cmpt}).}  The normalization of $Comptb2$, $N_{Com2}$
is low variable around its mean level of 0.15$\times L_{39}/D^2_{10}$. The X-ray 
%{\it Compton cloud} 
contribution from the Comptonizaton of the disk seed photons 
%coming from the disk 
%into X-ray emission of GX~340+0 
is weaker by factor 2 than that related to the NS seed photons.
% coming from the NS surface},  
%NS$-$boundary layer 
%{\it neutron star} surface emission
 during all spectral states (Fig.~\ref{index_temp_cmpt}).
We should point out 
%It is worth noting 
that the indices  $\alpha$,
 %$\alpha_{1/2}$,
  shown in the bottom panels of 
 Figures  11-12 in STF13
 %\ref{lc_2000}$-$\ref{lc_2001}, 
 only  slightly vary with time around 1(or the corresponding
  $\Gamma$ varies around 2).
 % (or $\Gamma\sim 2$). 

%On the other hand, the normalization of the $blackbody$ component $N_{BB}$ is almost constant 
%except for  the {\it flaring branch}  episodes, when $N_{BB}$ increases from 0.04 to 0.08 
%(see $black$ points in Figure \ref{lc_norm_2000}    at MJD=51774 and 51776.5 and 
%in Figure \ref{lc_2001}    at MJD=51921.4 and 51922.6 for 2000 and 2001 sets, respectively).

%Moreover  these  spectral variability  transitions are related to 
%a noticeable  increase of soft flux in the 3-10 keV energy range  and decrease of hard flux 
%that takes place  in the 10-50 keV energy range (see the second panels from above in Figs. \ref{lc_1998}-\ref{evolution_lc_all}).

%The spectral  index $\alpha_1$ ($\alpha=\Gamma-1$) is presented in the bottom panels of 
% Figures \ref{lc_2000}$-$\ref{lc_2001}. 
% %One can see that  

\subsubsection{Comptonized emission}

Note that the Comptonization  fraction $f_1$  changes  from 0 to 0.5 as shown on Fig.~\ref{T_e_vs_f_comp} 
({\it red triangles}). Thus, in most  cases of HB and NB states, the soft radiation  of GX~340+0 dominates 
{and is only slightly  modified  by reprocessing in the  
TL ($f_1=0.1-0.2$).}
%Compton cloud  and in NS boundary layer.
 % {\it neutron star} surface. 
Then, during the transition  to FB,
% the Comptonization  fraction 
%the fraction 
$f_1$ monotonically increases 
from 0.2 to 0.5 {when
% the TL electron temperature 
$kT^{(1)}_e$ decreases from 6 keV to 3 keV.}
 Consequently, the Earth observer  directly sees a smaller 
fraction of disk emission  ($1-f_1$) in   FB.
% is directly seen by the Earth observer. 
We remind a reader 
that the Comptonization  fraction $f_2$, {related to the NS surface seed photons upscattered off the TL electrons, is
about 1.}
%boundary layer 
We freeze 
%{\it neutron star} surface, 
%was frozen 
 $f_2=1$ 
 %(or   $\log(A_2)=2$)
%[$f=A/(1+A)$, where ] 
 on the assumption %because 
that  NS is completely surrounded by its {TL}
%{\it boundary  layer} 
(see Fig.~\ref{geometry} for the model geometry).

\subsection{Timing properties during HB-NB-FB evolution \label{transitions}}
%\subsection{Timing properties during HB-NB-FB transitions \label{transitions}}

%4444Sect4.3
%Timing 2+3

%In the present Paper 
We analyze the {\it RXTE} light curves
% have been analyzed 
using the {\it powspec} task from
FTOOLS 5.1.  We perform the timing analysis {\it RXTE}/PCA data 
%was performed 
using   the {\it event} mode in the 13 -- 30 keV  range.  The event mode time resolution  is 1.2$\times 10^{-4}$ s. We obtained
%generated 
PDSs  in  the 0.1 -- 500 Hz frequency range
with 0.001-second 
%time 
resolution. 
%We subtracted 
The contribution due to Poissonian statistics and Very Large Event Window for all PDSs were subtracted. 
%We used QDP/PLT plotting package to model PDSs.

% Proper timing
%We ?nd a similar timing behavior of 4U 1728-34 in our data
%set along with the energy spectra. 
%In particular, 

 We demonstrate an evolution of  the 3 -- 150 keV flux  during the 2000 ($R6$) 
%transition 
events (see {\it top panel} of Figure~\ref{PDS}). 
Here $red/blue$ points A, B, and C mark moments at MJD = 51773.2/51774.7, 51773.6/51774.1 and 51772.2/51774.6 related to different phases of  {\it Z}-state evolution.  In addition, 
$green$ points $D_1/D_2$ indicate moments at MJD=51774.27/51776.68 of FB where we find $\sim$6 Hz QPO at $flaring$ branch of GX~340+0 (see also Fig.~\ref{PDS_FBO} and details below).  
%A, B, and C mark moments at MJD = 51773.2 and 51774.7, 51773.6 and 51774.1,  51772.2 and 51774.6, respectively, related to different 
%transition phases. 

%{\bf [I find the previous statement, along with the related figure, unclear. Which is the meaning of different colors in the figure? What is blue? What is red? Which transition phases? What's A? What's B? What's C? What's D?
%Also the next paragraph appears unclear to me. It would be better to talk about.]}

% of transition. %and after X-ray outburst), respectively.  

% of this Figure 
We present   PDSs for 15-30 keV range ($left$ column) plotted along 
with  spectral diagrams $E*F(E)$ ($right$ column) 
for A ($red$, top), B ($blue$, middle) and C ($blue$, bottom) points  of the 
%X-ray 
light curve in the {\it bottom panel}.  
A typical exposure time   for each PDS is 3-4 ksec. In particular, the accumulation time for PDS during 50016-01-02-16 observation (see Fig.~\ref{PDS_FBO}) is 3876 sec.
%$Red$ histograms corresponds to aforementioned MJD moments, whereas $blue$ histograms refer to adjacent observations  to illustrate continuous fast evolution of PDSs. 

 An evolution of timing properties over the {\it Z} track is clearly seen in this segment of the light curve (see {\it top panel} of  Fig. \ref{PDS}). 
Strong 
%horizontal branch 
oscillations  with centroid frequency  at   50 Hz 
are observed  at HB (see Fig. \ref{PDS}, left panel, C $red$ PDS related to 
%50016-01-01-000, 
MJD=51772.2).
%$Horizontal$ brunch, 
While  broad oscillations ranging in 2 -- 10 Hz (B $red$, 50016-01-01-06, MJD=51773.6; C $blue$, 50016-01-02-04, MJD=51774.6) 
and $\sim$50 Hz (A $blue$, 50016-01-02-12, MJD=51774.7)
 are detected during middle and hard apex of $normal$ branch (NB), respectively. 
Relatively strong noise  component (LFN) is present  (up to 10\% rms) %break at 1 -- 3 Hz and broad QPOs centered at 7 -- 10 Hz are
and NB (up to 2\% rms) but none of these features  is seen   during FB
%at the most of $Flaring$ branch, 
when a very low frequency noise
(VLFN) component  arises (B $blue$, 50016-01-02-18, MJD=51774.1). 
% before and after burst (see panels A1, C1) but none of this is seen at the X-ray flare peak (see panel B1).  
 We show the $E*F(E)$  diagrams (A2, B2, C2 panels) corresponding  to the related PDSs 
(A1, B1, C1 panels) on the right.  We present the data by black points whereas we demonstrate   
 the spectral components using  dashed red, green, blue and  purple lines for $Comptb1$, $Comptb2$,  $Blackbody$ and $Gaussian$ respectively.

%We also detected $\sim$6 Hz quasi-periodic oscillations at the end of {\it flaring branch} 
%($D_1$ and $D_2$ points, see also Fig.~\ref{PDS_FBO}), which are not detected 
%previously in GX~340+0, but present in power spectra of all other Z sources during this state.

We also reveal  $\sim$6 Hz QPOs
% quasi-periodic oscillations 
in FB 
%the flaring branch state (FBO) 
during 
two ``dipping/flaring'' events ($R5$ data set), which are not detected 
previously in GX~340+0, but present in power spectra of all other Z sources during this state.
 More specifically, during first dipping phase of light curve 
[for example, see point  {\it B} $blue$ (MJD 51774.08) in Fig.~\ref{PDS}] 
we find  only strong VLFN component (see $blue$ histogram,
% 50016-01-02-18, 
in the middle panel of left column) without 
QPO feature. 
While during the last part ({\it exit phase}) of the intensity dip we find a broad QPO Lorentzian  centered  at 6 Hz (FWHM=$11.7\pm4.5$ Hz, rms=$6.3\pm1.0$\%) %, $\chi^2$=131 for 102 dof for a 67\% confidence level) 
in addition to VLFN  (rms=4.1\%$\pm$0.4\%, $\alpha_{LF}$=1.4$\pm$0.3, $\chi^2$= 139 for
102 d.o.f;  we present all parameter errors corresponding to 1$\sigma$ confidence level). 
The related point of the light curve is indicated by $green$ point $D_1$  observed on August 18, 2000 (50016-01-02-16, MJD=51774.27) seen in  Fig.~\ref{PDS}. We make the corresponding   
$\nu\times power$ diagram for this event   in 0.01 -- 150 Hz range (see  $left$ panel 
of Fig.~\ref{PDS_FBO}). 
We should point out  that the energy spectrum at point $D_1$ ($right$ panel of Fig.~\ref{PDS_FBO}) is  a typical FB spectrum [e.g., see also Z3 panel  of Fig. 7 in STF13
% \ref{Zsp_compar_RXTE}) 
with a relatively low contribution of the hard emission at 70$-$ 150 keV (hard tail), while at HB and NB the hard tail becomes stronger ($red/blue$ points A and C in Fig.~\ref{PDS}].

%Blue solid line shows a smoothed power spectrum. Broad QPO cetered at $\nu_{FBO}\sim$6 Hz 
%and %Power spectrum of GX~340+0 typically consist of three components: the broad-band noise with break $\nu_b$, 
%low frequency QPOs ($\nu_{sl}$, $\nu_l$) and 
%very low frequency noise component are presented in the power spectrum of GX~340+0 during flaring branch state (see text). 
%Similarly 
During another FB event (see $green$ point $D_2$) we detected the same behavior of PDS, 
characterized by the absence of QPO  at the beginning and middle of  FB but
the presence of broad centered QPO at frequency  about $\sim$6 Hz in  the {\it bottom phase} 
of FB  (50016-01-02-01, MJD=51776.68) in addition  to strong VLFN component. Note that the energy 
spectra for both points ($D_1$, $D_2$)  are very  similar (their best-fit parameters are mostly identical).
%the same parameters. Such similarity allows to conclude about the same source (system) configuration. 
%We also detected 6 Hz quasi-periodic oscillations at the dip %end part of an X-ray intensity dip 
%of X-ray light curve, 
%(1.5 days of total duration) 
%in GX~340+0, 
%which correspond to the {\it flaring branch} state in GX~340+0. 
%The  quasi-periodicc oscillations have properties that are different from those of flaring 
%branchh quasi-periodic oscillations seen in other Z sources, which are often ascribed 
%to oscillations in a radial inflow of matter. 
%The fact that the 6 Hz QPOs
%occur during a dip of the light curve  suggesting  that they arise in an 
%medium (
%area that partially  screened but relate to NB configuration of GX~340+0. %obscured the X-ray emission. 

 % Fig.17 inner  disk. . 
Note that the energy spectra of $D_{1/2}$ points are similar to that indicated by  B (blue) point (and generally typical to FB spectra, see  Z3 diagram of Fig.  7 in STF13
%\ref{Zsp_compar_RXTE}) 
but the related power spectra are completely  different.  In fact, we reveal  a broad QPO, centered at $\nu_{FBO}\sim$6 Hz, 
%has been revealed 
for the first time in  the PDS of GX340+0
% for this reason this power spectrum is shown separately 
(see Fig.\ref{PDS_FBO}).

Thus, we find  that the evolution of spectral and  timing characteristics is consistent  with  previous 
analysis of GX~340+0 using  EXOSAT, $Ginga$, {\it RXTE} and XMM-$Newton$ data 
[see  e.g. \cite{Penninx91}, \cite{Jonker00}].  
In addition to known timing features of  the {\it Z} pattern of GX~340+0 we reveal  $\sim$6 Hz FBO, which was  not found 
previously, while it is usually  observed in other {\it Z} sources such as Sco~X-1 and GX~17+2.
%It is worth to note that 
%In these %the 
%sources %Sco~X-l and GX~17+2 
This kind of  QPOs are seen in the lower FB, 
%whose frequency is also 
%associated with the same centroid frequency $\nu_{FB}$  %connects 
and also seen in  NB of those {\it Z-}sources. 
% $\nu_{NBO}$ ($\sim$6 Hz). %smoothly with the NBO . 
The frequency starts to increase from $\nu_{FB}\sim 6$ Hz to  $\nu_{NB}\sim$~20 Hz 
when these particular sources progress   from   FB  to NB.
%from the $\sim$6 Hz characteristic of NBO up to $\sim$~20 Hz. 
 In the lower FB,  this QPO frequency  $\sim 6$ Hz 
becomes very broad and  then vanishes in the noise. %No FBO have been reported from the other Z sources.
%Since the FBO are closely connected to the NBO frequency they probably arise from a related 
%process. We will discuss it further in conclusion section. %When a Z source reaches the FB, the mass accretion rate is thought to become super-
%Eddington. At this stage the radial inflow of matter becomes unstable and photohydrodynamic 
%oscillationss may be excited which cause the FBO (see e.g. Lamb 1989). 

Follow to \cite{KK96} and \cite{tlm98}, we can  interpret the appearance %new 
of this  QPO 
%quasi-periodic oscillations 
phenomenon during  FB % flaring branch state 
in terms of  the {\it Transition Layer}  (TL) 
model involving a radial flow and a thick, torus-like CC (see illustration of this interpretation in  Fig.~\ref{geometry_eclipse}).
To conclude the above analysis, we should  point out once again  that the {\it identification} of CCD states  of GX~340+0  along 
$Z-$track is made using  timing (presented here)  along  with spectral analysis (see \S 3).

\subsection{Comparison of spectral and timing  characteristics of Z source GX~340+0 and {\it atoll} sources GX~3+1 and 4U~1728-34} 
%as a function of mass accretion rate  \label{disc}}

In this Paper, we study the correlations between spectral, timing characteristics, and $\dot M$
% mass accretion rate 
seen  in
X-ray  spectral range  for 
% the Galactic bright LMXBs and 
 GX 340+0  during its evolution %transition 
across HB-NB-FB track  %between hard and soft states.
in order to 
%further 
investigate  a common behavior of  {\it Z} and {\t atoll} sources.  In this way we  proceed with  a comparative analysis for three sources: {\it Z-}source GX~340+0 and  {\it atolls},   GX~3+1 and  4U~1728-34 using  the same spectral 
model which consists of low temperature {\it Blackbody}, {\it Comptonized} continuum and 
{\it Gaussian} line components.

\subsubsection{Quasi-constancy of the photon index}

{\it Z-}source GX~340+0 and $atolls$, GX~3+1,  4U~1728-34 show 
%demonstrate 
an identical  behavior of $\Gamma$ vs $\dot M$ 
%mass accretion rate 
(or COMPTB normalization),  namely  the index $\Gamma$  is  strongly concentrated around 2 (see Fig. \ref{index_norm} for that in GX~340+0). 
% and  almost identical long-term variations of ASM mean count rate  (see also ST11). %Fig.~\ref{outburst_index_temperature2}).
According to FT11, ST11 and ST12, this result  can point out  that the cooling flux of the soft
disk photons is much less than the gravitational energy release 
in the {\it transition layer}  for all of  these three sources.

%This presumably can provide, according to FT11\&ST11, that the energy release in the transition layer for  these 
%two sources  is much higher than cooling flux of the disk photons (see FT11 and ST11 for details of  X-ray 
%spectral formation in TL). 

%    The relative variations of the COMPTB normalization  for GX 3+1 (from 0.04 to 0.15) and for  4U 1728-34
% (from 0.02 to 0.08)  are also similar.   Although variations of inferred electron temperature $kT_e$ 
% %as the best-fit parameter of COMPTB spectral component 
% is much smaller in the case of  GX 3+1 (from 2.5  to 4.5 %3.7 
% keV)  than that for 4U 1728-34  (from 2.5  to 15 keV).  
% Whereas variation of Comptonization fraction $f$  is a factor of 4 larger in the case of  GX 3+1 (from 0.1  to 0.9)  than that   for 4U 1728-34 (from 0.5  to 1).  
% % different for these while show different behavior of 
%%the electron temperature $T_e$ and COMPTB Normalization during state transition 

\subsubsection{The difference and similarity of the electron temperature $kT_e$ ranges in GX~340+0, GX~3+1 and 4U~1728-34}

%The difference in the ranges of the electron temperature $kT_e$ variations

%In % \ref{temperature_vs_hc}, , \ref{index_temperature12} 
One can see, using  Figure~\ref{T_e_vs_f_comp},  the ranges of 
the TL electron temperature $kT_e$ for  state evolutions of these three sources. 
As clearly seen in this Figure (see also Fig. \ref{norm_temp_3obj})
%{norm_temp_3ob}), 
$kT_e$ of {\it Z-}source GX~340+0 ranges   from  3  to  21 keV,  while that of  $atoll$ 
%source 
4U~1728-34  is slightly more  narrow,  from 3  to 15 keV and  $kT_e$ of  {\it atoll} GX~3+1 just varies
% within much narrow range at the temperature 
around 3 keV. 
Note that all objects have a common temperature interval, 3 -- 4 keV.
%where these sources behave in  the same time scales (see below), but show  quite different 
%timing properties.

 As it is shown in Table 5  variations of  of {\it blackbody},  and seed (NS and disk)  photon temperatures  are comparable 
%similar 
for all of these  three sources:  $kT_{BB}\simeq$0.6 keV and
$kT_s=1.1-1.7$ keV, respectively.
 %On the other hand, 
 But variations of $kT_e$ are different.
%of the electron temperature  are %quite 
%drastically different. 
%The electron temperature kTe changes in a wide range kTe = 2.5-15 keV for
%4U 1728-34, while for $atoll$ source GX~3+1 and Z source GX~340+0 $kT_e$ varies in a narrow range from
%2.3 to 4.5 keV (see Figures 8 and 9). 
%The reason for
 This  difference of $kT_e-$ ranges for all of these three sources  is quite evident. 
Whereas  GX~340+0 and  4U~1728-34 show a complete cycle of state evolution, 
namely, HB-NB-FB track for GX~340+0 and 4U~1728-34 evolves  from the extreme island state (EIS) to the upper banana (UB) state (see Di Salvo et al. 2001; ST11), GX~3+1 demonstrates only a short LB (lower banana)-UB track on the CCD. It is clear  from  Figure~\ref{T_e_vs_f_comp} which  shows the track of  GX 3+1 just for  a part of the full track
% according to the standard $atoll$-{\it Z} scheme 
[see definition of a state sequence in Hasinger \& van der Klis (1989)].

%On the other hand, 
But GX~340+0 shows almost the same  $kT_e$ range as that in $atoll$ 4U~1728-34, while it demonstrates  a different timing evolution. 
How different 
%The  differences between
 {\it Z} and {\it atoll} 
sources are one can clearly  see from Fig.~\ref{norm_temp_3obj}, which shows  the luminosity in terms 
$N_{Com}\times {D^2}_{10}$ versus $T_e$. 
 {\it Z-}source GX~340+0 ($pink$ points) is much brighter than {\it atolls}, 4U~1728-34 ($blue$ points) 
and GX~3+1 ($red$ points).

\subsubsection{Comparison of spectral evolution as a function of the
$luminosity$
% ($Comptb$ Normalization) 
for $\it Z$source GX~340+0 and $atoll$ sources GX~3+1 and 4U~1728-34}
% (Na Vashe ysmotrenie ????)}

Now we can make a  comparative analysis of  spectral parameter evolution for 
%{\it Z} source 
GX~340+0 and 
%$atoll$ sources 
GX~3+1, 
 4U~1728-34 based on $luminosity$ of soft disk photons,   which is proportional to 
% the $Comptb1$ normalization in the case of  GX~340+0, 
%({\bf Which Comptb? 1 or 2?})
%and, consequently, 
to disk mass accretion rate  and inversely proportional to square  of distances  to these sources (see Eq. \ref{comptb_norm} and Table 5).
We should remind a reader that  {\it atoll} sources 4U~1728-34 and GX~3+1 demonstrate
only one Comptonized ($Comptb$) component in their  X-ray spectra [\cite{ST11},
\cite{ST12}], while  {\it Z}-source GX~340+0 reveals two Comptonized components ($Comptb1$ and
$Comptb2$). The first $Comptb1$ component is
formed by up-scattering processes in TL for seed photons of temperature  $T_{s1}\le
1$ keV coming from the disk, while the second $Comptb2$ component is formed  as a
result of up-scattering processes  in TL for seed photons of temperature $T_{s2}\sim
1.5$ keV coming from the neutron star. In contrast with GX~340+0, the 
spectra of  {\it atolls},  4U~1728-34 and GX~3+1 contain only one Comptonized
($Comptb$) component, which is related to up-scattering processes  in TL for seed
photons of temperature $T_{s}\lax 1$ keV, presumably coming from the disk. Therefore we should
compare spectral components of the  same origin for considered {\it Z} and
{\it atoll} sources.  Namely, $Comptb1$ component (with $f_1$ illumination parameter) for
GX~340+0  and $Comptb$ components (with $f$ illumination parameter) for 4U~1728-34
and GX~3+1.

% taking into account that the distances to these sources are different (see 
%Table 5). 
Specifically, for GX~340+0 the distance is about 10 kpc (Fender \& Henry 2000), whereas 
for 4U~1728-34 and GX~3+1  the distances are 4.5 kpc~ and %in the range of 
4.2$-$6.4 kpc, respectively [see \cite{par78} and \cite{kk00}]  
%  [van Paradijs (1978)] 
%whereas for 4U~1728-34 it is 4.5 kpc~\citep{par78}.
%[Kuulkers \& van der Klis (2000)]. 

  %(bottom panels) 
We present the $luminosities$ 
% measured 
of % $\Gamma$ 
%COMPTB normalization 
these % two {\it atoll}
sources as a function of %COMPTB normalization (left), the Comptonization fraction $f$ (center) and 
%the electron temperature 
$kT_e$ in Fig. \ref{norm_temp_3obj}. For  
%{\it Z-}source 
GX~340+0 ($pink$), we show a variable  $N_{Com1}$ normalization only. 
%(right). 
%The last parameter is
%extremely suitable to compare spectral properties of different objects because of
%$kT_e$ is independent from source distance to Earth observer $D$. 
%As well as,
%parameter $kT_e$ is suitable to trace the source spectral states, particularly for
%NSs, because $kT_e$ decreases when sources move from the low state to the high state during mild
%variability as a result of a more efficient electron cooling by the increased seed
%photon supply. Moreover, the electron temperature $kT_e$ is a directly measurable
%quantity. 
GX~340+0 demonstrates much higher luminosities by factor 5 than  that 
 %for $atoll$ sources 
 in 4U~1728-34 and GX~3+1 and  %while 
%the electron temperature 
$kT_e$ 
%of TL
%{\it Compton cloud}
in GX~340+0 widely varies, 
%in the wide range 
from 3 to 21 keV.
% (similar to that in 4U~1728-34). 
%On the other hand, $atoll$ sources 4U~1728-34 and GX~3+1 
%demonstrate  the same range of  luminosity, but their CC electron temperatures  are   different.  
%Thus, individual properties of sources are defined by combination of many parameters.
%a wider  range of COMPTB
%normalization (presumably proportional to mass accretion rate) by  factor of 2  than that for  4U~1728-34 
%(along vertical axis of Fig.~\ref{outburst_index_temperature2}), while the electron temperature $kT_e$ varies 
%only from 2.5 to 4.5 keV (along horizontal axis of Fig.~\ref{outburst_index_temperature2}).
%varies , resting close to  3 keV.
According to FT11, ST11 and the presented study, 
%the electron temperature 
$kT_e$ for  {\it atoll} and {\it Z}-sources varies  from 2.5 to 25 keV. 
Specifically, a common range of $kT_e$ for GX~340+0, GX~3+1 and 4U~1728-34 is around 3 %3.7 
keV and the  low limit of $kT_e$  around  2.5 keV occurs 
%takes place
at the  peak luminosity for 4U~1728-34 (see ST11) and during a  rise of luminosity for GX~3+1 
and  during FB stage for GX~340+0 (see Figs. \ref{T_e_vs_f_comp}, \ref{norm_temp_3obj}). %, i.e. during UB state of $mild$ variability.
%It is worth  noting 

We should note  that  common states in $L$ vs $T_e$ diagram for all these sources take place during rapid transitions in FB and LB -- UB  for GX~340+0  and   GX~3+1, 4U~1728-34, respectively.   These states can be  possibly traced  within a narrow $kT_e$ interval   from 2.5 to  4.5 keV.
%We remind a reader that along the track  of atoll sources 4U~1728-34 and GX~3+1 there are  a sequence of
%CCD states listed according to standard atoll-{\it Z}
%classification \citep{hasinger89}. They are an  island state (IS),  extreme island state (EIS), lower banana state (LB) and upper banana state (UB).
%IS --  island state,
%LB -- lower banana state  and 
%UB -- upper banana state) which  are listed according to standard atoll-Z
%scheme~\cite{hasinger89}.

% Note also that $kT_e$, collected by FT11
% for a number of  NS sources, is concentrated around  $kT_e\sim$~3 keV, 
%which, as will be shown further, indicate the presence of $banana$ states for these sources.
%Possibly  this value of the electron temperature is related with crucial
%value for stable transfer of matter from the accretion disk onto NS through
%transition layer.

\subsubsection{The difference and similarity of   time scales for state evolution for GX~340+0, GX~3+1 and 4U~1728-34}

We point out  that all these three objects show a full range evolution %transition 
track but in  different 
time scales. Specifically, 
{\it Z-}source  GX~340+0 and 
{\it atoll}  
GX~3+1 demonstrate  
electron temperature and $luminosity$ %normalization 
variations occur on the time scales of hours-days, while {\it atoll}  4U~1728-34 
show a full range of the model parameter variations on significantly larger time scale (days-months).  
However, 
%it should be noted that  
the low electron temperature phase  of {\it atoll}  4U~1728-34  occurs
% takes place 
at the same short  time scale as that for GX~340+0 and GX~3+1. 
% Consequently, one can argue that low values 
%of the electron temperature of CC  can be an indicator of short % small 
%time scales for CCD stage  evolution in  NS X-ray binaries.

\subsubsection{Correlation of illumination parameter $f$ with  electron temperature $kT_e$ and its relation with different states in the color-color diagram}

%The Comptonization fraction parameter $f$ traces different ranges of states on color-color diagram}

%As one can see from , %the bottom panels of %Finally, middle of  Figure \ref{outburst_index_temperature2} 
Using Table 5 one can notice that 
the variation  of Comptonization fraction $f$ is  larger   for GX~3+1 ($0.2-0.9$) than that for 4U 1728-34 ($0.5-1$) 
and GX~340+0 ($0.01-0.5$).  Thus it is obvious that  illuminations   of the Transition 
Layer by soft photons  are different for these three sources. 
%For 4U~1728-34 and GX~340+0 the solid angle viewed from NS changes by factor 2  
%whereas in  GX~3+1 the illumination of TL by photons coming from NS surface  and  thus the solid angle  
%changes by factor 4. 
%However the photon index  $\Gamma$ is almost constant, around %2 for these three sources which means the energy  release in %the transition layer for  these three  sources  is much higher than cooling flux of the disk photons 
%(see FT11, ST11 and ST12 for details of  X-ray spectral 
%formation in TL). 

In  Figure~\ref{T_e_vs_f_comp} we plot the electron temperature $T_e$ versus Comptonized fraction $f$
%$f=A/(1+A)$  
for {\it Z-}source 
GX~340+0,  and {\it atolls},  GX~3+1 and 
4U~1728-34 during different spectral states.
% transitions. % $$mild$  variability. 
{\it Red}, $green$  and {\it blue} points correspond to 
{\it RXTE} %/{\it Beppo}SAX 
observations of GX~340+0, GX~3+1 and 4U~1728-34 respectively. 
%{\it Pink/bright blue}  and {\it blue/green} points correspond to 
%{\it RXTE}/{\it Beppo}SAX 
%observations of GX~3+1 and 4U~1728-34 respectively. 
On the left hand side of the  Figure one can see  a sequence of CCD states 
HB, NB, FB   (the horizontal, normal and flaring   branches respectively)  
%NB --  the Normal branch and 
%FB -- Flaring branch) 
which  are presented  according to the standard {\it Z-}scheme~\citep{hasinger89}.
%On the right-hand side of the  Figure we show  a sequence of CD states 
%(EIS -- the extreme island state, 
%IS --  island state,
%LLB -- lower left banana state,
%LB -- lower banana state and 
%UB -- upper banana state) which  are listed according to standard atoll-Z scheme~\cite{hasinger89}.
%One can see that  
%which here matched by 
%the electron temperature 
Based on the obvious  correspondence  between $kT_e$ and spectral states (Fig.~\ref{temperature_vs_hc}) 
and timing evolution of GX~340+0, 
%[see {\it incorporated panel} ({\it top center})], 
$kT_e$  is directly related to 
the sequence of CCD states  in  GX~340+0. Note that recently ST12 revealed one-to-one correspondence between 
spectral states, 
%the electron temperature 
$kT_e$ and  $f$ for {\it atolls},  GX~3+1 and 4U~1728-34.  
% with the digitalization (calibration) on left vertical axis. 
The direction in which the inferred $\dot M$  increases is indicated by arrows.

In Fig.~\ref{T_e_vs_f_comp} one can see  three different tracks  on the plot
of $kT_e$ versus $f$ for GX~340+0,   GX~3+1 and 4U~1728-34 and how these tracks  are related 
to  the standard sequence of CCD states.
When 
%the fraction 
$f$ rises, 
%the electron temperature 
$kT_e$ monotonically decreases  from 
%approximately 
21 keV to $\sim$~3 keV for GX~340+0, and  from 4.5 keV to $\sim$~2.3 keV 
for GX~3+1, while 4U~1728-34 shows  a more complex example
%complicated 
%behavior pattern. 
At high temperature state (EIS), 
when  $kT_e$ decreases,  $f$ slightly changes from 0.9 to 1.  Further as $kT_e$  drops  from
12 keV to 4 keV, $f$ decreases from 0.9 to 0.5. Ultimately, at  the low-temperature state (LB-UB) %state transition 
$f$ surges from 0.5 to 1. 
Thus, one can see that the CCD state evolution can be observable 
as  a relation between $kT_e$ and $f$ too. 

\subsection{Comparison of spectral hardness diagrams for atoll, {\it Z} and BHC binaries \label{ccd}}

\begin{figure}[ptbptbptb]
\includegraphics[scale=1.1,angle=0]{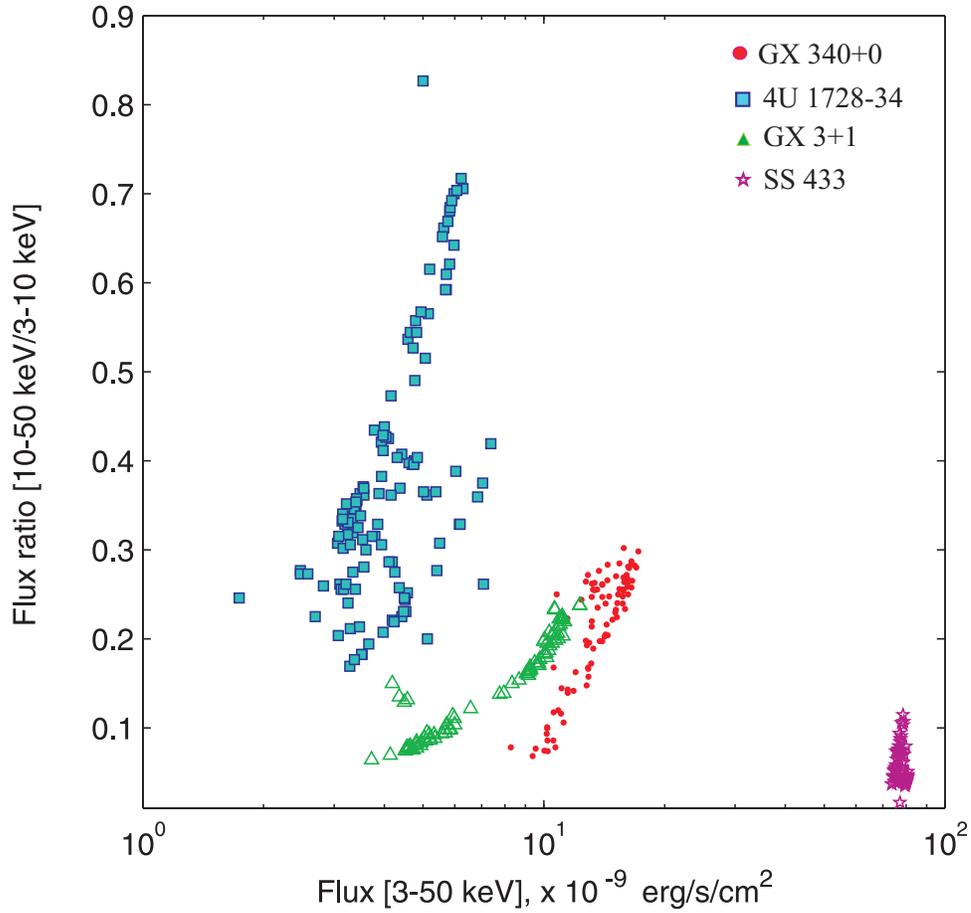}
\caption{
Spectral hardness (10 -- 50 keV/3 -- 50 keV) vs flux in  3 -- 50 keV range  of Z source GX~340+0 
($red$), $atoll$ sources 4U~1728-34 ($blue$, taken from ST11) and GX~3+1 ($green$, taken from ST12), and 
BHC SS~433 ($violet$, taken from ST10) for {\it RXTE} data.
%
%{\it Left:} The ``color-color diagram'' %[$\frac{3-10 keV}{10-50 keV}$ %
%[30 -- 10/10 -- 50 keV flux ratio versus 10 -- 50/3 -- 50 keV flux ratio] of GX~3+1 ($pink$) and 
%4U~1728-34 ($blue$) during faint-bright %LS -- HS 
%transitions (long-term variability). 
%{\it Right:} Fragment of ASM light curve of GX~3+1 %is shown 
%for (schematic) demonstration (illustration) of two kinds 
%of flux variability, wherein the long-term time trend (faint-bright) %LS -- HS) %with high amplitude
%associated with COMPTB normalization changing   and the scale %direction 
%of short-term time variation (LB -- UB) %with low amplitude 
%associated with the electron temperature of Compton cloud changing
% are indicated. The blue line shows 
%a mean count rate and indicates said long-term variability of GX~3+1 soft flux.
%$power$ diagram of GX~3+1  observed on September 1, 2010 
%(94307-05-01-00 {\it RXTE} observation, MJD=55441).
}
\label{HID_4object}
\end{figure}

\begin{figure}[ptbptbptb]
\includegraphics[scale=0.9,angle=0]{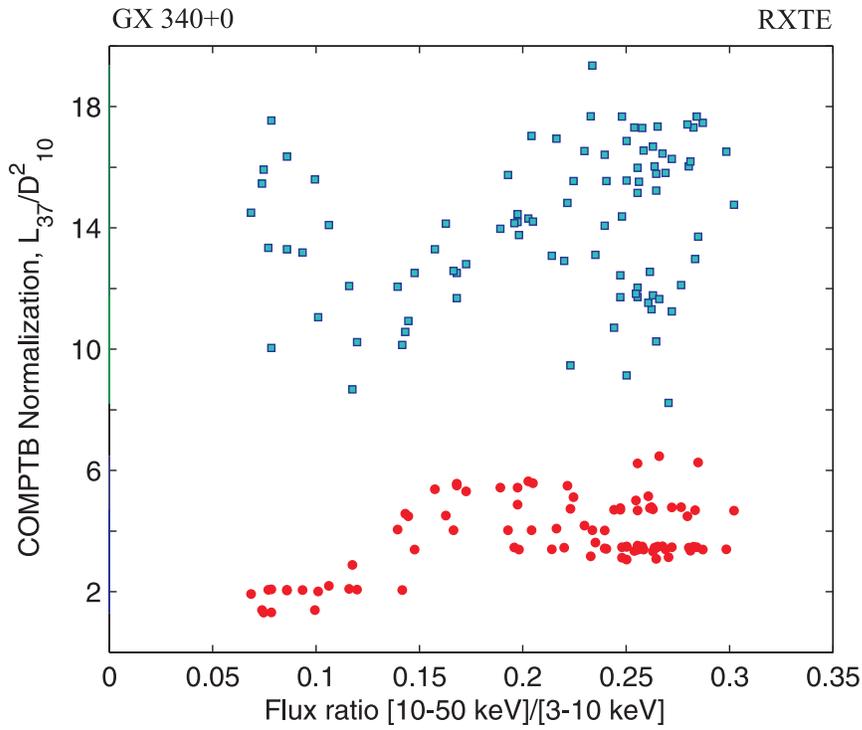}
\caption{
The normalizations $N_{com1}$ ($red$) and $N_{com2}$ ($blue$) of $Comptb1$ and $Comptb2$  components, respectively, 
are plotted versus flux ratio [10 -- 50 keV/3 -- 10 keV]
using  our spectral model 
$wabs*(blackbody+Comptb1+Comptb2+Gaussian)$ during evolution %transition 
events. 
%($Left:$)  photon index $\Gamma$ is  plotted versus electon temperature $T_e$ (in keV) of $Comptb1$ component  
%%
%%COMPTB normalization 
%%$N_{COMPTB}=L_{39}/D^2_{10}$ 
%and  versus
%Comptonized fraction $f=A/(1+A)$  on the ($right$) using  our spectral model 
%$wabs*(blackbody+Comptb1+Comptb2+Gaussian)$ during transition events (see Tables 3, 4). 
%$Red$ and $blue$ points correspond to {\it RXTE} and {\it Beppo}SAX  observations of GX~340+0 
%respectively. 
%%Photon index plotted versus electon temperature $T_e$ (in keV) of $Comptb1$ component 
% in the frame of  our spectral model 
%$wabs*(blackbody+Comptb1+Comptb2+Gaussian)$ during transition events (see Tables 3, 4). 
%$Blue$ and $pink$ points correspond to {\it RXTE} and {\it Beppo}SAX  observations of GX~340+0 
%respectively. 
}
\label{index_temp_cmpt}
\end{figure}

To compare evolution %transition 
properties of GX~340+0 with {\it atolls} and BHC sources in terms of flux (or luminosity), 
 %define the hard color (HC) as a ratio of the fux in the 10-50 keV to that in the 3-50 keV energy band. 
%, and the soft color (SC) as a ratio of the flux in the 3-10 keV to that in the 10-50 keV energy band. 
we plot HC (10-50 keV/3-50 keV) versus the flux in the 3-50 keV range  %SC,  
to create our hardness-intensity diagram (HID) for four sources (see Fig.~\ref{HID_4object}):
{\it Z-}source GX~340+0 ($red$), {\it atolls}, GX~3+1 ($green$) and 4U 1728-34 ($blue$), 
and BHC SS~433 ($crimson$). We should remind a reader that  \cite{ST10}, hereafter ST10,  found a correlation of the index with mass  accretion rate $\dot M$   
and a clear signature of the index saturation when 
$\dot M$ increases. Thus ST10  argue that SS~433 should be a black hole because the index changes and saturates with $\dot M$ 
%mass accretion rate 
in BHs only.  

As it appears from  Figure \ref{HID_4object},
NS ($atoll$ and {\it Z})  binaries trace a specific elongated %$\varepsilon$- or C-shaped 
or diagonal track in this HID over a larger range of luminosity and (hard) colors than that of BHC SS~433. However, 
$atoll$ sources systematically are fainter than   {\it Z}  and BHC sources and  show much harder spectra when they are in the low state  [see e.g. \cite{vdKl94apj} for details of comparison of NS and BH sources]. In fact, when {\it atolls} in the soft state, their spectra are pretty similar to the spectra of 
{\it Z-}sources. 
%On the other hand  
However {\it Z-}sources are usually softer than $atolls$
% sources, 
{\it but not in the case of GX~340+0}  which  
%It is worth noting  that 
%GX~340+0 
is an {\it intermediate case} between $atolls$ and BHs in terms of their luminosities. This fact  possibly reflects 
{\it intermediate} rate of mass transfer in GX~340+0 among %Z 
$atoll$ and BH 
sources [see also  a review by van der Klis (1994)].

The resulting luminosity  in GX 340+0 is  in the range of (0.3 -- 0.7) $L_{\rm Edd}$.  This estimate can be derived using the NS seed photon temperature  which varies  in the 1-1.7 keV range (see Table 3).  Type-I X-ray bursts  are not observed  in GX 340+0. This observational fact   can be a consequence    
of  high  $\dot M$ 
%mass accretion rate 
in this source.  Usually  type-I X-ray bursts are seen in low-hard states of NS when the resulting radiation pressure between outbursts is quite low. Thus nothing can stop to accumulate plasma in the NS surface until the plasma column reaches its critical weight which leads to the explosion (X-ray burst).  Thus a certain level of the surface weight should be attained in order to start an X-ray burst.
%Generally, in the nuclear burning model for X-ray bursts, a well-determined surface  weight  must exist for an X-ray burst to occur. 
%\cite{nh02}, hereafter NH02,   performed a stability analysis of accumulating fuel on the surface of a  compact object.  In Fig. 1 of their paper
%\cite{nh02}, hereafter NH02   showed for solar composition material accreting on two kinds of compact objects  (NS and BH). They  considered a range of accretion rates, parameterized by the ratio $L_{acc}/L_{\rm Edd}$, where they  took the Eddington luminosity to be  $L_{Edd}= 4\pi GMc/k_{es}$ with $k_{es}=0.4 cm^2 g^{-1}$. Their results correspond to three choices of the temperature at the base of the accreted layer: $T_{in}= 10^{8.5}$, $10^8$, and $10^{7.5}$ K. 
\cite{nh02}
% hereafter NH02 
% Their calculations shown in Figure 1 in NH02  
demonstrated  
%NSs are unstable 
that X-ray bursts (instability)  can occur for a wide range of the temperature 
% for a wide range of the temperature at the base 
of the accreted flow  $T_{in}$ (and corresponding emergent luminosity $L$).
% changes in  wide limits. 
Although this instability is less probable   
% But 
%the width of the instability strip 
%(as a function of $L_{acc}$) 
%is less 
for higher  $T_{in}$.
% which corresponds to higher values of $\dot M$ (and consequently to high values
%of  the emergent  luminosity $L$).
% mass accretion rate. 
%The reason for this is clear from the analysis of \cite{p83}
%Paczynski (1983)  
%who showed that when the flux escaping from the stellar core into the accretion layer increases (which happens when 
%$T_{in}$ increases), bursting behavior is restricted to a smaller range of the surface density.   
Probably this is precisely  our case of GX 340+0 where  $L$
%the emergent luminosity 
is order of Eddington or little bit less ($L\gax 0.5L_{\rm Edd}$). 
%In fact, in our case of GX 340+0   the NS seed photon temperature is quite higher and it changes from 1 to 1.7 keV, i.e. the emergent luminosity is of order 0.3-0.7 L_{Edd}. 
%NH02 also  argue  that if Black Hole Candidates (BHCs) had surfaces, they would be expected to exhibit X-ray bursts. Thus  the lack of type I X-ray bursts in BHCs can be considered as indication for an event horizon. 

%It is possible to add some text in relation with "LB, UB etc" in the caption of
%corresponding Figure:

%Along the track  of atoll sources 4U~1728-34 and GX~3+1 we indicated  a sequence of
%CD states 
%(EIS -- the extreme island state, 
%IS --  island state,
%LB -- lower banana state  and 
%UB -- upper banana state) which  are listed according to standard atoll-Z
%scheme~\cite{hasinger89}.

%The exact causes of the spectral and timing
%variability is still unknown, but it is thought that the differences between the two classes result from a higher rate
%of mass transfer in Z sources than atoll sources (see van der Klis 1995, for a review).

\section{Discussion}

%In this Paper 
To further study 
 the similarities between  {\it Z} and 
{\it atoll} sources  we have investigate  how 
%the relations between 
X-ray spectral, timing characteristics 
%properties, 
and mass  accretion rate $\dot M$ are related 
in 
%observed in X-rays from the Galactic bright 
 {\it Z-}source GX 340+0  during the evolution %transition 
across HB-NB-FB track. %between hard and soft states.
 This investigation  helps us to  constrain the models related  to X-ray spectral  formation in  compact objects. 
%under binary condition at the late evolution stges. %the different QPOs. 

%\subsection{A monotonic increase of mass accretion rate along Z-track}
\subsection{Does mass accretion rate monotonically  increase   during HB$\to$NB$\to$FB evolution?}

%From one hand, obtained in this Paper 
We observe a quasi-constancy of $N_{Com2}$ % and decreasing of $N_{Com2}$ 
%(more exactly, low variability of the normalization of $Comptb$ component near its quasi-constant level)
during HB-NB-FB evolution, %transitions, 
which can be presented as an argument against  the commonly adopted view that
$\dot M$ monotonically increases along the {\it  Z-}track, %(increasing in the direction HB$\to$NB$\to$FB), 
since $\dot M$
%the mass accretion rate 
is related to the normalization of the  Comptonized component.
% $N_{com}$.  
In fact, $N_{com}$ 
%this normalization of the  Comptonized component 
(see Eq. \ref{comptb_norm}) is proportional to the soft photon luminosity which is  its turn proportional to $\dot M$.
%mass accretion rate.    
%$normalization$ parameter $N_{Com1}$. 

We detect  spectral and timing evolution of GX~340+0 similar  to %corresponding 
that of {\it dipping} {\it Z-}source  Cyg~X-2, where we  see that the spectral  features of the source are   strongly 
influenced  by its  high inclination angle 
(Vrtilek et al., 1999, Titarchuk et al. 2007).  Cowley et al. (1979) reported that  Cyg~X-2 is observed at an inclination between 
65$^\circ$ -- 75$^\circ$. 
\cite{F09} analyzing $Beppo$SAX observations of Cyg~X-2 and using a  sum of two 
COMPTB components, successfully described  the spectral behavior
%evolution 
of the source from its HB 
to NB. Consequently we can suggest that  modulation  of  
the flux from the TL outer part  and  TL innermost part  
%(NS boundary layer) 
(related to $N_{Com1}$ and $N_{Com2}$, respectively)  is a result of  geometric  effects.  

%In fact, we find only low encreasing of the integral normalization of spectral components 
%($N_{BB}+N_{Com1}+N_{Com2}$) during HB$\to$NB$\to$FB transition, which indicate on increasing 
%of mass accretion rate $\dot M$. 
 In fact, there is no  direct information on  the binary inclination angle $i$ of GX~340+0, based on radial velocity analysis.  
 The only indication on inclination angle $i$ of 35$^\circ$ for GX~340+0 is  deduced using the so called 
``relativistically smeared out'' profile of 
Fe~K$_{\alpha}$ line.  \cite{d'Ai09} suggest  that  the iron line  profile can be produced as  a result of  reflection of  X-ray  hard radiation,  forming in Comptonization region, from relatively cold NS and disk surfaces.  
But the spectral features due to reflection should be seen only if the spectral index of the incident flux onto NS or disk surfaces is essentially less than 1 \citep{LT07} not like in our case of GX 340+0 (see Figs. \ref{index_norm}-\ref{index_temperature12}).

%}Laurent & Titarchuk 2007). In our case the spectral index is around one which means one cannot see so called the reflection bump in the emergent spectrum.  
An  assumption of high inclination angle of GX~340+0  leads us to suggest  that  $\dot M$ can be greater at 
FB, when  we can possibly  see only some relatively small part of X-ray emission from  NS surface  due to Comptonization region screening effect (see Fig.~\ref{geometry_eclipse} for explanation of geometrical constraints). 
%Comptonizing effects of full NS surface etc].
% caused by (possible) high inclination of GX~340+0 and puffed up CC configuration during flaring events.
%Note that the spectral evolution of ~X-2, which demonstrate similar to GX~340+0 spectral and timing properties  can be provided by two Comptonized areas.

Furthermore, in the case of low inclination angle we would also see strong BB component and monotonic increase of the 
normalization of Comptonized component %$\dot M$ 
along HB -- FB track. However, our spectral analysis indicates low BB contribution during all {\it Z-}states (see Sect.~4.2.1). 
This fact can be considered as an additional argument in favor of high inclination angle for binary GX~340+0 
(see discussion in Sect.~5.3). In turn, the influence of high inclination on the soft X-ray flux
% from the object 
can result in some distortion 
of real $normalization$ value. Thus, the normalization parameter $N_{com}$ is not a reliable indicator for $\dot M$ in the GX~340+0 case. 
%$normal$ branches. 
%with growth of so called {\it hard tail} at $upper$ HB.

%Furthermore, in previous simultaneous radio (VLA) and X-ray ($Ginga$) observations of GX~340+0~\cite{oost94} 
%it was found decreasing of radio emission (at 1.4 GHz and 4.8 GHz wavelength) during transition 
%from $Normal$ to {\it Flaring branch}. %on {\it color-color diagram} of GX~340+0. 
%This transition episod (event) is asociated with decreasing of [1.2 -- 37 keV] X-ray flux registered 
%by LAC/$Ginga$ detector, which proves in favor of a global radio/X-ray correlation on time scale of week. 
%Furthermore, ~\cite{oost94} stated % which proves %in favour of %a positive 
%correlation between X-ray and radio flux on a time scale of day for GX~340+0 during $normal$ branch. 
%However revealed $positive$ correlation  is inconsistent with well known radio/X-ray $anti-correlation$ 
%observed for many BHC and NS hosting sources during X-ray state transitions acompanied by radio flares. 
%%In particular, Z source GX~5-1 demonstrate clear radio/X-ray anti-correlation in the $normal$ branch (Tan et al., 1992).
%This fact can be considered as indication of possible attenuation of X-ray flux by above mentioned screening effect, 
%which can modify real X-ray flux evolution and own mass accretion rate indication.

\subsection{Detection of QPOs in  the  Flaring Branch of GX~340+0 
%(Na Vashe ysmotrenie ????)
}

In this study we report a discovery of 6 Hz  QPO in the FB of GX~340+0.  Specifically,  when the 
source enters the upper part of  FB, a QPO peak appears with a central frequency of 
about 6 Hz. This part of  FB corresponds to a dip (more specifically, the final part of the dip) in the light curve. 
%Although these QPOs are u observed in  FB, 
Note
%It is worth noting 
that these QPOs  exhibit a number of clear differences with  respect to
FBOs that were observed in Sco~X-1 and GX~17+2.  In these two sources, there is a continuous 
evolution %transition 
between the NB  and FB and a value of the QPO frequency steeply rises near the NB-FB state evolution, %transition, 
but never disappears (see Dieters \& van der Klis 2000, Penninx et al. 1990). But  in our data the 6 Hz QPO  is clearly  seen in NB (which becomes broad in  lower FB), but there is  no QPO in the beginning and middle of FB. 
%Only in the upper FB do QPO appear once again. 

%In Sco~X-1 and GX~17+2  QPOs appear in the lower FB
%~10\% of the FB 
%(see Dieters \& van der Klis 2000, Penninx et al. 1990), in Cyg~X-2 in the upper FB 
%$\sim$20\% of the 
%FB. 
The QPO frequency seen in Cyg~X-2 is 26 Hz, is the highest ever seen in FB of  any 
{\it Z-}source, with a FWHM of only 3 Hz (Wijnands et al 1998).
%({\bf it should be reference};
In Sco~X-l and GX~17+2 FBOs can be observed up to
%to frequencies 
 $20$ Hz as the source moves along  FB, but at that point the QPO  usually 
 becomes very broad (with the width of $\sim$10 Hz, see e.g. Dieters \& van der Klis 2000), and further up in 
the FB, this QPO drops  into the background noise components (Middleditch \& Priedhorsky 1986). The FBOs, 
which are  seen before,  have been related to a hard energy spectrum, whereas the QPOs that we reveal, are associated with a soft spectrum. We should note that Hasinger et al. (1990) reported some excess of power between 5 and 20 Hz in GX 340+0 during a dip   in FB, which they interpreted as a possible FB QPO. 
%This excess may be due to the broad NBO in the lower FB, which contaminates the power spectra at these frequencies. 
In view of the differences in the FB  properties, we suggest that the 6 Hz QPO which  we see 
in  FB, may be caused by a similar mechanism as the usual N/FBO.  It is remarkable that the 
QPO disappears when the source enters into the dip and appears once again when GX~340 leaves  
%exiting from 
the dip. 
This can suggest that during the dip the emergent emission is suppressed because  
%the introduction into 
the line of sight to the source is affected  by an absorbing or scattering medium.
% that at the same time 
%attenuates 
%The observed flux is attenuated  and causes the QPO by oscillating near 6 Hz. 
%This interpretation is strengthened by the fact that the energy spectra related to the QPO event and that during the  dip are  the same.  
%According to the model for the N/FBO by 
Fortner et al. (1989) suggest that  NBO arises due to a feed-back 
loop between the radiation force and the accretion rate  in an approximately radial inflow fed by 
matter that was initially in the accretion disk, but has lost its angular momentum due to photon 
drag. In this model, FBOs are also  assumed to originate as a result of  oscillations in this radial flow, 
at super-Eddington accretion rates. 

%(One possible explanation for the increase in QPO frequency up the 
%FB is that when Lx ~ ZrEdd> the flow is no longer stable against matter/radiation separation, 
%andd the infall velocity and therefore also the QPO frequency in the matter-dominated parts 
%off the flow increase. In this regime the coherency parameter Q of the oscillations diminishes 
%rapidlyy due to the inhomogeneity of the flow.) 
%Independently of this idea of 
%
In this way, as it has been figured out by a number of authors 
(e.g.  van der Klis et al. 1987, Alpar et al. 1992 and  Titarchuk \& Osherovich 2001,),
near  the Eddington limit
% the innermost part of the accretion flow
the radiation-pressure dominated region of the disk 
might  be puffed  up 
%to a Compton cloud (torus-like) shape 
which for certain inclinations could partially obscure the innermost part of the X-ray emitting 
 regions, and their  oscillations.
% in such a torus could provide an %alternative 
%explanation for N/FBO 

%An argument against 
%this is that this would 
%%work only for a rather restricted range of geometrical configurations where the observer, the edge 
%(or semi-transparent part) of the torus and the emitting region are on one line (Lamb 1989). 
%We submit that the Fortner et al. (1989) radial-flow model explains the usual N/FBO, but 
%that oscillations in a thick, torus-like inner disk that partially obscures the emitting region 
%explains the 26Hz QPO that we have found in the upper FB of CygX-2. This type of QPO 
%would be expected to be relatively rare and usually not very persistent because of the required 
%line-up of observer, torus and emitting region; this is in accordance with our observation of only 
%one such a QPO peak during 800s in all EXOSAT CygX-2 observations (a total of 4xl05 s) 2. 

\subsection{Possible indication of a  high inclination of binary GX~340+0}
% (Na Vashe ysmotrenie ????)}

The fact that the intensity shows a dip, when the QPO appears, can be an indication that  the torus 
enters into the line of sight.
% to the source. 
%The QPO frequency of 26 Hz corresponds to the 
%Keplerr frequency at a distance of ~190 km from the neutron star, consistent with the position 
%of the outer regions of the radiation pressure dominated region (see Frank et al. 1992) and of 
%the radial flow region (see Lamb 1989). 
%Combining the Fortner et al. (1989) radial-flow model with
Adopting a torus-like shape for the innermost part of the source
when the source becomes near-Eddington luminosity regime we can explain the differences in the shape of 
 FB in  HID between different {\it Z-}sources and the correlated differences that are observed 
in N/FBO properties (see Kuulkers et al. 1994). All these differences can be due to differences  
in binary inclination angles (Fig.~\ref{geometry_eclipse}). 
We suggest that the sequence of events, that is observed,  takes place when the source 
approaches the Eddington limit.  It is possible that  
%is mostly due to two gradual processes: 
a gradual increase in the geometrical thickness of the inner region (Compton cloud) is due 
to a gradual increase  of total $\dot M$
%total mass accretion rate
%arried by the radial flow 
(Fortner et al. 1989). 
%Because  the presence of the 
%torus-like Compton Cloud,  the region in which the radial flow takes place is not spherical, such as, for example, cone-like 
%in shape. The opening angle of the cone decreases when $\dot M$ increases. 
What the observer sees strongly depends on the binary inclination angle $i$. At low $i$, the torus never comes 
 to the line of sight and thus during FB  the X-ray intensity (or flux) always increases,  
 the N/FBO are seen all the way between 6 and 20 Hz (e.g. Sco X-1, GX 17+2). But at  
higher inclinations $i$,  FB is seen out by the Earth observer   when  the flux increases and thus the Compton cloud (torus) shielding the NS enters into the  line of sight, similar to that in case of Cyg X-2 (see illustration of this event in 
Fig. \ref{geometry_eclipse}). 

In general, sources observed at angles $> 65^0$, the so-called dipping sources, show regular dipping, modulated at the orbital period by the presence of an extended bulge. 
%This may not be  the case 
The observational situation can be different for GX 340+0  where achromatic dipping seems  to relate to some turbulent accretion episodes at the inner edge of the disk. For this geometry, one can argue that 
inclination should not be so important.  However if the inclination angles  $i$ are quite small  one can see the direct BB component of the NS while in the case  of GX 340+0 this NS BB component is not observed and it can be an additional argument that the Earth observer sees  GX 340+0 at relatively high inclination angles. 

% the FB turns around to lower flux as we seen for GX 340+0. 
%At even higher $i$ (CygX-2), the  FB would be directed towards lower fluxes exactly as seen for GX~340+0. 

In fact, if  the transition layer (or Comptonization  region) is puffed up then the NS BB direct radiation is not seen and in this case there is no dependence 
on the inclination angle. But if the luminosity less than Eddington  then there is a difference between the low  
an high inclination cases. At the former case one can see the direct BB component of the NS while in the latter  case the BB component is not seen. The NS BB radiation is up-scattered off Compton cloud plasma 
(the TL innermost part). 
%of the Transition Layer). 
This can be  our case, because independently of the mass 
accretion rate (or spectral state), the illumination factor $f_2$ of the NS soft photons is always about 1 
(see Sect.~4.2.1). It means that we always 
see this source at high inclination angle independently of $\dot M$.
% mass accretion rate.   

\section{Conclusions \label{summary}} 

In this Paper, we demonstrate the results of our study of 
% of  the correlations between 
spectral-timing correlations and  their dependences on mass accretion rate observed in
 GX~340+0  with the {\it RXTE} 
 %{\it Rossi} X-ray Timing Explorer  
 and {\it Beppo}SAX.
 % satellites.
% from the Galactic bright LMXB and
 %Z source   during the transition across HB-NB-FB track.  %between hard and soft states.
We study  
%investigate 
the similarities between {\it Z} and 
{\it atoll} sources to make constraints and  figure out models which can explain  X-ray spectral formation in these compact objects. 
%under binary condition at the late evolution stges. %the different QPOs. 
%For our  analysis 
In our study we use   the broad  energy band 
%spectral coverage 
and  adequate spectral   resolution of the  $Beppo$SAX detectors  in the 0.1$-$200 keV range
%from 0.1 to 200 keV 
combined   with the {\it extensive}  {\it RXTE} observations  in the 3$-$150 keV range.

  We find that the X-ray broad-band energy 
spectra of GX 340+0  
%during all %these 
%{\it Z}-states %transitions %spectral states 
can be properly 
%adequately 
reproduced by  an additive model, composed % sum % composition 
of  low-temperature $Blackbody$ component, 
two Comptonized %({\it COMPTB}) 
components with different  {\it seed} photon temperature ($T_{s1/s2}$=1/1.5 keV).
% which is presumably related to {\it Compton cloud}    
%and second component [{\it Comptb2},  $T_{s2}$=1.5 keV], which associated with 
%neutron star (NS) surface,   
%a $hard$ component ({\it Comptb1}, photon index $\Gamma_1\approx$2) 
%with ``seed'' photon temperature $T_{s1}$=1.2 keV %with turnover at high energies 
%and {\it soft thermal} component ({\it Comptb2},  $\Gamma_2=$1.7 -- 3.1) 
%with characteristic color temperature $\sim$2 keV, 
We should also  add the iron-line ({\it Gaussian}) component 
%in order 
to fit the {\it Beppo}SAX and {\it RXTE} data of GX 340+0.
Spectral analysis using our  model demonstrates   that the photon 
%power-law 
indices $\Gamma_1$ and $\Gamma_2$ of these Comptonized components only slightly vary around %value  are almost constant 
2, namely $\Gamma_1$=1.99$\pm$0.02 and  $\Gamma_2$=2.00$\pm$0.02. While 
 the electron temperature $kT^{(1)}_e$ of the outer part of the transition layer (TL)  
changes from 21 to 3 keV  and the electron temperature of  the TL  innermost part
% of the Transition (boundary) Layer 
$kT^{(2)}_e$ varies around 3 keV 
%during %these 
%spectral transitions 
along {\it Z}-track 
from HB 
% {\it horizontal branch} %(HB) %via {\it normal brach} 
to FB.
%{\it flaring branch}. % (FB). 
%We find that changes for the electron temperature $kT^{(2)}_e$ of NS surface %(layer) 
%is close to that of CC ($kT^{(1)}_e$), 
%%the electron temperature $kT^{(1)}_e$ of CC, %Compton cloud, 
%except for {\it flaring branch}, 
%where $kT^{(1)}_e$ monotonically decreases from 3 keV to 1.5 keV 
%while $kT^{(2)}_e$ show non-monotonic (irreguliar) behaviour in the range of 1.5 -- 4.5 keV. 

Based on{ \it RXTE} data we %also 
detected $\sim$6 Hz quasi-periodic oscillations at the  {\it flaring branch} (FB), which were not detected 
previously in GX~340+0, but this type of QPO frequency  have been revealed in PDSs
%  in power spectra 
of all other $Z-$sources at FB. 

%We also detected $\sim$6 Hz quasi-periodic oscillations at the dip %end part of an X-ray intensity dip 
%of X-ray light curve, 
%%(1.5 days of total duration) 
%in GX~340+0, 
%which corresponds to the {\it flaring branch} state in GX~340+0. 
%The  quasi-periodicc oscillations have properties that are different from those of flaring 
%branchh quasi-periodic oscillations seen in other Z sources, which are often ascribed 
%to oscillations in a radial inflow of matter. 
%%%The fact that the 6 Hz quasi-periodic 
%%%oscillations occur during a dip suggests that they arise in a medium (area) that partially 
%%%screened but relate to NB configuration of GX~340+0. %obscured the X-ray emission. 
%Follow to \cite{KK95}, 
%We interpret the appearance %new 
%of quasi-periodic oscillations phenomenon at the FB % flaring branch state 
%in terms of {\it Transition layer} % (TL) 
%model involving a radial flow and a thick, torus-like CC. %inner  disk. . 

We %also 
explain  this {\it stability} of the indices $\Gamma$ %$\Gamma_1$ and $\Gamma_2$ of  both 
of Comptonozed components 
around a value of 2. %{\it Comptb1/Comptb2})
%in the framework of our model.
% in which the spectrum is dominated by the strong thermal Comptonized 
%component formed  at the surface of {\it neutron star}.  
%in the transition layer (TL),  
%when the  energy release in the TL is much higher than the flux  coming to the TL from the accretion disk.  
%The Comptonized emission from NS surface 
%is due to Comptonized component of CC and has the same spectral characterictics, except ``seed'' photon temperature. 
% located between the  accretion disk and NS %neutron star surface. 
%The index quasi-stability takes place when the  energy release in the TL is much higher than the flux  coming to the TL 
%from the accretion disk. 
The index stability now found  for Comptonized spectral components  
%({\it COMPTB1}) 
of the {\it Z} source GX~340+0 
during the state %spectral 
evolution of the source 
%from the low to high luminosity states 
is similar to that  previously established  in   {\it atolls},  4U~1728-34, GX~3+1 and 
 shown for quite a few  of other LMXBs
 % low mass X-ray neutron star 
%LMXB NS 
%binaries  
(see Farinelli \& Titarchuk, 2011).

We also find in the  {\it Beppo}SAX  spectra   that there are two blackbody components,  
%sources of  emission, 
one, with relatively low color temperatures 0.6-1 keV, is probably  related to  the  disk emission  and another one of color temperatures  about  1.5 keV probably corresponds 
to the  NS surface emission.
%or which  of soft photons are  . 
%
%and 1.5 keV, respectively.  

%We demonstrate that analysis of spectral and timing properties in X-ray from atoll source GX~3+1 allow to 
%distinguish (separate) $mild$ and long-term variabilities, and link them with LB -- UB state transitions and transitions 
%between $bright$ and $faint$ phases in luminosity, respectively. In this waw we described $mild$ flux evolution of GX~3+1 
%between LB and UB states on time scale of hours -- days. in terms of two key (basic) parameters 
%(the electron temperature $kT_e$ and Comptonized fraction $f$) of energy spectra.
%$%as brighter -- fainter phase % high -- low state 
%transitions and LB -- UB transitions, respectively, 
%which are associated with COMPTB normalization and the electron temperature of Comptonized layers changings, correspondingly.

\begin{figure}[ptbptbptb]
\includegraphics[scale=0.99,angle=0]{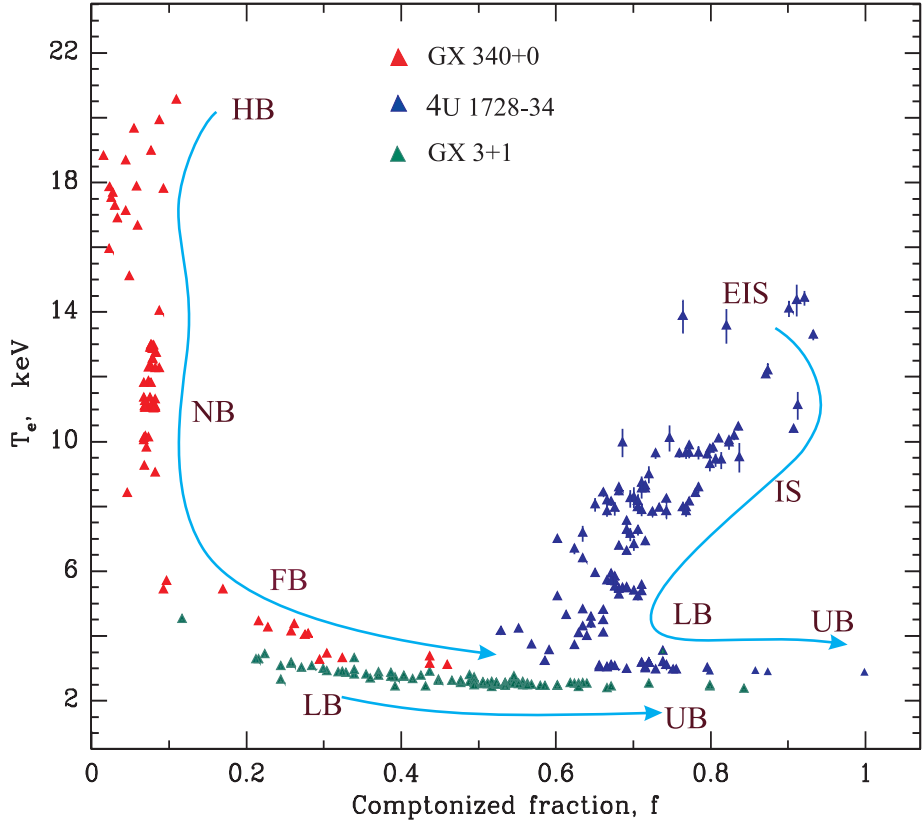}
\caption{%{\bf [I would change the name of the x axis: $f_1$ instead of the generic $f$.]}
Electron temperature $kT^{(1)}_{e}$ (in keV)  plotted versus illumination  fraction $f$
%=A/(1+A)
for atoll sources GX~3+1 and 
4U~1728-34 as well as for {\it Z-} source 
GX~340+0 (using a variable $f_1$ factor)   during different states.
% transitions. % $$mild$  variability. 
{\it Red}, $green$  and {\it blue} points correspond to 
{\it RXTE} %/{\it Beppo}SAX 
observations of GX~340+0, GX~3+1 and 4U~1728-34 respectively. 
%{\it Pink/bright blue}  and {\it blue/green} points correspond to 
%{\it RXTE}/{\it Beppo}SAX 
%observations of GX~3+1 and 4U~1728-34 respectively. 
Along the track of {\it Z} source GX~340+0 we show a sequence of CCD states: HB,
NB and FB. Whereas along the tracks of atoll sources 4U~1728-34 and GX~3+1 we  indicate a sequence of CCD states for these particular sources.  
%On the left-hand side of this  Figure we show  a sequence of CCD states:
%HB, NB and FB 
%which  are  the horizontal,   normal and flaring  branches respectively
% and FB -- Flaring branch) 
%and on  the right hand side IS, EIS, LLB, LB and UB
% which are island state,
%which  are listed according to the standard {\it Z-}scheme [see definition of these states in   \cite{hasinger89}]. 
%On the right hand side of this  Figure we also  show  other  sequence of CCD states, 
%(EIS -- the extreme island state, 
%IS --  island state,
%LLB -- lower left banana state,
%LB -- lower banana state and 
%UB -- upper banana state) which  are listed according to standard atoll-Z scheme~\cite{hasinger89}.
%One can see that  
%which here matched by 
%the electron temperature 
$kT_e$  is  related to the sequence of CCD states for {\it Z-}source GX~340+0.
% with the digitalization (calibration) on left vertical axis. 
The directions in which inferred $\dot M$  increases are indicated by arrows.
% with the digitalization (calibration) on left vertical axis. 
%Along the track  for GX~340+0  points of $T_e-f$  correlation are %indicated where kHz QPOs are found.   
%indicated branches (HB, NB and FB) referring to the discussion in Sect.~4.2.6.
%Power spectra of GX~340+0 % measured in $L^{soft}_{39}/D^2_{10}$ units versus Comptonized fraction $f$ is 
%are plotted in the {\it incorporated panel} ({\it top center}) typical for corresponding (indicated) branches. %
%for GX~3+1 using  our spectral model 
%$wabs*(blackbody+COMPTB+Gaussian)$ during long-term ($slow$) variability (see Table 4).
}
%\label{outburst_index_temperature1}
\label{T_e_vs_f_comp}
\end{figure}

\begin{figure}[ptbptbptb]
\includegraphics[scale=0.8,angle=0]{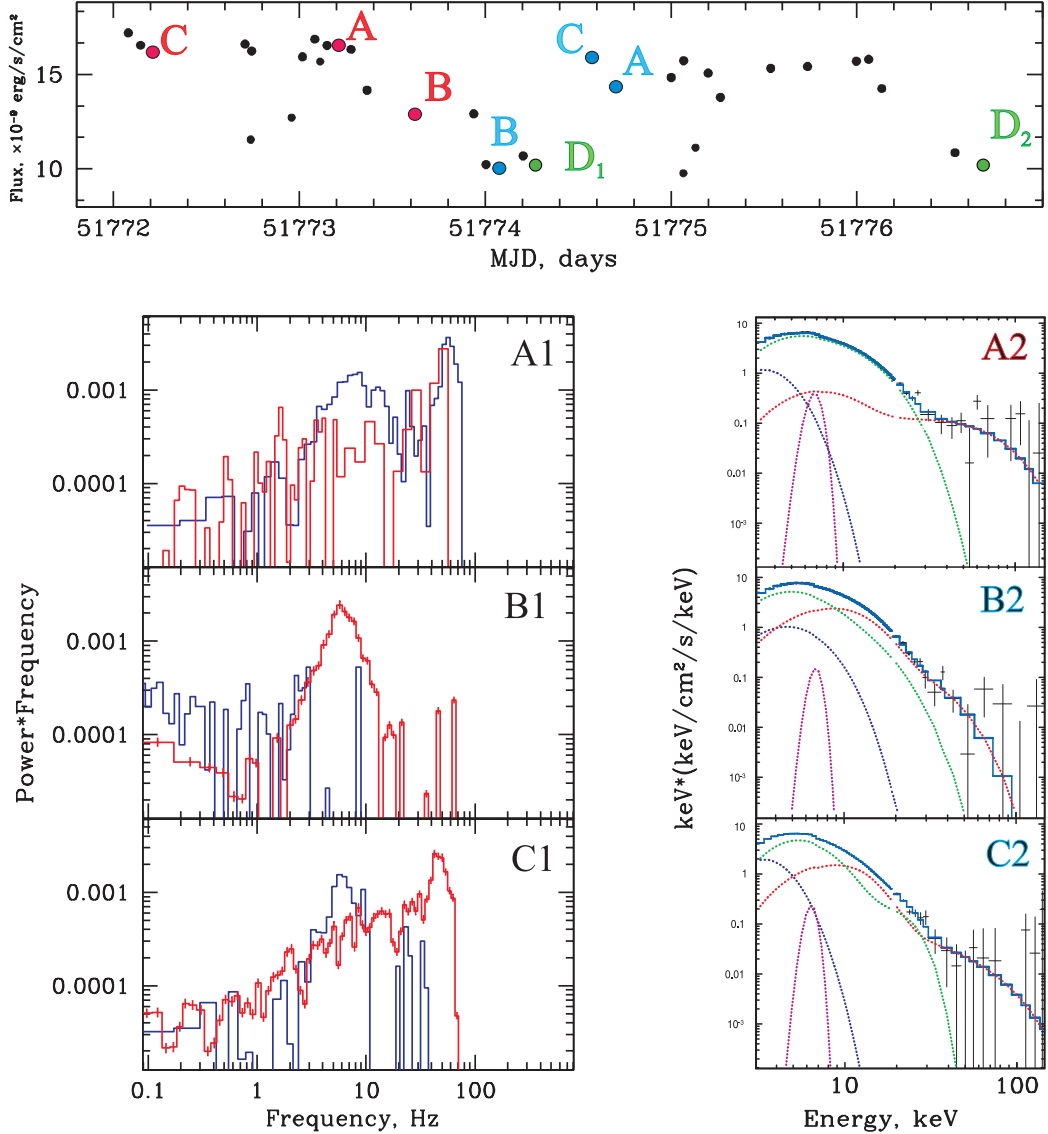}
\caption{
{\it Top}: evolution of the 3 -- 150 keV flux during the 2000 ($R5$) evolution %transition 
events. Red/blue points 
A, B, and C mark moments at MJD = 51773.2/51774.7, 51773.6/51774.1 and 51772.2/51774.6 related to different %transition 
phases of Z-state evolution.  
%$Flaring$ branch (B $blue$) is mainly present along  with VLFN (less than $\sim$1 Hz) %component 
%and  FBOs at $\sim$6 Hz ($green$ points) indicated by $D_1$, $D_2$ (see also Fig.~\ref{PDS_FBO}). 
 % of transition.
%and after X-ray outburst), respectively.  
$Bottom$: 
PDSs for 15-30 keV  band ($left$ column) are plotted along with the $E*F(E)$ diagram ($right$ column) 
for A, B and C  points  of X-ray light curve.
%($red$, top), 
%B ($blue$, middle) and C ($blue$, bottom)   
%$Red$ histograms corresponds to aforementioned MJD moments, whereas $blue$ histograms refer to adjacent observations  to illustrate continuous fast evolution of PDSs. 
%The strong HBOs and broad NBOs are seen at 20 -- 50 Hz/2 -- 10 Hz during {\it HB} (C $red$)/{\it NB} 
%(B $red$, C $blue$), respectively. 
%On the right we present 
$E*F(E)$ diagrams (panels A2, B2, C2) are related to the corresponding power spectra 
(panels A1, B1, C1).  The data are shown by black points and  
 the spectral model components are displayed  by red, green, blue and dashed purple lines for $Comptb1$, $Comptb2$,  $Blackbody$ and $Gaussian$ respectively.
}
\label{PDS}
\end{figure}

\begin{figure}[ptbptbptb]
\includegraphics[scale=0.9,angle=0]{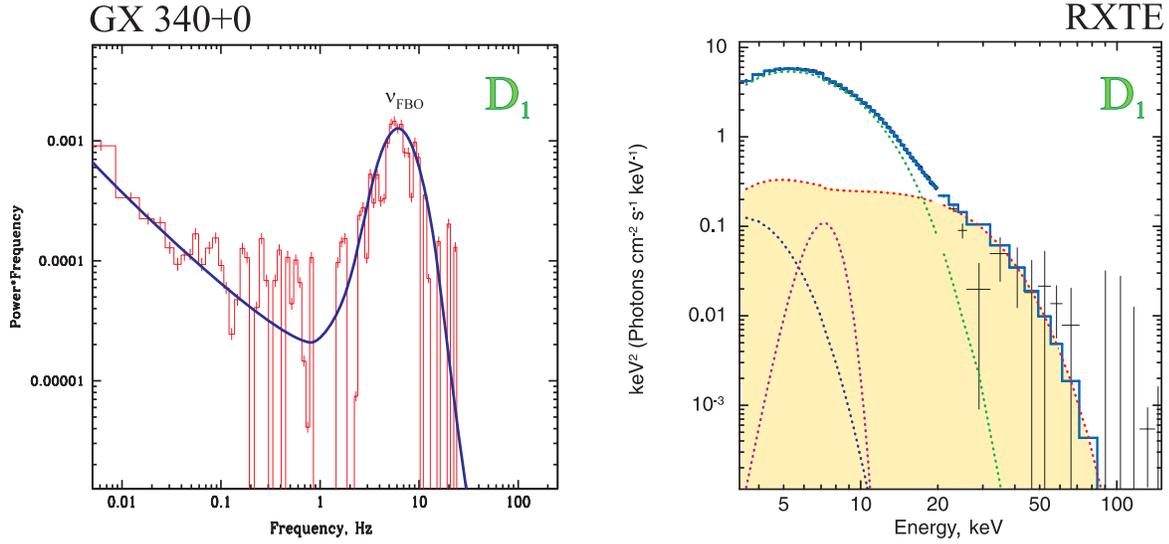}
\caption{%The $\nu\times power$ diagram of GX~340+0 in 0.01 -- 150 Hz range,  observed on August 18, 2000 
%(50016-01-02-16, MJD=51774.47). The PDS accumulation time  is 3876 sec.
%Blue solid line shows a smooth power spectrum. Broad QPO is centered at $\nu_{FBO}\sim$6 Hz 
%and %Power spectrum of GX~340+0 typically consist of three components: the broad-band noise with break $\nu_b$, 
%low frequency QPOs ($\nu_{sl}$, $\nu_l$) and 
%very low frequency noise component (VLFN) are present  
%in the power spectrum of GX~340+0 
%during flaring branch (FB) (see details in the text).
{\it Left panel:} The $\nu\times power$ diagram of GX~340+0 in 0.01 -- 150 Hz range,  observed on August 18, 2000 (50016-01-02-16, MJD=51774.47, point $D_1$ on Fig.~\ref{PDS}). The PDS accumulation time  is 3876 sec. Blue solid line shows a smooth power spectrum. Broad QPO is centered at $\nu_{FBO}\sim$6 Hz and very low frequency noise component (VLFN) are present during flaring branch (see details in the text). {\it Right panel:} $E*F(E)$ diagram is related to the corresponding power spectrum (point $D_1$). The data are shown by black points and the spectral model components are displayed by $red$, $green$, $blue$ and $purple$ dashed lines for $Comptb1$, $Comptb2$, $Blackbody$ and $Gaussian$ respectively. Yellow shaded area demonstrates relatively low contribution of $Comptb1$ component at point $D_1$.
%which is related to moderate up-scattering processes in TL for seed photons of temperature $T_{s1}\le 1$ keV coming from the disk.
}
\label{PDS_FBO}
\end{figure}

\begin{figure}[ptbptbptb]
\includegraphics[scale=0.8, angle=0]{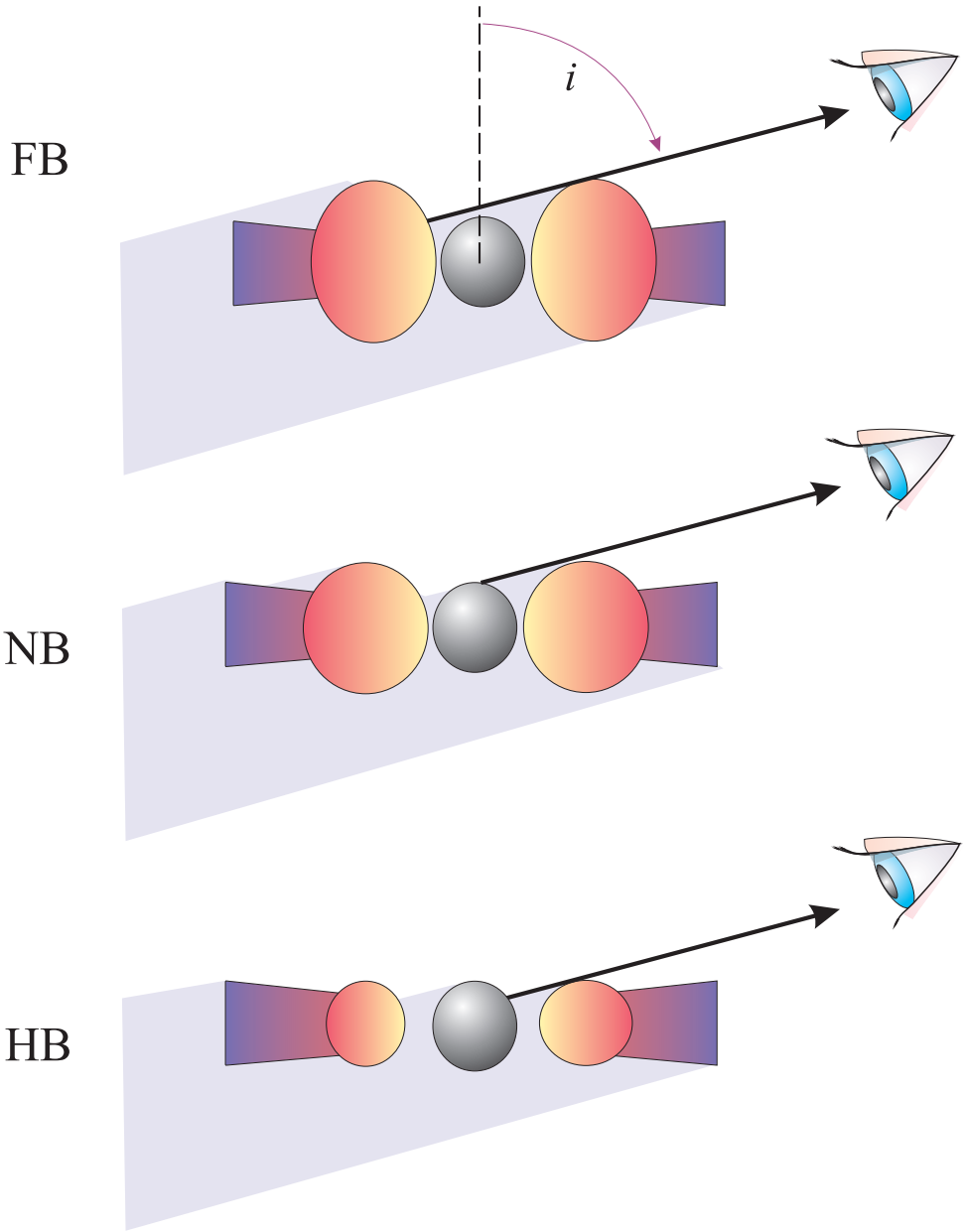}
\caption{
A suggested  geometry of GX~340+0 during state evolution. %transitions.   
As GX~340+0 moves from  HB, through  NB, to  FB.  CC puffs up, 
which affects   a visibility of the central area (NS). In the upper panel (FB) we schematically indicate a
probable inclination of GX~340+0 as inferred from various observations (see text). In particular, 
the innermost part of the source  is hidden by  CC
% from our view, 
which causes dips in intensity (``dipping FB''). 
The successive  decrease  of screening effect during next stage (NB) is 
schematically shown in the middle panel. CC shrinking provides  visibility of central region (NS). 
% Some fraction of inner parts now is seen by the Earth observer. 
Finally, the lower (HB) panel demonstrates 
the minimal screening effect of central area, providing a proper contribution of NS  emission. 
}
%\label{norm_temp_3obj}
\label{geometry_eclipse}
\end{figure}

\begin{figure}[ptbptbptb]
\includegraphics[scale=0.9, angle=0]{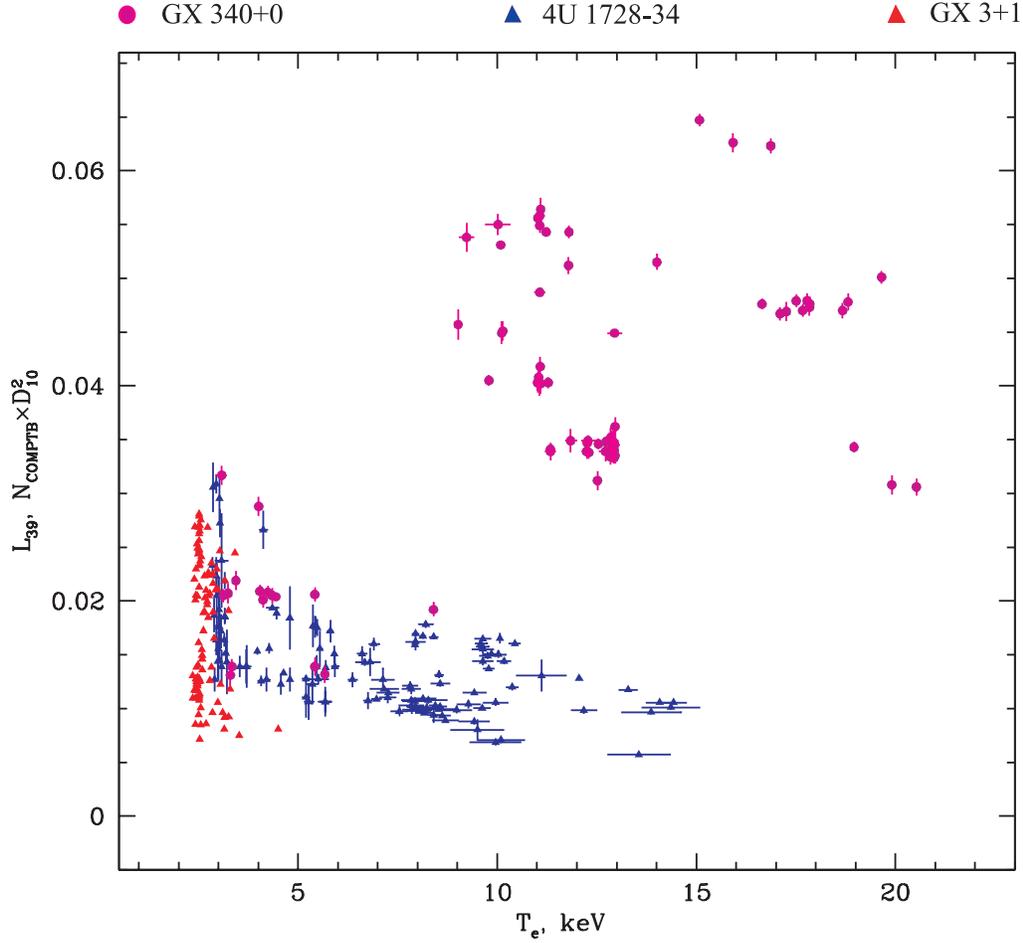}
\caption{
$L_{39}$ measured in $N_{COMPTB}\times D^2_{10}$ units
%
%COMPTB normalization measured in $L_{39}/D^2_{10}$ units 
versus electron temperature $T_e$ (in keV) obtained % using the best-fit  spectral model 
%$wabs*(blackbody+Comptb+Comptb+Gaussian)$ 
for Z source GX~340+0 ($pink$), we present a variable  $N_{com1}$ normalization for this source; {\it atoll} sources GX~3+1 ($red$, taken from ST12) and 
4U~1728-34 ($blue$, taken from ST11)  for  {\it RXTE} data. 
Here $D_{10}$ is the distance to the source (see  Table 5).
%  and  
%%whereas 
%% {\it bright blue} and $green$ points for $Beppo$SAX data. 
%A suggested  geometry of GX~340+0 during state transitions.   
%As GX~340+0 moves from the HB, through the NB, to the FB, the inner disk part (CC) becomes to puff up, 
%influencing the visibility of central area (NS). In the upper panel (FB) we schematically indicate the 
%probable inclination of GX~340+0 as inferred from various observations (see text). In particular, 
%the central area is hidden by inner torus (CC) from our view, which causes dips in intensity (``dipping FB''). 
%The succesive decreasing of screening effect during further (next) state (NB) is 
%schematically shown in the middle panel. The shrinking of CC provides particular (local) visibility of cental area. 
% Some fraction of inner parts now is seen by the Earth observer. Finally, the lower (HB) panel demonstrates 
%the minimal screening effect of central area, providing proper contribution of NS surface emission. 
}
\label{norm_temp_3obj}
%\label{geometry_eclipse}
\end{figure}

We establish   %This model allow to argue %Spectral analysis using this model provides 
that 
%electron temperature 
$kT_e$ 
%of Compton cloud 
monotonically arises in the 3$-$21 keV range
%increases 
%from 3 keV to 21 keV 
when GX~340+0 evolves  %transits 
from FB to HB (see Fig. \ref{T_e_vs_f_comp}). %{\it upper banana} to {\it lower banana}, 
%associated with low flux variabilities on time scale of hours-days.  
%Specifically, we
%We also detected, at least, two noise components (VLFN \& HFN) %demonstrate 
%and its evolution (alternative behavior) during LB -- UB transitions: %, associated with low flux variability on time of hours, 
%the power spectrum of X-ray signal in {\it upper banana} are dominated by %a very low frequency noise 
%VLFN with 
%peaked noise component 
%the breake at around 20 Hz, whereas in {\it lower banana} the power spectrum are dominated by %a high frequency noise 
%HFN in 1 -- 100 Hz range and acompanied by reduced VLFN below $\sim$ 1 Hz.

In ST11 we show
%  strong theoretical arguments  
that if 
% the dominance of 
the energy release $Q_{cor}$ in the Transition Layer is much higher than that in the disk $Q_{disk}$
%  thermal Comptonized component formed in the 
%transition layers 
% with respect to the soft  flux $Q_{disk}$ coming from the accretion disk, 
i.e.  $Q_{cor}/Q_{disk}\gg 1$,  then the resulting photon index of the Comptonized component of the spectrum, % is leads
%to almost   constant photon  index 
$\Gamma\approx2$.
We also suggest  that {\it the index stability effect   is an intrinsic signature of  {\it atolls},  for example
%sources  
%such as 
GX~3+1,  4U 1728-34, whereas in BHs $\Gamma$ monotonically increases with $\dot M$ and finally  saturates at high $\dot M$ values } (see ST09).
% It is worth to remind a reader  the index correlation vs mass accretion rate for a number of 
% BH sources and how  the index depends on mass accretion rate in NSs GX~3+1 and 4U 1728-34. 
%Photon 
The indices  of BHCs 
%candidates  
(GRO~J1655-40, GX~339-4, GRS~1915+105, H1743-322, SS~433,  Cyg X-1, 4U 1543-47, 
 XTE J1550-564)   demonstrate   strong  
correlation with $\dot M$
%mass accretion rate  
or with the soft  (disk) photon normalization $N_{com}=L_{39}/D^2_{10}\propto \dot M$. 
%which is proportional to mass accretion rate.   
%This index-$\dot M$ correlation  is followed by the index saturation 
When $\dot M$ exceeds a certain level (a few times of the Eddington mass accretion rate) the index 
$\Gamma$ starts to saturate.  
%in comparison to $atoll$ NS source (GX~3+1) sample 
In this  Paper we show that  the behavior of the index  $\Gamma$ for   NS GX~340+0 is 
the same  to that in {\it atolls}, 4U 1728-34,  GX~3+1,    and thus is drastically  different  
  from that in   BHCs.   %The photon index $\Gamma\approx2$
  % independently of any model parameter 
%  while mass accretion rate changes by factor 4.

%  ADDITIONS:

%The theoretical and 
Our  observational results and their theoretical explanation  
%evidence  
supports the idea  that the X-ray spectral evolution 
of NS LMXBs, 
%including their transient hard X-ray tails, 
%can be explained by 
reveal the unique 
%We also interpret  
quasi-stability of the photon  indices $\Gamma$ %$\Gamma_1$ and $\Gamma_2$ of  both 
of each Comptonized components 
near a value of 2. %{\it Comptb1/Comptb2})
This can be explained 
%in the framework of 
by the model in which   
% in the X-ray emergent spectrum is dominated by ,
% and at {\it neutron star} surface. %,  
%where 
the  energy release in the transition layer   is much higher than 
that in the accretion  disk and consequently the NS blackbody and  thermal Comptonized  components   %formed  in the Transition Layer 
are dominant  in the X-ray emergent spectrum.
%at high photon energies. 
% flux  coming to the TL from disk.  
%the interplay between thermal and bulk motion Comptonization. 
%The applying %introduction 
%of a new XSPEC Comptonization model, $Comptb$, %including thermal and bulk Comptonization, 
%has provided additional support to this interpretation.

We acknowledge productive discussion of the paper 
%content
 %and editing  of the paper content 
 with Chris Shrader. Also we should point out a very deep analysis of our paper by the referee who suggests many important modifications of the paper. 
%{\it
%We are very grateful to the referee whose constructive suggestions help us to improve the paper quality.
%}

\newpage

%
% Table 1
%
\begin{deluxetable}{lcccc}
%%%%%\rotate
\tablewidth{0in}
\tabletypesize{\scriptsize}
%  \begin{center}
    \tablecaption{The list of $Beppo$SAX observations of GX~340+0  used in our analysis.}
    \renewcommand{\arraystretch}{1.2}
%    \begin{tabular}[h]
%      \hline
\tablehead{
Number of set & Obs. ID& Start time (UT)  & End time (UT) &MJD interval }
%Satellite&Obs. ID& Start time (UT)  & End time (UT)}
%%%%%Obs.  &ID           & time (UT)& time (UT)& of state& }
\startdata
S1 & 20261005  & 1997 Sep. 4 11:53:58 & 1997 Sep. 4 19:33:21 &50695.5-50695.9 \\
S2 & 20261006  & 1997 Oct. 2 20:32:50 & 1997 Oct. 3 04:16:33 &50723.8-50724.4 \\
S3 & 21240001  & 2000 Aug. 16 05:50:34 & 2000 Aug. 18 05:56:11 &51772.2-51774.2$^1$ \\
S4 & 212400011 & 2000 Aug. 18 06:16:17 & 2000 Aug. 20 14:21:16 &51774.2-51776.9$^1$ \\
S5 & 212400012 & 2001 March 9 19:27:20 & 2001 March 11 11:13:47 &51977.8-51979.9$^1$\\
S6 & 21375002  & 2001 Aug. 9 05:01:20 & 2001 Aug. 10 08:56:49 &52130.2-52131.9$^2$ \\
      \enddata
%      \hline
%      \end{tabular}
   \label{tab:table}
% \end{center}
Reference. 
(1) \citet{iaria06}; 
(2) \citet{lavagetto04}
%{ooster01} 
%Piraino et al., (2000)
\end{deluxetable}

%begin{deluxetable}{cccc}
%%%%%\rotate   
%\tablewidth{0in}
%\tabletypesize{\scriptsize}
%  \begin{center}
 %   \tablecaption{The list of $Beppo$SAX observations of 4U~1728-34  used in analysis.}
  %  \renewcommand{\arraystretch}{1.2}
%    \begin{tabular}[h]
%      \hline
%\tablehead{
%Obs. ID& Start time (UT)  & End time (UT) &MJD interval}
%Satellite&Obs. ID& Start time (UT)  & End time (UT)}
%%%%%Obs.  &ID           & time (UT)& time (UT)& of state& }
%\startdata
%20674001& 1998 Aug. 23 19:15:27 & 1998 Aug. 24 09:14:15 &51048.8-51049.4$^1$ \\
%20889003& 1999 Aug. 19 02:01:32 & 1999 Aug. 20 04:54:32 &51409.1-51410.2$^2$\\
%20889003& 1999 Aug. 19 02:01:32 & 1999 Aug. 20 04:54:32 &51409.1-51410.2$^1$& \cite{piraino00}\\
 %     \enddata
%      \hline
%      \end{tabular}
%   \label{tab:table}
% \end{center}
%Reference
%$(1)$ \cite{disalvo2000a}, $(2)$ \cite{piraino00} 
%Piraino et al., (2000)
%\end{deluxetable}

%%%%%%%%%%%%%%%%%
%\end{document}

%
% Table 2
%
\newpage
\begin{deluxetable}{lllll}
%\rotate
\tablewidth{0in}
\tabletypesize{\scriptsize}
%  \begin{center}
    \tablecaption{The list of groups of {\it RXTE} observation of GX~340+0}
    \renewcommand{\arraystretch}{1.2}
%    \begin{tabular}[h]
%      \hline
\tablehead{Number of set  & Dates, MJD & RXTE Proposal ID&  Dates UT & Rem.  \\
                          &            &                 &           &     }
 \startdata
R1  &    50555       & 20054        & Apr. 17 13:40:48 -- 13:08:00, 1997 &           \\
R2  &    50605-50609 & 20059        & June 6 -- 10, 1997                 &           \\
R3  &    50712-50756 & 20053$^1$    & Sept. 21 -- Nov. 4, 1997           & $Beppo$SAX\\
R4  &    51130-51131 & 30040        & Nov. 13 -- 14, 1998                &          \\
R5  &    51772-51776 & 50016        & Aug. 16 -- 20, 2000                &          \\
R6  &    51920-51924 & 50016        & Jan. 11 -- 15, 2001                &          \\
R7  &    52908-52915 & 80020        & Sept. 26 -- Oct. 3, 2003           &          \\
R8  &    53210-53262 & 80020        & July 24 -- Sept. 14, 2004          &          \\
R9  &    53874-53907 & 91152        & May 19 -- June 21, 2006            &          \\
R10 &    54297       & 93046        & July 16 01:14:56 -- 01:29:20, 2007 &          \\
R11 &    54893-54894 & 94312        & March 3 -- 4, 2009                 &          \\
      \hline
      \enddata
%      \hline
%      \end{tabular}
    \label{tab:par_bbody}
%  \end{center}
References:
(1) \cite{church06} %, \citet{ooster01} 
%Strohmayer et al. 1996; 
%(2) Ford \& van der Klis (1998);
%(3) van Straaten et al. (2002); 
%(4) Di Salvo et al. (2001); 
%(5) Mendez, van der Klis \& Ford (2001); 
%(6) Migliari, van der Klis \& Fender (2003); 
%(7) Jonker, Mendez \& van der Klis (2000);
%(8) TS05
\end{deluxetable}

\begin{deluxetable}{clcccccc}
%\rotate
\tablewidth{0in}
\tabletypesize{\scriptsize}
%  \begin{center}
    \tablecaption{Best-fit parameters of spectral analysis of Beppo$SAX$
observation of GX~340+0 in 0.3 -- 150~keV energy range in three additive models$^{\dagger}$: 
{\it wabs*(Blackbody + Comptb + Gaussian}), % [{\it ``single CompTB}''] and  
{\it wabs*(Blackbody1+Blackbody2+Comptb+Gaussian}) and  
{\it wabs*(Blackbody + Comptb1 + Comptb2 + Gaussian}). %[{\it ``double CompTB''}]. 
%for selected (**, in Table 1) 
%for observations with numbers 90401-NN-NN-NN. 
%Parameter errors correspond to 1$\sigma$ confidence level.
%%%%    Errors are given at the 90\% confidence level.
}
    \renewcommand{\arraystretch}{1.2}
%    \begin{tabular}[h]
%      \hline
\tablehead{
%Model & Parameter & 90401-01-01-00 & 90401-01-01-01 & 90401-01-01-02 & 90401-01-01-03 & 90401-01-02-00 & 90401-01-02-01 & 
%90401-01-03-00 & 90401-01-03-01 & 90401-01-03-02 }
Model & Parameter & 00-20261005 & 00-20261006 & 00-21240001 & 00-212400011 & 00-212400012 & 00-21375002 }
 \startdata
%wabs &                   &          &        &        &         &         &          \\
wabs     & N$_H$ (cm$^{-2}$) & 6.38$\pm$0.09  & 6.18$\pm$0.16& 6.52$\pm$0.13& 6.56$\pm$0.09 & 6.53$\pm$0.09  & 6.29$\pm$0.09  \\
bbody    & kT$_{BB}$ (keV)    & 0.99$\pm$0.02   & 0.98$\pm$0.03 & 0.69$\pm$0.02 & 0.98$\pm$0.03  & 0.68$\pm$0.01  & 0.60$\pm$0.03   \\
     & N$_{BB}^{\dagger\dagger}$ & 7.67$\pm$0.01 & 7.41$\pm$0.06 & 10.43$\pm$0.07 & 7.56$\pm$0.02 & 7.41$\pm$0.05 & 10.41$\pm$0.07 \\
%comptb &                    &        &         &        &         &         &             \\
comptb  & $\alpha=\Gamma-1$  & 1.02$\pm$0.01& 1.00$\pm$0.05 & 1.09$\pm$0.04 & 1.02$\pm$0.03 & 1.03$\pm$0.01 & 1.08$\pm$0.06\\
     & kT$_{s}$ (keV)   & 1.56$\pm$0.04& 1.49$\pm$0.05  & 1.76$\pm$0.06& 0.48$\pm$0.04 & 1.43$\pm$0.04  & 1.40$\pm$0.04 \\
     & logA$$          & -1.29$\pm$0.03& -0.86$\pm$0.02& -0.98$\pm$0.03&-0.86$\pm$0.03 & -0.72$\pm$0.07 & -0.86$\pm$0.05  \\
     & kT$_{e}$ (keV)   & 6.21$\pm$0.07 & 3.98$\pm$0.08 & 4.89$\pm$0.08& 3.62$\pm$0.06 & 3.98$\pm$0.08 & 4.98$\pm$0.09 \\
     & N$_{Com}^{\dagger\dagger}$ & 7.47$\pm$0.05& 10.02$\pm$0.01 & 15.19$\pm$0.09 & 10.02$\pm$0.07 & 14.20$\pm$0.04 & 13.61$\pm$0.05 \\
%Gaussian &               &       &         &              &         &         &           \\
Gaussian     & E$_{line}$ (keV)  & 6.87$\pm$0.06 & 6.80$\pm$0.08& 6.74$\pm$0.07 & 7.00$\pm$0.06 & 6.80$\pm$0.06 & 6.71$\pm$0.05  \\
     & $\sigma_{line}$ (keV)& 0.65$\pm$0.05 & 0.60$\pm$0.01& 0.81$\pm$0.06  & 0.69$\pm$0.04 & 0.91$\pm$0.06 & 0.74$\pm$0.05 \\
     & N$_{line}^{\dagger\dagger}$ & 0.45$\pm$0.02 & 0.43$\pm$0.06&0.55$\pm$0.07& 0.26$\pm$0.07 & 0.39$\pm$0.07 & 0.35$\pm$0.08 \\
      \hline
     & $\chi_{red}^2$ (d.o.f.) & 3.2 (336)& 1.29 (330)  & 1.59 (332)& 1.3 (332)&1.43(332) & 2.7 (330)  \\
      \hline
wabs     & N$_H$ (cm$^{-2}$) & 5.40$\pm$0.07  & 6.29$\pm$0.06& 6.43$\pm$0.09& 6.43$\pm$0.09 & 6.35$\pm$0.07  & 5.70$\pm$0.09  \\
%bbody &                  &          &        &        &         &         &          \\
bbody1    & kT$_{BB1}$ (keV)    & 0.60$\pm$0.08   & 0.59$\pm$0.04 & 0.61$\pm$0.05 & 0.52$\pm$0.06  & 0.58$\pm$0.08  & 0.60$\pm$0.02   \\
%     &  (keV)            &          &        &        &         &         &          \\
     & N$_{BB1}^{\dagger\dagger}$ & 2.1$\pm$0.1 & 7.58$\pm$0.03 & 7.98$\pm$0.09 & 7.81$\pm$0.06 & 8.2$\pm$0.1 & 3.70$\pm$0.09 \\
bbody2    & kT$_{BB2}$ (keV)    & 1.10$\pm$0.03  & 1.50$\pm$0.06 & 1.51$\pm$0.02 & 1.47$\pm$0.04  & 1.51$\pm$0.02  & 1.50$\pm$0.04 \\
%     &  (keV)            &          &        &        &         &         &          \\
     & N$_{BB2}^{\dagger\dagger}$ & 2.00$\pm$0.07 & 6.42$\pm$0.5 & 9.95$\pm$0.08 & 7.96$\pm$0.09 & 6.49$\pm$0.08 & 4.75$\pm$0.09 \\
%     & EW$_{BB}$         &  1.4(2)  & 1.4(3) &1.4(5)       & 1.4(5) & 1.4(1) & 1.4(2)   \\
%     &  (keV)            &          &            &          &       &        &           \\
%comptb &                    &        &         &        &         &         &             \\
comptb     & $\alpha=\Gamma-1$& 1.00$\pm$0.01& 1.06$\pm$0.04 & 0.99$\pm$0.03 & 1.02$\pm$0.04 & 1.05$\pm$0.06 & 1.10$\pm$0.09\\
     & kT$_{s}$ (keV)   & 1.21$\pm$0.09& 1.20$\pm$0.02  & 1.26$\pm$0.07& 0.18$\pm$0.06 & 1.20$\pm$0.08  & 1.19$\pm$0.08 \\
     & logA$$          & -1.02$\pm$0.06& -0.55$\pm$0.02& -0.21$\pm$0.03&-0.29$\pm$0.02 & -1.08$\pm$0.06 & 2.0$^{\dagger\dagger\dagger}$  \\
     & kT$_{e}$ (keV)   & 4.66$\pm$0.08 & 3.51$\pm$0.09 & 3.54$\pm$0.04& 3.35$\pm$0.03 & 4.33$\pm$0.09  & 3.32$\pm$0.02 \\
     & N$_{Com}^{\dagger\dagger}$ & 10.78$\pm$0.09& 8.67$\pm$0.05 & 6.24$\pm$0.09 & 7.16$\pm$0.08 & 7.63$\pm$0.06 & 5.43$\pm$0.07 \\
%Gaussian &               &       &         &              &         &         &           \\
Gaussian     & E$_{line}$ (keV)  & 6.75$\pm$0.08 & 6.78$\pm$0.09& 6.80$\pm$0.07 & 6.81$\pm$0.05 & 6.90$\pm$0.07 & 6.75$\pm$0.05  \\
     & $\sigma_{line}$ (keV)& 0.95$\pm$0.04 & 0.90$\pm$0.06& 0.89$\pm$0.07  & 0.79$\pm$0.04 & 0.91$\pm$0.05 & 0.94$\pm$0.03 \\
     & N$_{line}^{\dagger\dagger}$ & 2.60$\pm$0.05 & 0.66$\pm$0.09&0.65$\pm$0.08& 0.63$\pm$0.06 & 0.53$\pm$0.07 & 0.65$\pm$0.09 \\
%     & EW$_{laor}$ (eV) &         & 290$\pm$18     &  378$\pm$10 & 256$\pm$15&150$\pm$15     & 180$\pm$10    \\
%     &  (eV)        & & &         &   & &    &   & &    \\
%Gaussian2 &               &         &       &          &       &        &         \\
%     & E$_{line2}$ (keV)  & 9.07(5) & -     & 8.6(2)   &9.2(1) & 9.1(2) & 9.5(2)  \\
%     & N$_{line2}^{\dagger\dagger}$ &0.26(5)& - &0.1(2)    &2.3(9) & 0.2(2) & 0.1(1) \\
%Flux$^{\dagger\dagger\dagger}$ &    &            &          &       &        &        &         \\
%     & 3 - 60 keV        & 5.18     & 5.48       & 5.01     & 4.76  & 7.79   & 5.51    \\
%     & 13 - 150 keV      & 2.84     & 2.99       & 2.69     & 2.30  & 4.65   & 3.02    \\
      \hline
     & $\chi_{red}^2$ (d.o.f.) & 3.28 (334)& 1.07 (328)  & 1.53 (330)& 1.65 (330)&1.63(330) & 2.18 (327)  \\
      \hline
%
%Model & Parameter           & 00-20261005    & 00-20261006  & 00-21240001   & 00-212400011 & 00-212400012 & 00-21375002 }
wabs     & N$_H$ (cm$^{-2}$) & 5.7$\pm$0.1    & 6.0$\pm$0.1  & 6.01$\pm$0.03 & 6.3$\pm$0.1  & 6.4$\pm$0.1  & 5.4$\pm$0.3  \\
bbody    & kT$_{BB}$ (keV)    & 0.56$\pm$0.08  & 0.57$\pm$0.04& 0.56$\pm$0.03 & 0.53$\pm$0.05& 0.43$\pm$0.04& 0.57$\pm$0.01\\
     & N$_{BB}^{\dagger\dagger}$&0.92$\pm$0.09& 0.9$\pm$0.1  & 0.90$\pm$0.05 & 0.91$\pm$0.04& 4.47$\pm$0.02& 2.01$\pm$0.06\\
comptb1&$\alpha_1=\Gamma_1-1$& 0.99$\pm$0.04  & 1.00$\pm$0.08& 1.02$\pm$0.07 & 1.01$\pm$0.07& 0.99$\pm$0.01& 1.04$\pm$0.06\\
     & kT$_{s1}$ (keV)       & 0.70$\pm$0.03  & 0.6$\pm$0.1  & 0.69$\pm$0.02 & 0.70$\pm$0.03& 0.83$\pm$0.04& 1.03$\pm$0.01\\
     & logA$_1$              & -0.32$\pm$0.06 &-0.30$\pm$0.07& -0.30$\pm$0.05&-0.30$\pm$0.03&-1.42$\pm$0.06&-1.33$\pm$0.07\\%$^{\dagger}$  \\
     & kT$^{(1)}_{e}$ (keV)  & 5.38$\pm$0.02  & 5.42$\pm$0.03& 5.60$\pm$0.04 & 5.41$\pm$0.04& 6.04$\pm$0.03& 19.50$\pm$0.07\\
     & N$_{Com1}^{\dagger\dagger}$&1.36$\pm$0.07&1.86$\pm$0.04& 0.57$\pm$0.06& 1.53$\pm$0.04& 9.96$\pm$0.06& 3.97$\pm$0.07\\
%        &                   &                &              &               &              &              &              \\
comptb2&$\alpha_2=\Gamma_2-1$& 1.01$\pm$0.03  & 1.04$\pm$0.05& 1.01$\pm$0.03 & 1.05$\pm$0.08& 1.06$\pm$0.05& 0.99$\pm$0.02\\
     & kT$_{s2}$ (keV)       & 1.0$\pm$0.1    & 1.00$\pm$0.09& 1.71$\pm$0.05 & 1.06$\pm$0.09& 1.50$\pm$0.02& 1.51$\pm$0.04\\
%     & logA$_2$             & 2.0$^{\dagger}$ & 2.0$^{\dagger}$ & 2.0$^{\dagger}$ &-0.28$\pm$0.05 & 2.0$^{\dagger}$ & 0.19$\pm$0.06\\
     & kT$^{(2)}_{e}$ (keV)  & 1.71$\pm$0.09  & 1.90$\pm$0.02 & 2.04$\pm$0.08& 2.05$\pm$0.03& 1.52$\pm$0.03& 2.76$\pm$0.09\\
     & N$_{Com2}^{\dagger\dagger}$&10.84$\pm$0.07&12.17$\pm$0.06&14.26$\pm$0.05&13.11$\pm$0.04&11.79$\pm$0.05&8.20$\pm$0.06\\
Gaussian & E$_{line}$ (keV)  & 6.85$\pm$0.08  & 6.68$\pm$0.03& 6.68$\pm$0.05 & 6.70$\pm$0.04& 6.71$\pm$0.03& 6.80$\pm$0.06\\
%     & $\sigma_{line}$ (keV)& 2.60$\pm$0.04  & 2.56$\pm$0.07& 2.79$\pm$0.08 & 2.64$\pm$0.07& 0.90$\pm$0.07& 1.54$\pm$0.06 \\
& N$_{line}^{\dagger\dagger}$& 1.07$\pm$0.09  & 0.89$\pm$0.08& 0.68$\pm$0.09 & 0.77$\pm$0.09& 0.89$\pm$0.02& 0.7$\pm$0.1\\
%%Flux$^{\dagger\dagger\dagger}$ &    &            &         &       &       &        &        \\
%%     & 3 - 60 keV        & 5.18     & 5.48       & 5.01    & 4.76  & 7.79  & 5.51    \\
%%     & 13 - 150 keV      & 2.84     & 2.99       & 2.69    & 2.30  & 4.65  & 3.02    \\
      \hline
   & $\chi_{red}^2$ (d.o.f.) & 0.97 (332)     & 0.98 (326)   & 0.99 (328)    & 1.38 (328)   & 1.49 (328)   & 1.28 (326) \\
      \hline
      \enddata
%      \hline
%      \end{tabular}
    \label{tab:BeppoSAX_fit_table}
%  \end{center}

$^\dagger$ Errors are given at the 90\% confidence level.
$^{\dagger\dagger}$ %The spectral model is  $wabs*(blackbody + COMPTB + COMPTB + Gaussian)$,
The normalization parameters of $Blackbody$ and $CompTB$ components are in units of 
$L_{37}^{soft}/d^2_{10}$ 
$erg/s/kpc^2$, 
where $L_{37}^{soft}$ is the soft photon  luminosity in units of 10$^{37}$ erg/s, 
$d_{10}$ is the distance to the source in units of 10 kpc 
and $Gaussian$ component is in units of $10^{-2}\times total~~photons$ $cm^{-2}s^{-1}$ in line;
$^{\dagger\dagger\dagger}$ when parameter $\log(A)\gg1$, it is fixed to a value 2.0 
for the model $wabs*(Blackbody1+Blackbody2+Comptb+Gaussian)$, 
parameter $\log(A_2)$ is fixed at 2.0 for the model $wabs*(Blackbody+Comptb1+Comptb2+Gaussian)$ (see comments in the text); 
$\sigma_{line}$ of $Gaussian$ component is fixed to a value 0.60 keV for the model $wabs*(Blackbody+Comptb1+Comptb2+Gaussian)$ (see comments in the text), 
$N_H$ is units of 
%was fixed at value of 6.2
$10^{22}$ cm$^{-2}$. % (Iaria et al., 2006). 
%$^{\dagger\dagger\dagger\dagger}$spectral fluxes (F$_1$/F$_2$) in the (3 -- 10)/(10 -- 60) keV energy ranges, correspondingly, 
%in units of $\times 10^{-9}$ ergs/s/cm$^2$.
%----------
%%Parameters not listed in the table were fixed
%$^\dagger$this parameter is fixed (see details in the text), 
%$^{\dagger\dagger}$ normalization parameters of $blackbody$ and $COMPTB$ components are in units of 
%$L_{37}^{soft}/d^2_{10}$, where $L_{37}^{soft}$ is the soft photon  luminosity in units of 10$^{37}$ erg/s, 
%$d_{10}$ is the distance to the source in units of 10 kpc 
%and $Gaussian$ component is in units of $10^{-2}\times total~~photons$ $cm^{-2}s^{-1}$ in line.
%%$^{\dagger\dagger\dagger}$ spectral flux 
%%in the 3-- 150 energy range 
%%in units of $\times 10^{-10}$ erg/s/cm$^2$, 
%%$\sigma_{line2}$ for Gaussian2 component was fixed to value 0.01 keV.
%%$^{\dagger\dagger}$spectral flux density in radio band centered at 15 GHz in
%%units of mJy.
\end{deluxetable}

%
% Table 4 (RXTE)
%
\newpage
\bigskip
\begin{deluxetable}{lccccccccccccc}
%\begin{deluxetable}{lcccccccccccccc}
%\begin{deluxetable}{cccccccccccccc}
\rotate
\tablewidth{0in}
\tabletypesize{\scriptsize}
%  \begin{center}
    \tablecaption{Best-fit parameters of spectral analysis of PCA+HEXTE/{\it RXTE} 
observations of GX~340+1 in 3 -- 150~keV energy range$^{\dagger}$. Adopted model: {\it wabs*(Blackbody + Comptb1 + Comptb2 + Gaussian)}.  
Parameter errors are given at 90\% confidence level.}
%\vspace{1em}
    \renewcommand{\arraystretch}{1.2}
%    \begin{tabular}[h]
%      \hline
%ID               & day  & & &   $\Gamma-1$          &           &$L_{39}/d^2_{10}$& keV & keV   &  &  keV        &  & & & }
%% \tablehead
%%{Observational & MJD, & $\alpha=$  & $T_e,$ & log(A) & N$_{COMPTB}^{\dagger\dagger}$ & $T_s$, & $N_{Bbody}^{\dagger\dagger}$ & E$_{line}$,& $\sigma_{line},$ & $N_{line}^{\dagger\dagger}$ &  $\chi^2_{red}$ (d.o.f.)& F$_1$/F$_2^{\dagger\dagger\dagger}$ \\
%%ID             & day  & $\Gamma-1$ & keV    &                           &                                      & keV    &                               &  keV       &       keV              &                    &                         &                                                & }
%% \startdata%   id     MJD    alf     T_e       log     norm_COMPTB    kT_bb     N_bb    E_line      N_line   Xi_2(dof)  Flux3-10 Fl10-60
 \tablehead                                                                                                          
%     1            2         3            4             5                6                                7                                   8                9                 10                          11                  12                               13                    14                                                  16     
{Observational & MJD, & $\alpha_1=$  & $kT^{(1)}_e,$ & $\log(A_1)$& $N_{Com1}^{\dagger\dagger\dagger}$ & $N_{Bbody}^{\dagger\dagger\dagger}$ & $\alpha_2=$  & $kT^{(2)}_e,$ & N$_{Com2}^{\dagger\dagger\dagger}$ &E$_{line}$,& $N_{line}^{\dagger\dagger\dagger}$ &  $\chi^2_{red}$ (d.o.f.)& F$_1$/F$_2^{\dagger\dagger\dagger\dagger}$ \\
ID             & day  & $\Gamma_1-1$ & keV        &           &                                    &                                     & $\Gamma_2-1$ & keV        &                                    & keV       &                                    &                         &                                     }
 \startdata%   id     MJD    alf     T_e       log     norm_COMPTB    kT_bb     N_bb    E_line      N_line   Xi_2(dof)  Flux3-10 Fl10-60
20054-04-01-000 & 50555.570 & 0.99(1) & 17.96(5) & -1.19(4) & 4.72(9) & 1.03(9) & 1.01(1) &  2.72(7) & 11.77(3) & 6.74(9) & 1.05(2) & 1.17(67) & 9.89/3.49 \\ %13.38 3.04  3.57 5.24 2.99 3.30 3.75 12.39 
20059-01-01-000 & 50605.255 & 0.99(2) & 5.43(2)  & -0.54(2) & 2.06(6) & 0.92(3) & 1.02(2) &  1.63(4) & 13.29(5) & 6.45(8) & 2.04(3) & 1.46(67) & 9.07/1.14 \\ %10.21 1.07  3.92 5.06 2.09 1.43 2.99 10.05
20059-01-01-01  & 50605.904 & 1.03(3) & 18.77(5) & -1.32(2) & 4.68(4) & 1.04(8) & 0.99(2) &  2.74(1) & 11.72(6) & 6.67(9) & 1.03(5) & 0.93(67) & 9.87/3.39 \\ %13.26 3.01  3.59 5.22 2.96 3.27 3.71 12.34
20059-01-01-02  & 50606.471 & 1.03(2) & 20.40(9) & -0.90(3) & 3.14(9) & 1.06(5) & 1.00(1) &  2.91(6) & 13.4(9)  & 6.50(7) & 1.02(4) & 1.08(67) & 11.61/3.59\\ % 15.14 3.18  4.25 6.23 3.40 3.55 4.34 14.26
20059-01-01-03  & 50606.996 & 1.04(3) & 12.32(8) & -1.09(4) & 3.45(5) & 1.01(6) & 1.03(4) &  2.49(1) & 16.03(4) & 6.5(1)  & 2.09(2) & 1.19(67) & 12.36/3.50\\ % 15.87 3.17  4.57 6.68 3.56 3.59 4.59 15.03
20059-01-01-04  & 50607.329 & 1.02(2) & 12.34(9) & -0.06(3) & 3.50(6) & 1.07(4) & 1.00(2) &  2.48(4) & 16.45(9) & 6.4(1)  & 2.03(2) & 1.31(67) & 12.61/3.55\\ % 16.16 3.24  4.66 6.82 3.63 3.66 4.67 15.32
20059-01-01-05  & 50608.008 & 0.99(1) & 15.08(5) & -1.29(3) & 6.47(5) & 1.09(5) & 1.02(3) &  2.70(2) & 11.65(9) & 6.40(9) & 2.83(6) & 1.08(67) & 10.86/3.53\\ % 14.39 3.09  3.96 5.79 3.21 3.41 4.07 13.41
20059-01-01-06  & 50609.062 & 1.01(2) & 11.28(8) & -1.05(1) & 4.03(6) & 1.08(6) & 1.01(1) &  2.27(1) & 17.03(9) & 6.5(1)  & 2.29(8) & 0.94(67) & 11.70/2.71\\ % 14.41 2.45  4.59 6.43 3.11 2.86 4.16 13.78
20053-05-01-00  & 50712.048 & 1.02(3) & 4.01(9)  & -0.42(5) & 2.88(9) & 0.92(7) & 0.98(3) &  2.18(3) & 8.67(8)  & 6.74(3) & 0.75(7) & 0.84(67) & 10.58/1.56\\ % 10.58 1.73  4.09 5.51 2.52 2.09 3.45 11.49  
20053-05-01-01  & 50714.179 & 1.06(4) & 3.12(4)  & -0.11(2) & 2.05(6) & 0.92(2) & 1.01(1) &  1.85(2) & 10.1(1)  & 6.41(8) & 2.31(6) & 0.97(67) & 10.00/1.88\\ % 11.86 1.40  3.83 4.98 2.16 1.71 3.01 10.24
20053-05-01-02  & 50716.066 & 1.02(2) & 16.65(4) & -1.20(3) & 4.76(5) & 1.03(9) & 1.02(3) &  2.69(1) & 12.55(4) & 6.7(1)  & 1.03(9) & 1.09(67) & 10.52/3.47\\ % 13.99 3.12  3.84 5.57 3.14 3.41 3.94 13.09
20053-05-01-03  & 50920.197 & 1.05(3) & 12.96(9) & -1.08(4) & 3.62(9) & 1.01(8) & 1.01(5) &  2.59(7) & 13.11(6) & 6.4(1)  & 2.1(1)  & 0.94(67) & 10.51/3.12\\ % 13.62  2.81  3.90 5.63 3.04 3.12 3.89 12.84 
20053-05-02-00  & 50753.947 & 0.99(1) & 4.43(5)  & -0.56(1) & 2.04(2) & 0.92(1) & 1.04(3) &  1.56(8) & 16.35(1) & 6.62(2) & 0.70(4) & 1.12(67) & 9.42/1.14 \\ %10.56  1.09  3.97 5.25 2.25 1.49 3.19 10.43 
20053-05-02-01  & 50754.091 & 0.99(2) & 10.09(8) & -1.14(3) & 5.31(3) & 1.07(5) & 1.02(2) &  2.26(2) & 12.80(9) & 6.57(9) & 2.28(5) & 1.07(67) & 10.82/2.29 \\ %13.12  2.14  4.30 5.97 2.81 2.52 3.78 12.63 
20053-05-02-02  & 50754.825 & 1.02(2) & 12.30(9) & -1.08(4) & 3.38(6) & 1.0(1)  & 1.01(3) &  2.48(1) & 15.98(8) & 6.36(9) & 2.09(9) & 0.93(67) & 12.37/3.39 \\ %15.78  3.12  4.62 6.69 3.53 3.53 4.57 14.99
20053-05-02-03  & 50755.133 & 1.03(2) & 11.34(8) & -1.09(3) & 3.41(5) & 1.07(5) & 1.03(4) &  2.45(2) & 15.54(7) & 6.47(8) & 2.08(4) & 0.89(67) & 12.04/3.19 \\ %15.24  2.93  4.53 6.54 3.39 3.34 4.41 14.50
20053-05-02-05  & 50756.680 & 1.01(1) & 17.85(4) & -1.21(2) & 4.76(7) & 1.04(8) & 1.03(5) &  2.71(2) & 11.71(3) & 6.8(1)  & 1.06(5) & 0.89(67) & 9.89/3.28  \\ %13.17 2.98  3.61 5.24 2.96 3.25 3.72 12.34
20053-05-02-04  & 50756.685 & 1.05(4) & 12.94(5) & -1.08(4) & 3.40(8) & 1.0(1)  & 0.99(6) &  2.54(1) & 15.81(4) & 6.42(9) & 2.04(9) & 0.91(67) & 12.23/3.57 \\ %15.80 3.28  4.51 6.58 3.56 3.67 4.56 14.97
30040-04-01-00  & 51130.995 & 1.04(2) & 16.87(4) & -1.46(3) & 6.23(7) & 1.04(6) & 1.00(2) &  2.69(3) & 12.03(8) & 6.45(2) & 2.18(2) & 0.86(67) & 10.19/3.39 \\ %13.58 3.02  3.76 5.37 3.03 3.30 3.80 12.68
30040-04-01-06  & 51131.435 & 1.02(2) & 17.78(9) & -0.99(6) & 4.79(6) & 1.09(8) & 1.02(3) &  2.71(2) & 11.31(5) & 6.70(9) & 1.03(4) & 1.02(67) & 9.70/3.48  \\ %13.19 2.94  3.59 5.09 2.89 3.19 3.62 12.12
30040-04-01-01  & 51131.601 & 1.02(3) & 18.67(8) & -1.34(2) & 4.70(8) & 1.08(7) & 1.02(3) &  2.78(3) & 10.71(4) & 6.70(8) & 1.03(2) & 1.06(67) & 9.18/3.24  \\ %12.42 2.84  3.36 4.82 2.76 3.07 3.45 11.51
30040-04-01-020 & 51131.877 & 1.00(1) & 19.65(7) & -1.24(4) & 5.01(6) & 1.09(8) & 1.01(1) &  2.70(1) & 11.83(3) & 6.70(9) & 1.05(3) & 0.96(67) & 9.99/3.38  \\ %13.37 2.96  3.66 5.27 2.99 3.24 3.74 12.44
30040-04-01-03  & 51132.596 & 1.02(2) & 12.73(2) & -1.05(2) & 3.48(3) & 1.03(7) & 1.01(2) &  2.39(2) & 16.86(4) & 6.47(6) & 2.24(2) & 0.96(67) & 12.91/3.32 \\ %16.23 2.96  4.95 7.00 3.56 3.39 4.67 15.41
30040-04-01-05  & 51132.829 & 1.05(3) & 12.91(8) & -1.08(4) & 3.45(9) & 1.04(9) & 1.03(3) &  2.50(1) & 16.03(5) & 6.40(1) & 2.08(1) & 1.00(67) & 13.27/3.72 \\ %16.99 3.38  4.95 7.14 3.82 3.83 4.89 16.11
30040-04-01-04  & 51132.881 & 1.01(1) & 12.93(1) & -1.07(1) & 3.33(5) & 1.03(8) & 1.02(4) &  2.51(2) & 16.68(9) & 6.40(6) & 2.14(2) & 0.96(67) & 12.67/3.49 \\ %16.16 3.11  4.81 6.79 3.67 3.52 4.61 15.27
50016-01-01-11  & 51772.077 & 0.99(2) & 12.91(4) & -1.08(4) & 3.40(7) & 1.06(3) & 0.99(3) &  2.49(1) & 16.51(5) & 6.4(1)  & 2.95(3) & 0.71(67) & 13.24/3.96 \\ %17.19 3.52  4.76 7.13 3.92 3.96 5.02 16.20  
50016-01-01-12  & 51772.144 & 1.04(4) & 11.8(1)  & -1.10(7) & 3.5(1)  & 1.01(4) & 1.01(2) &  2.4(1)  & 17.29(9) & 6.41(7) & 2.18(4) & 1.03(67) & 13.11/3.42 \\ %16.54 3.11  4.94 7.10 3.71 3.56 4.82 15.76
50016-01-01-000 & 51772.210 & 0.99(2) & 12.72(9) & -1.04(2) & 3.39(9) & 1.04(9) & 1.03(4) &  2.45(3) & 16.55(8) & 6.70(9) & 0.80(5) & 1.02(67) & 12.69/3.43 \\ %16.18 3.11  4.75 6.85 3.62 3.53 4.68 15.31  
50016-01-01-01  & 51772.708 & 1.01(2) & 12.28(8) & -1.09(2) & 3.49(5) & 1.04(8) & 0.98(4) &  2.39(1) & 17.34(5) & 6.4(1)  & 2.62(6) & 1.12(67) & 13.08/3.52 \\ %16.60 3.22  4.79 7.11 3.77 3.66 4.89 15.79  
50016-01-01-10  & 51772.742 & 1.01(1) & 12.96(2) & -1.08(4) & 3.35(7) & 1.02(3) & 0.99(1) &  2.54(2) & 16.19(8) & 6.42(9) & 2.89(2) & 1.01(67) & 12.50/3.73 \\ %16.23 3.34  4.60 6.70 3.67 3.73 4.68 15.31  
50016-01-01-07  & 51773.012 & 1.02(2) & 11.06(9) & -1.06(7) & 4.0(1)  & 1.01(6) & 1.01(2) &  2.29(1) & 19.4(1)  & 6.53(8) & 2.29(5) & 0.76(67) & 12.83/3.10 \\ %15.93 2.83  4.92 7.00 3.53 3.28 4.65 15.25 
50016-01-01-02  & 51773.082 & 1.04(3) & 12.9(1)  & -1.08(6) & 3.48(9) & 1.04(9) & 1.01(1) &  2.45(5) & 17.31(6) & 6.40(7) & 2.16(7) & 0.76(67) & 13.11/3.75 \\ %16.86 3.39  4.77 7.09 3.82  3.82 4.92 15.94  
50016-01-01-03  & 51773.146 & 1.02(2) & 12.3(1)  & -1.02(3) & 3.39(4) & 1.32(5) & 1.05(4) &  2.45(9) & 17.46(5) & 6.70(8) & 2.94(5) & 0.87(67) & 12.72/3.81 \\ %16.53 3.45  4.67 6.82 3.73  3.84 4.75 15.59  
50016-01-01-04  & 51773.211 & 1.01(1) & 12.26(9) & -1.10(2) & 3.47(5) & 1.33(9) & 1.04(3) &  2.46(7) & 17.67(8) & 6.64(9) & 2.90(4) & 0.76(67) & 12.76/3.77 \\ %16.54 3.36  4.70 6.86 3.72  3.77 4.76 15.58 
50016-01-01-05  & 51773.281 & 0.99(2) & 13.0(1)  & -1.06(2) & 4.49(3) & 1.02(8) & 1.05(4) &  2.46(1) & 17.41(9) & 6.40(5) & 2.79(5) & 1.16(67) & 12.61/3.71 \\ %16.32 3.29  4.65 6.78 3.66  3.70  4.69 15.37 
50016-01-01-09  & 51773.362 & 1.02(3) & 11.09(3) & -1.13(1) & 5.6(1)  & 1.12(6) & 1.01(1) &  2.32(2) & 14.31(9) & 6.54(9) & 2.09(2) & 1.07(67) & 11.47/2.69 \\ %14.16 2.45  4.47 6.29 3.06  2.84  4.09 13.54
50016-01-01-06  & 51773.619 & 0.99(1) & 10.0(2)  & -1.14(9) & 5.5(1)  & 1.05(2) & 1.00(3) &  2.22(3) & 12.51(3) & 6.6(1)  & 2.15(3) & 1.06(67) & 10.65/2.23 \\ %12.89 2.03  4.27 5.89 2.75 2.41 3.72 12.39
50016-01-01-08  & 51773.939 & 1.05(4) & 9.2(1)   & -1.13(4) & 5.4(1)  & 1.1(1)  & 1.03(4) &  2.13(9) & 13.3(1)  & 6.5(1)  & 2.39(9) & 1.24(67) & 10.82/2.09 \\ %12.91 1.91  4.42 6.02 2.70 2.29 3.71 12.46
50016-01-02-00  & 51774.005 & 1.02(2) & 5.42(4)  & -0.99(3) & 1.39(8) & 0.92(4) & 0.99(2) &  1.52(2) & 15.5(1)  & 6.87(9) & 0.56(3) & 1.39(67) & 9.25/0.98  \\ %10.23 0.89  3.91 5.23 2.14 1.31 3.09 10.09  
50016-01-02-18  & 51774.075 & 1.05(6) & 3.30(5)  & -0.32(2) & 1.31(9) & 0.92(9) & 0.99(1) &  1.52(3) & 15.92(8) & 6.89(4) & 0.56(3) & 1.28(67) & 9.04/0.99  \\ %10.03 0.92  3.83 5.13 2.08 1.32 3.01 9.92
50016-01-02-05  & 51774.138 & 1.01(2) & 2.8(1)   & -1.08(2) & 3.5(1)  & 1.1(1)  & 0.99(3) &  1.55(1) & 13.05(7) & 6.39(5) & 1.90(4) & 1.26(67) & 9.56/1.04 \\ %15.89 3.21  4.56 6.67 3.59 3.63 4.61 15.08  
50016-01-02-03  & 51774.204 & 0.99(1) & 3.67(6)  & -0.97(6) & 1.32(8) & 0.93(8) & 1.02(2) &  1.51(2) & 17.5(1)  & 6.89(4) & 0.59(5) & 1.27(67) & 9.64/1.04  \\ %10.69 0.95  3.99 5.46 2.29 1.41 3.28 10.56 
50016-01-02-16  & 51774.270 & 1.01(3) & 3.34(3)  & -0.11(3) & 1.39(7) & 0.91(4) & 1.01(1) &  1.70(8) & 15.60(9) & 6.67(5) & 0.57(9) & 1.22(67) & 8.89/1.32  \\ %10.20 1.20  3.81 4.94 2.11 1.54 2.97 9.98
50016-01-02-04  & 51774.574 & 1.01(2) & 12.9(1)  & -1.08(2) & 3.5(1)  & 1.1(1)  & 0.99(4) &  2.46(1) & 15.23(6) & 6.40(6) & 2.86(4) & 0.99(67) & 12.38/3.51 \\ %15.89 3.21  4.56 6.67 3.59 3.63 4.61 15.08  
50016-01-02-12  & 51774.704 & 0.99(1) & 11.04(3) & -1.05(4) & 4.08(9) & 1.01(7) & 1.01(3) &  2.36(2) & 16.94(5) & 6.33(5) & 4.28(5) & 0.95(67) & 11.47/2.87 \\ %14.34 2.61  4.41 6.28 3.17 2.99 4.14 13.68
50016-01-02-17  & 51775.001 & 1.02(2) & 11.07(9) & -1.13(5) & 5.49(7) & 1.06(3) & 1.00(1) &  2.4(1)  & 14.82(6) & 6.58(6) & 2.94(8) & 0.98(67) & 11.88/2.94 \\ %14.83  2.69  4.50 6.52 3.28 3.09 4.33 14.15
50016-01-02-02  & 51775.067 & 1.07(6) & 11.1(1)  & -1.08(2) & 4.02(8) & 1.9(1)  & 1.01(2) &  2.4(1)  & 16.41(8) & 6.38(7) & 3.10(1) & 0.97(67) & 12.54/3.18 \\ %15.72  2.93  4.74 6.84 3.50 3.35 4.59 15.02
50016-01-02-07  & 51775.200 & 1.02(2) & 11.08(9) & -1.05(4) & 4.18(9) & 1.07(9) & 1.00(3) &  2.3(1)  & 16.53(8) & 6.37(4) & 2.02(3) & 1.03(67) & 12.02/3.05 \\ %15.07  2.76  4.52 6.59 3.36 3.17 4.42 14.36  
50016-01-02-06  & 51775.267 & 1.04(3) & 11.1(1)  & -1.12(3) & 4.87(2) & 1.00(1) & 1.02(2) &  2.39(9) & 14.45(6) & 6.56(3) & 2.08(2) & 1.02(67) & 11.17/2.62 \\ %13.78  2.39  4.36 6.08 3.04 2.78 4.01 13.24  
50016-01-02-15  & 51775.533 & 1.00(1) & 12.9(1)  & -1.09(1) & 3.49(9) & 1.02(3) & 1.05(5) &  2.45(6) & 15.56(3) & 6.42(7) & 2.06(3) & 0.97(67) & 11.99/3.32 \\ %15.32  2.97  4.47 6.49 3.42 3.36 4.44 14.51
50016-01-02-08  & 51775.734 & 1.05(4) & 12.87(8) & -1.08(5) & 3.41(5) & 1.03(2) & 0.99(3) &  2.49(1) & 15.52(8) & 6.45(4) & 2.88(4) & 0.99(67) & 12.02/3.40 \\ %15.42  3.06  4.48 6.48 3.46 3.44 4.47 14.60  
50016-01-02-10  & 51775.998 & 1.00(1) & 12.9(1)  & -1.07(4) & 3.52(9) & 1.07(5) & 1.06(4) &  2.39(2) & 15.15(8) & 6.40(5) & 2.49(5) & 0.94(67) & 12.31/3.39 \\ %15.70  2.99  4.56 6.69 3.52 3.41 4.57 14.84
50016-01-02-09  & 51776.064 & 1.01(2) & 12.84(9) & -1.06(5) & 3.34(7) & 1.0(1)  & 0.99(3) &  2.41(1) & 17.31(9) & 6.40(6) & 2.69(3) & 0.99(67) & 12.42/3.37 \\ %15.79  3.08  4.56 6.74 3.59 3.50 4.65 15.03  
50016-01-02-13  & 51776.131 & 1.02(3) & 11.07(8) & -1.13(7) & 5.58(9) & 1.06(4) & 1.01(5) &  2.31(3) & 14.21(9) & 6.6(1)  & 2.75(4) & 1.20(67) & 11.53/2.72 \\ %14.25  2.46  4.42 6.33 3.14 2.86 4.17 13.63
50016-01-02-11  & 51776.529 & 0.99(2) & 3.24(2)  & -0.38(6) & 2.07(9) & 0.92(6) & 1.03(4) &  1.94(8) & 10.23(7) & 6.43(7) & 2.31(9) & 0.92(67) & 9.10/1.59  \\ %10.86  1.45  3.81 5.08 2.19 1.78 3.05 10.34
50016-01-02-01  & 51776.680 & 1.01(2) & 4.12(4)  & -0.46(3) & 2.01(7) & 0.9(1)  & 1.02(3) &  1.8(1)  & 11.1(1)  & 6.35(6) & 2.26(3) & 1.16(67) & 8.85/1.34  \\ %10.19  2.74  4.58 6.71 3.42 3.17 4.50 14.56  
50016-01-04-01G & 51920.397 & 1.01(1) & 18.81(9) & -1.81(5) & 4.78(8) & 1.02(4) & 1.01(2) &  2.9(1)  & 11.24(9) & 6.7(1)  & 1.08(5) & 0.87(67) & 9.28/3.61  \\ %12.88 3.19  3.26 4.83 2.92 3.39 3.58 11.87
50016-01-04-00  & 51920.705 & 1.03(4) & 15.92(5) & -1.64(3) & 6.26(9) & 1.08(9) & 0.99(1) &  2.69(9) & 13.71(9) & 6.41(4) & 2.10(3) & 1.03(67) & 11.64/3.78 \\ %15.42 3.41  4.16 6.21 3.51 3.76 4.41 14.46
50016-01-03-11  & 51921.001 & 0.99(2) & 12.94(7) & -1.07(9) & 3.47(9) & 1.03(4) & 1.01(2) &  2.49(1) & 16.27(5) & 6.46(5) & 2.14(4) & 1.12(67) & 12.47/3.61 \\ %16.08  3.27  4.58 6.71 3.65 3.68 4.67 15.23
50016-01-03-12  & 51921.317 & 0.99(3) & 11.79(4) & -1.08(5) & 5.12(8) & 1.06(1) & 1.00(1) &  2.33(4) & 15.54(4) & 6.53(5) & 2.29(3) & 0.69(67) & 12.23/2.98 \\ %15.21  2.74  4.58 6.73 3.41 3.17 4.50 14.56
50016-01-04-02  & 51921.384 & 1.02(2) & 4.35(2)  & -0.45(4) & 2.05(7) & 0.93(3) & 1.01(2) &  1.65(5) & 13.2(1)  & 6.50(2) & 2.08(5) & 0.97(67) & 8.92/1.24  \\ %10.16  1.13  3.81 4.97 2.11 1.48 2.99 9.96
50016-01-03-02  & 51921.700 & 0.98(2) & 3.44(3)  & -0.36(2) & 2.19(9) & 0.92(4) & 0.98(2) &  1.68(1) & 14.09(8) & 6.49(5) & 2.11(2) & 1.12(67) & 9.79/1.41  \\ %11.20  1.31  4.16 5.43 2.35 1.64 3.29 10.99
50016-01-03-01  & 51922.039 & 1.01(1) & 3.08(9)  & -0.07(5) & 3.17(9) & 1.04(5) & 0.97(4) &  2.42(2) & 17.68(5) & 6.46(4) & 2.00(3) & 1.18(67) & 10.01/2.09 \\ %15.11  2.79  4.57 6.52 3.36 3.19 4.38 14.38
50016-01-03-13  & 51922.312 & 1.03(5) & 11.3(1)  & -1.14(9) & 3.39(8) & 1.07(6) & 1.03(4) &  2.62(8) & 12.51(9) & 6.54(7) & 1.01(2) & 0.79(67) & 10.84/1.96 \\ %12.79  1.83  4.54 5.97 2.69 2.21 3.69 12.46 
50016-01-03-00  & 51922.379 & 1.05(4) & 5.42(2)  & -0.69(3) & 2.06(7) & 0.95(5) & 1.01(2) &  1.58(1) & 13.34(7) & 6.7(1)  & 0.73(5) & 1.48(67) & 8.52/1.02  \\ %9.54   0.95  3.71 4.79 1.95 1.28 2.80 9.42
50016-01-03-14  & 51922.648 & 1.04(3) & 8.40(3)  & -1.32(6) & 1.92(6) & 0.96(2) & 0.99(1) &  1.55(4) & 14.50(8) & 6.71(7) & 0.77(8) & 1.28(67) & 8.47/0.91  \\ %9.38   0.84  3.66 4.75 1.94 1.21 2.81 9.29
50016-01-03-03  & 51922.702 & 1.05(4) & 4.04(9)  & -0.41(5) & 2.09(5) & 0.92(3) & 0.98(4) &  1.84(2) & 12.08(9) & 6.42(5) & 2.31(2) & 1.05(67) & 9.49/1.54  \\ %11.03 1.43 3.92 5.28 2.33 1.79 3.24 10.72
50016-01-03-040 & 51922.733 & 1.01(1) & 11.80(3) & -1.14(2) & 5.43(6) & 1.04(7) & 1.02(3) &  2.29(3) & 14.19(2) & 6.50(5) & 2.03(3) & 0.83(67) & 11.37/2.62 \\ %13.99  2.39  4.36 6.25 3.09 2.80 4.11 13.42 
50016-01-03-05  & 51923.383 & 1.09(2) & 17.67(7) & -1.55(6) & 4.7(1)  & 1.06(5) & 0.99(2) &  2.80(1) & 12.43(7) & 6.73(4) & 1.03(1) & 0.90(67) & 10.30/3.28 \\ %13.99  3.28  3.70 5.43 3.15 3.53 3.92 12.99
50016-01-03-06  & 51923.628 & 1.03(3) & 19.91(2) & -1.02(3) & 3.08(9) & 3.07(3) & 1.00(1) &  2.74(2) & 10.25(4) & 6.40(5) & 3.90(2) & 0.97(67) & 9.23/3.51  \\ %12.73  3.06  3.30 4.83 2.85 3.26 3.52 11.72
50016-01-03-070 & 51923.694 & 1.02(2) & 17.26(3) & -1.51(4) & 4.69(8) & 1.05(3) & 1.02(2) &  2.76(1) & 12.97(3) & 6.72(6) & 1.01(4) & 1.02(67) & 10.65/3.76 \\ %14.41  3.34  3.30 4.83 2.85 3.26 3.52 11.72
50016-01-03-09  & 51924.299 & 1.05(6) & 17.10(6) & -1.34(5) & 4.67(6) & 1.06(2) & 1.03(2) &  2.66(3) & 14.76(7) & 6.71(5) & 0.98(3) & 1.16(67) & 11.80/4.01 \\ %15.81  3.59  4.13 6.29 3.63 3.92 4.54 14.77
50016-01-03-10  & 51924.365 & 1.01(1) & 12.52(6) & -1.07(3) & 3.12(9) & 1.03(4) & 1.02(2) &  2.44(1) & 17.67(6) & 6.40(5) & 2.17(3) & 1.00(67) & 12.44/3.29 \\ %15.73  2.99  4.68 6.73 3.53 3.41 4.58 14.98
50016-01-03-08  & 51924.583 & 1.03(3) & 17.51(8) & -1.59(6) & 4.79(6) & 0.99(20 & 0.99(1) &  2.8(1)  & 12.11(8) & 6.71(7) & 1.08(2) & 0.79(67) & 10.03/3.67 \\ %13.71  3.25  3.58 5.28 3.09 3.49 3.84 12.69
80020-02-01-00  & 52908.041 & 1.01(1) & 12.92(9) & -1.07(8) & 3.40(5) & 1.03(9) & 1.01(2) &  2.44(1) & 13.08(9) & 6.41(2) & 2.97(3) & 0.98(67) & 10.32/2.84 \\ %13.16 2.56  3.84 5.62 2.92 2.90 3.81 12.47
80020-02-01-01  & 52908.109 & 1.00(1) & 12.25(6) & -1.05(3) & 3.39(7) & 1.07(8) & 1.03(4) &  2.42(1) & 13.76(6) & 6.43(2) & 2.98(2) & 0.94(67) & 10.09/2.63 \\ %12.72 2.36  3.83 5.52 2.78 2.69 3.66 12.07
80020-02-02-00  & 52914.565 & 1.02(2) & 4.24(6)  & -0.53(5) & 2.08(6) & 0.92(5) & 1.01(1) &  1.67(2) & 10.04(5) & 6.24(9) & 2.31(5) & 0.89(67) & 7.23/1.04  \\ %8.27  0.89  3.18 4.01 1.64 1.16 2.35 8.04
80020-02-02-01  & 52914.941 & 1.05(4) & 12.93(8) & -1.08(4) & 3.45(5) & 1.02(8) & 0.99(2) &  2.48(6) & 12.91(4) & 6.42(9) & 2.94(3) & 1.07(67) & 10.24/2.92 \\ %13.16 2.59  3.86 5.54 2.89 2.92 3.75 12.41
80020-02-02-02  & 52915.008 & 1.03(2) & 11.02(4) & -1.13(7) & 5.56(4) & 1.01(7) & 0.99(1) &  2.31(7) & 11.68(6) & 6.50(9) & 2.28(2) & 1.02(67) & 9.35/2.23  \\ %10.58 2.00   3.67 5.14 2.48 2.32 3.31 11.04
80020-02-04-00  & 53210.048 & 1.01(2) & 12.94(7) & -1.08(7) & 3.47(6) & 1.05(6) & 0.98(2) &  2.53(6) & 14.38(4) & 6.45(3) & 2.82(7) & 1.33(67) & 11.26/3.29 \\ %14.54 2.91  4.18 6.09 3.22 3.28 4.16 13.69
80020-02-05-00  & 53222.037 & 1.07(5) & 9.79(5)  & -1.12(1) & 4.05(4) & 1.03(7) & 1.05(4) &  2.12(2) & 12.06(8) & 6.51(9) & 2.96(3) & 1.06(67) & 9.61/1.85  \\ %11.46 1.70   4.00 5.32 2.36 2.03 3.25 11.07
80020-02-06-00  & 53238.305 & 1.04(4) & 18.96(8) & -1.08(5) & 3.43(5) & 1.04(8) & 1.01(3) &  2.51(4) & 14.07(7) & 6.40(9) & 2.73(2) & 1.36(67) & 11.07/3.18 \\ %14.25 2.82  4.21 5.96 3.12 3.17 4.03 13.43
80020-02-07-00  & 53245.879 & 1.03(2) & 10.14(9) & -1.13(2) & 4.51(9) & 1.08(7) & 1.04(5) &  2.24(2) & 14.14(5) & 6.53(9) & 2.01(2) & 1.20(67) & 9.85/2.16  \\ %12.00 1.95  3.98 5.43 2.52 2.29 3.41 11.51 
80020-02-08-00  & 53255.234 & 0.98(5) & 20.53(2) & -0.91(3) & 3.06(8) & 1.07(6) & 1.02(2) &  3.00(1) & 9.13(6)  & 6.39(7) & 1.67(3) & 1.37(67) & 7.43/3.32  \\ %10.76 2.77  2.59 3.84 2.38 2.86 2.90 9.64
80020-02-09-00  & 53262.321 & 0.99(2) & 11.23(8) & -1.13(4) & 5.43(4) & 1.04(9) & 1.02(3) &  2.31(1) & 13.97(9) & 6.51(8) & 2.25(5) & 1.10(67) & 10.76/2.51 \\ %13.28 2.32  4.23 5.89 2.87 2.69 3.82 12.73
94312-01-01-00  & 54893.348 & 1.01(4) & 17.84(8) & -1.63(9) & 4.73(8) & 1.0(1)  & 1.01(1) &  2.95(2) & 9.46(9)  & 6.70(8) & 0.93(4) & 1.14(67) & 8.48/2.96  \\ %11.44  2.64  3.25 8.27 2.46 2.83 3.08 10.63
94312-01-01-02  & 54894.766 & 0.99(2) & 14.01(9) & -1.02(5) & 5.15(7) & 1.06(9) & 1.02(4) &  2.69(3) & 11.53(8) & 6.46(6) & 2.86(5) & 1.16(67) & 10.53/3.46 \\ %13.99  3.08  4.01 5.52 3.02 3.34 3.83 13.05
94312-01-01-04  & 54894.895 & 1.10(9) & 12.54(8) & -1.07(3) & 3.46(5) & 1.1(1)  & 1.01(5) &  2.44(1) & 14.15(4) & 6.40(7) & 2.69(2) & 1.17(67) & 10.38/2.60 \\ %12.98  2.37  4.19 5.56 2.73 2.69 3.60 12.37
93046-04-01-00  & 54297.052 & 1.02(5) & 11.02(9) & -1.05(2) & 4.03(9) & 1.02(6) & 1.00(2) &  2.39(1) & 12.58(7) & 6.3(1)  & 0.9(1)  & 1.13(67) & 9.07/2.21  \\ %12.89  1.99  3.72 4.87 2.33 2.26 3.11 10.73
91152-03-01-000 & 53874.858 & 0.99(2) & 9.02(8)  & -1.04(6) & 4.6(1)  & 1.05(9) & 1.02(6) &  2.28(4) & 10.57(9) & 6.58(5) & 2.72(6) & 1.30(67) & 9.57/1.90  \\ %11.47  1.73  4.05 2.26 2.32 2.04 3.18 11.04
91152-03-02-00  & 53907.403 & 1.05(4) & 10.11(1) & -1.10(5) & 4.5(1)  & 1.07(4) & 1.01(5) &  2.21(1) & 10.9(1)  & 6.53(9) & 2.23(1) & 1.16(67) & 9.14/1.92  \\ %11.06  1.64  3.91 5.04 2.21 1.93 4.73 10.57
91152-03-02-01  & 53907.680 & 1.01(1) & 11.0(1)  & -1.06(9) & 4.03(7) & 1.06(5) & 1.03(4) &  2.29(2) & 15.74(5) & 6.43(8) & 2.38(3) & 0.98(67) & 10.25/2.56 \\ %12.81  2.23  4.07 5.62 2.71 2.56 5.52 12.14
     \enddata%     \hline%      \end{tabular}                         
    \label{tab:fit_table_rxte}
%  \end{center}
$^\dagger$ The spectral model is  $wabs*(blackbody + Comptb1 + Comptb2 + Gaussian)$;% where $N_H$ is fixed at 
%a value 9.6$\times 10^{22}$ cm$^{-2}$ (Church et al., 2006); % (Oosterbroek et al., 2001)
%a value 6.8$\times 10^{22}$ cm$^{-2}$ (Oosterbroek et al., 2001); 
color temperature %$T_s$ and 
$T_{BB}$ of $Bbody$ component is fixed at %1.3 and 
0.6 keV and ``seed'' photon temperatures $T_{s1}$/$T_{s2}$ are fixed at 1.1/1.5 keV, respectively
 (see comments in the text); 
$^{\dagger\dagger}$ parameter $\log(A_2)$ is fixed at 2.0 (see comments in the text), 
%$^{\dagger\dagger}$ when parameter $\log(A)\gg1$, this parameter is fixed at 2.0 (see comments in the text), 
$^{\dagger\dagger\dagger}$ normalization parameters of $blackbody$ and $COMPTB$ components are in units of 
$L_{37}/d^2_{10}$, where $L_{37}$ is the source luminosity in units of 10$^{37}$ erg/s, 
$d_{10}$ is the distance to the source in units of 10 kpc 
and $Gaussian$ component is in units of $10^{-2}\times total~~photons$ $cm^{-2}s^{-1}$ in line 
%({\bf Lenochka please specify units for these components, look at XSPEC manual})
,  
$\sigma_{line}$ of Gaussian component is fixed to a value 0.6 keV (see comments in the text),
$N_H$ was fixed at value of 6.2$\times 10^{22}$ cm$^{-2}$ (Iaria et al., 2006),  
$^{\dagger\dagger\dagger\dagger}$spectral fluxes (F$_1$/F$_2$) in units of $\times 10^{-9}$ ergs/s/cm$^2$ for  (3 -- 10) and (10 -- 50) keV energy ranges respectively.  
%in units of $\times 10^{-9}$ ergs/s/cm$^2$.
%* this observations are  fitted with $bmc+Gaussian1+Gaussian2+bbody$ model, see values of the best-fit BB color temperature and EW in Table 2, 3 and 4.
\end{deluxetable}

%~~~~~~~~~~
\vspace{2.in}
\newpage
~~~~~~~~
%\vspace{2.in}
\begin{deluxetable}{llcccccccc}
%\begin{deluxetable}{cccccccccccccccc}
\rotate
\tablewidth{0in}
\tabletypesize{\scriptsize}
%  \begin{center}
    \tablecaption{Comparisons of the best-fit parameters  of {\it Z} source GX~340+0 and {\it atoll} sources GX~3+1$^1$ and 4U~1728-34$^2$}
%\vspace{1em}
    \renewcommand{\arraystretch}{1.2}
%    \begin{tabular}[h]
%      \hline
%ID               & day  & & &   $\Gamma-1$          &           &$L_{39}/d^2_{10}$& keV & keV   &  &  keV        &  & & & }
 \tablehead
{Source & Alternative & Class$^3$& Distance, & Presence of & $kT_e$,  & $ N_{comptb}$ &$kT_{BB}$ &  $kT_{s}$  & $f$ \\
  name  & name        &          & kpc       & kHz QPO     & keV         &  $L_{39}^{soft}/{D^2_{10}}$        & keV  & keV  &  }
 \startdata%   id     MJD      kT_Bbody     N_bb     [kT_s]    alf       T_e       log    norm_COMPTB E_line[Sigma_l] N_line Xi_2(dof)  Flux3-10 Fl10-60
4U~1642-45  & GX 340+0 & Z, Sp, B     & 10.5$^{4}$ &    +$^7$        & 3-21   &  0.08-0.2 & 0.6 & 1.1-1.5 & 0.01-0.5   \\
4U~1744-26  & GX 3+1   & Atoll, Sp, B & 4.5$^5$ &     none$^8$      & 2.3-4.5 &  0.04-0.15 & 0.6 & 1.16-1.7 & 0.2-0.9   \\
4U 1728-34  & GX~354-0 & Atoll, Su, D & 4.2-6.4$^6$ & +$^9$&  2.5-15& 0.02-0.09 & 0.6-0.7 & 1.3 & 0.5-1\\     
      \enddata%     \hline
%      \end{tabular}
    \label{tab:fit_table_comb}
%  \end{center}
%$^\dagger$ The spectral model is  $wabs*(blackbody + COMPTB + Gaussian)$,
%normalization parameters of $blackbody$ and $COMPTB$ components are in units of 
%$L_{37}^{soft}/d^2_{10}$ 
%$erg/s/kpc^2$, 
%where $L_{37}^{soft}$ is the soft photon  luminosity in units of 10$^{37}$ erg/s, 
%$d_{10}$ is the distance to the source in units of 10 kpc 
%and $Gaussian$ component is in units of $10^{-2}\times total~~photons$ $cm^{-2}s^{-1}$ in line.
%$^{\dagger\dagger\dagger}$when parameter $\log(A)\gg1$, it is fixed to a value 1.0 (see comments in the text). 
%$\sigma_{line2}$ of Gaussian2 component is fixed to a value 0.01 keV (see comments in the text), 
%$^{\dagger\dagger\dagger\dagger}$spectral fluxes (F$_1$/F$_2$) in the (3 -- 10)/(10 -- 60) keV energy ranges, correspondingly, 
%in units of $\times 10^{-9}$ ergs/s/cm$^2$.
References:
(1) ST12,
(2) ST11, 
(3) Classification of the system in the various schemes (see text): Sp = supercritical, Su = subcritical, 
B = bulge, D = disk,
(4) Fender \& Henry  (2000), Ford et al. (1998),  Christian \& Swank (1997),
(5) \citet{kk00}, Ford et al. (2000), 
(6) \citet{par78}, 
(7) \citet{Jonker98}, 
(8) \citet{stroh98},
(9) \citet{to99}  
\end{deluxetable}

\newpage

\vspace{10.in}

% 
% Figure 1
%

%\begin{figure}[ptbptbptb]
%\includegraphics[scale=0.9,angle=0]{f1.eps}
%\includegraphics[scale=0.9,angle=0]{asm_1.eps}
%\caption{  
% Evolution of ASM/{\it RXTE} count rate %, flux density $S_{8.46 GHz}$ at 8.46 GHz (VLA), 
%BMC normalization and 
%photon index $\Gamma$ 
%during  1997 -- 2009 observations of GX~340+0. 
% {\it Green} triangles show {\it Beppo}SAX NFI data, listed in Table 1 and  {\it blue} vertical strips (at {\it top of the panel}) indicate temporal distribution of the {\it RXTE} data 
%of pointed observations used in our
%analysis, and
%{\it bright blue} rectangles indicate
% (trace) 
%the {\it RXTE} data sets listed in Table 2. 
%Blue rectangles indicate the {\it RXTE} data of pointed observations and green triangles show {\it Beppo}SAX NFI data.
%Black dashed line illustrates  mean count rate of 
%of  Dwell type ASM light curve and indicates long-term quasi-periodic variability of mean soft flux 
%during  $\sim$ six years cycle.
%b. Model flux in 3-10 keV and 10-50 keV energy ranges (blue and crimson points respectively).
%c. Electron temperature $kT_e$ in keV.
%d. {\it COMPTB} and blackbody normalizations (crimson and blue points respectively). 
%e. In the last bottom panel we present variations of spectral index $\alpha=\Gamma-1$.
% data of pointed observations and  
%green triangles show BeppoSAX NFI data used for the analysis.
%}
%\label{variability_97-09}
%\end{figure}

%\newpage

% 
% Figure 2
%

\newpage

% 
% Figure 4
%

%newpage
%{index_norm}-\ref{index_temperature12}
% 
% Figure 5
%

\newpage

% 
% Figure 6
%

%\newpage

% 
% Figure 7
%

%\begin{figure}[ptbptbptb]
%\includegraphics[scale=1.0, angle=0]{f7.eps}
%\includegraphics[scale=1.0, angle=0]{sp_compar_sax_nb_hb.eps}
%\includegraphics[scale=1.0, angle=0]{Z3-states_4.eps}
%\caption{Three representative $EF_E$  diagrams for different states along Z-track of GX~340+0. 
%Data are taken from {\it RXTE} observations 
%50016-01-01-11 ($Z1$ panel, HB),
%{\it Horizontal branch}), 
%91125-03-01-000 ($Z2$ panel, NB), 
%{\it Normal branch}) 
%and 50016-01-02-18 ($Z3$ panel, FB).
%{\it Flaring branch}). 
%The data are shown by black crosses and  
% the spectral model components are displayed  by dashed red, green, blue and  purple lines for $Comptb1$, $Comptb2$, 
% $Blackbody$ and $Gaussian$ respectively. Yellow shaded areas demonstrate the evolution of $Comptb1$ component 
%during evolution %transitions 
%between the HB ($Z1$) and FB ($Z3$) states when the electron temperature $kT_e$ of the Compton
%cloud monotonically decreases from 21 to 3 keV (see also Fig.~\ref{index_temperature12}).
%Data are taken from BeppoSAX observations 
%2124100011 ($green$, {\it lower Normal branch}), 212400012 ($pink$, {\it upper Normal branch}) and 21375002 ($blue$, {\it Horizontal branch}).
%
%electron temperatures of plasma in Compton cloud
%[$kT_e=$2.9 keV ($red$), 3 keV ($blue$), 4 keV ($green$) and 5 keV ($voilet$)] in model $wabs*(Blackbody+COMPTB+Gaussian)$ 
%during $banana$ state of 4U~1820-30. Data are taken from RXTE observations 
%30057-01-04-01 ($green$), 70030-03-07-020 ($blue$), 70030-03-05-02 ($green$) and 70030-03-05-01 ($violet$).
%}
%\label{Zsp_compar_RXTE}
%\end{figure}

\newpage

% 
% Figure 8
%

%\newpage

% 
% Figure 9
%

%\begin{figure}[ptbptbptb]
%\includegraphics[scale=1.0,angle=0]{f9.eps}
%\includegraphics[scale=1.0,angle=0]{ccd_hid_3.eps}
%\caption{
%{\it From Top to Bottom:} 
%Color-color diagrams (CCD) ({\it left column}) and hardness-intensity diagrams (HID) ({\it right column}). 
%of GX~340+0. 
%The ordinate and the abscissa of CCDs indicate the flux ratios: 
%(a) [16-50 keV/3-16 keV] and  [3-5 keV/7-10 keV] respectively; 
%(c) [20-50 keV/3-10 keV] and  [10-20 keV/3-10 keV]; 
%(e) [10-20 keV/3-10 keV] and  [9-16 keV/3-5 keV]. 
%While HIDs demonstrate  flux ratios:  
%(b) [16-50 keV/3-16 keV];
%(d) [20-50 keV/3-10 keV];
%(f) [7-10 keV/3-5 keV] 
%versus flux (3-50 keV) measured in units of $10^{-9}$ erg s$^{-1}$cm$^{-2}$.
%}
%\label{ccd_evolution_sp rxte vs HB-NB-FB}
%\end{figure}

\newpage

% 
% Figure 10
%

\newpage

% 
% Figure 13
%

\newpage

% 
% Figure 14
%

\newpage

% 
% Figure 15
%

\newpage

% 
% Figure 16
%

\newpage

% 
% Figure 17
%

\newpage

% 
% Figure 18
%

\end{document}